%% file: MA.tex
\pdfoutput=1

\documentclass[12pt,twoside,final]		{hu}

\usepackage	[utf8]                  			{inputenc}
\usepackage [hidelinks]						{hyperref}

\usepackage	[T1]                    			{fontenc}
\usepackage                             		{color}
\usepackage                         		    {amsmath}
\usepackage									{amssymb}
\usepackage                         		    {graphicx}
\usepackage	[english]             	  		{babel}
\usepackage                    	       		{url}
\usepackage	[square,numbers]					{natbib}
\usepackage									{epsfig,bm, float}
\usepackage                             		{lmodern}
\usepackage                             		{algorithmicx}
\usepackage									{algorithm}
\usepackage	[noend]							{algpseudocode}
\usepackage									{enumitem}
\usepackage	[nottoc,notlof,notlot]			{tocbibind}
\usepackage									{makecell}
\usepackage									{tabularx}
\usepackage									{listings}
\usepackage									{fancyvrb}
\usepackage	[labelfont=bf,textfont=bf]		{caption}
\usepackage									{mathtools}
\usepackage									{textcomp}
\usepackage									{booktabs}
\usepackage									{threeparttable}
\usepackage									{adjustbox}
\usepackage									{multicol}
\usepackage									{subfigure}

\definecolor{darkblue}{rgb}{0.0,0.0,0.4}
\definecolor{darkgreen}{rgb}{0.0,0.4,0.0}

\graphicspath{{./}}

\renewcommand{\bibname}{Whatever floats your boat}
\renewcommand*{\bibfont}{\small}
\def\sline{\def\baselinestretch{1.0}\large\normalsize}
\makeatletter
\def\BState{\State\hskip-\ALG@thistlm}
\makeatother
\makeatletter
\def\hlinewd#1{%
\noalign{\ifnum0=`}\fi\hrule \@height #1 %
\futurelet\reserved@a\@xhline}
\makeatother
\DeclarePairedDelimiter\abs{\lvert}{\rvert}%

\begin{document}

  \include{titlepages-eng} 
  \include{abstracts}

  \include{dedication}  
  \tableofcontents
  
  \include{contents}


      
    \include{references} 

    \include{acknowledgments}

    \include{deposition}


\end{document}

%% file: titlepages-eng.tex

\thispagestyle{empty}
\begin{center}
  \renewcommand{\baselinestretch}{1.00}
    \Large\bfseries\sffamily
    Optimizing Communication by Compression for Multi-GPU\\
    Scalable Breadth-First Searches\\
  \par
  \vfill
  \renewcommand{\baselinestretch}{1.50}
  \large\mdseries\sffamily
  This Master thesis has been carried out by Julian Romera\\
  at the\\  
  Ruprecht-Karls-Universität Heidelberg\\
  under the supervision of\\
  JProf Dr Holger Fröning
\end{center}\par
\vspace{5\baselineskip}
\newpage

\thispagestyle{empty}
\begin{center}
   \renewcommand{\baselinestretch}{1.50}
   \Large\sffamily
   Ruprecht-Karls-Universität Heidelberg\\
   \large Institut für Technische Informatik\\
   \par\vfill\normalfont
   Master thesis\\
   submitted by\\
   Julian Romera\\
   born in Madrid, Spain\\
   2016
\end{center}

\renewcommand{\baselinestretch}{1.00}\normalsize

%% file: abstracts.tex
\thispagestyle{empty}
\begin{center}
  \begin{minipage}[c][0.48\textheight][b]{0.9\textwidth}
    \small
    \textbf{
  Optimizing Communication by Compression for Multi-GPU\\
    Scalable Breadth-First Search:
    }\par
    \vspace{\baselineskip}
    \input{abstract-ger}
  \end{minipage}\par
  \vfill
  \begin{minipage}[c][0.48\textheight][b]{0.9\textwidth}
    \small
    \textbf{
    Optimizing Communication by Compression for Multi-GPU\\
    Scalable Breadth-First Search:
    }\par
    \vspace{\baselineskip}
    \input{abstract-eng}

  \end{minipage}
\end{center}

%% file: abstract-ger.tex

Die Breitensuche, auch \emph {Breadth First Search} (BFS) genannt, ist ein fundamentaler Bestandteil vieler Graph Operationen wie \emph {spanning trees}, \emph {shortest path} oder auch \emph {betweenness centrality}. Ihre Bedeutung steigt stets an, da sie immer mehr zu einem zentralen Aspekt vieler populärer Datenstrukturen wird, welche intern durch Graph Strukturen repräsentiert werden.
Die Parallelisierung des BFS Algorithmus erfordert die Verteilung der Daten auf mehrere Prozessoren und wie die Forschung zeigt, ist die Leistungsfähigkeit durch das Netzwerk begrenzt \cite{compressingbitmap-sieve}. Aus diesem Grund ist es wichtig Optimierungen auf die Kommunikation zu konzentrieren, was schluessendlich die Leistung dieses wichtigen Algorithmus erhöht.
In dieser Arbeit wird ein alternativer Kompressionsalgorithmus vorgestellt. Er unterscheidet sich von existierenden Methoden dadurch, dass er die Charakteristika der Daten berücksichtigt, was die Kompression verbessert.
Ein weiterer Test zeigt zudem, wie dieser BFS Algorithmus von traditionellen instruktionsbasierten Optimierungen in einer verteilten Umgebung profitiert. Als letzten Schritt werden aktuelle Techniken aus dem Hochleistungsrechnen und andere Arbeiten in diesem Bereich betrachtet.

%% file: abstract-eng.tex

The \emph{Breadth First Search} (BFS) algorithm is the foundation and building block of many higher graph-based operations such as spanning trees, shortest paths and betweenness centrality. The importance of this algorithm increases each day due to it is a key requirement for many data structures which are becoming popular nowadays. These data structures turn out to be internally graph structures.  

When the BFS algorithm is parallelized and the data is distributed into several processors, some research shows a performance limitation introduced by the interconnection network \cite{compressingbitmap-sieve}. Hence, improvements on the area of communications may benefit the global performance in this key algorithm. \\

In this work it is presented an alternative compression mechanism. It differs with current existing methods in that it is aware of characteristics of the data  which may benefit the compression. 

Apart from this, we will perform a other test to see how this algorithm (in a distributed scenario) benefits from traditional instruction-based optimizations. 
Last, we will review the current supercomputing techniques and the related work being done in the area.


%% file: dedication.tex

{
\thispagestyle{empty}
  \thispagestyle{plain}
  \vspace*{.4in}
  \raggedleft
  
  \emph{Dedicated to Catren}\\
}
  

%% file: contents.tex
\chapter{Introduction}
\label{introduction}
\section{Introduction}
\label{subintroduction1}

The study of large Graphs datasets has been a very popular subject for some years. Fields such as Electronic engineering, Bioscience, the World Wide Web, Social networks, Data mining, and the recent expansion of \emph{``Big Data''} result in larger amounts of data each time, which need to be explored by the current and state-of-the-art graph algorithms. One result of this recent expansion of large-sized graph is the establishment of benchmarks that engage organisations in the development of new techniques and algorithms that solve this new challenge. \medskip

\emph{Graph 500}\footnote{\url{http://www.graph500.org}} is a new benchmark created in 2010  which serves this purpose. It uses the \emph{Breadth-first Search} (BFS) algorithm as a foundation. This benchmark contrasts with other well-known and more mature tests like LINPACK (used in the \emph{Top 500}\footnote{\url{http://www.top500.org/}} Challenge), in that the latter executes linear equations in a computation-intensive fashion. In contrast, the former makes use of graph based data-intensive operations to traverse the graph.

In this work we focus on possible optimizations of the \emph{Breadth-first Search} (BFS) algorithm.

\section{Motivation}
\label{motivation}

As stated before, as the growth of data volumes increases at a very high rate and the structures containing this data use an internal graph representation, the research in graph algorithms also increases.

Some of the current research done on the area of the \emph{BFS} algorithms is framed under the \emph{Graph 500} challenge. 

The work done on improving our previous implementation of the garph500 benchmarch (\texttt{``Baseline''}), as well as the research of other authors in this area, shows that there is a severe impact in the overall performance of the distributed version of the algorithm due to the latency introduced by the the data movement between processors.     

\section{Contributions}
\label{contributions}

We believe that the conclusions presented in this work may benefit a broad variety of Breadth-first Search implementations. Compressing the data movement may alleviate the main bottleneck of the distributed version of the algorithm: the communication. In this work we make use of a Compressed Sparse Row (CSR) matrix, 2D data partitioning and Sparse vector multiplications (SpMVM). By the analysis of the  transmitted data (a SpMV vector represented as an integer sequence) we have been able to successfully achieve over 90\% in terms of data transfer reduction and over an 80\% of communication time reduction. All this is achieved thanks to specific numerical proprieties the the data of the analyzed graph (Section \ref{graphcomputations}). \medskip

Furthermore, we will also show how different techniques to reduce the instruction overhead may improve the performance, in terms of Traversed Edges per Second \emph{TEPS} (Section \ref{graphcomputations500}) and time, of a \emph{graph500} implementation. Finally, we will discuss further optimizations. Here follows the main contributions of this work:

\begin{enumerate}

	\item As mayor goal, we have used a \emph{Bit Packing} compression scheme (Section \ref{compressionalgorithms} with \emph{delta compression} on top, optimized for the SIMD instruction set of Intel\texttrademark \space x86 architecture. The reasoning and criteria of the used algorithm are discussed in further detail on Section \ref{sectioncompressionintegration}. Also, a scalability study before and after integrating the compression has been performed in sections \ref{compression} and \ref{sectionfinalresults}.

	\item In addition to the added compression library (which contains several compression codecs), two more libraries have been integrated (partially). With this we enable our graph500 application to test more codecs for other different graphs and problem sizes. This is described in further detail in section \ref{implementationsandlibraries}. The newly added implementations are not fully operative yet. These new packages offer benefits such as new compression codecs, different input vector sizes (which would allow bigger problem sizes and the avoidance of the performance penalty imposed by the conversion of the full vector, required by the current library), and a GPU-based implementation. This converts our graph500 implementation on a modular compression test-bench.

	\item As part of this work, the instruction overhead has been reduced. With this, we have tested the effects of some common instruction transformation techniques at different scales in our BFS implementation. The used techniques have been described more in detail on Section \ref{sectionoptimizinginstructions} (\emph{Optimizing  instruction overhead})

	\item An external instrumentation (\emph{Score-P}) has been added to help with the detection of the bottlenecks, optimization of our code and validation of the final results. The reasoning for the profiler selection criteria is discussed in Section \ref{sectioninstrumentation}. As a summary, the selected profiler provides a fine-grain way to gather a high level of details with low overhead \footnote{\url{http://www.vi-hps.org/projects/score-p/}}.
	
	\item Some other tasks have been performed to our \emph{Graph 500} implementation. Briefly, these have been: 

	\begin{itemize}[label={--}]
		\item The application's built process has been improved to reduce the building time. Also, a pre-processing of the compiled code allows a more fine-grain tuning. For this purpose we have chosen Maketools \footnote{\url{https://www.gnu.org/software/make/}}. As result, we detect the maximum capabilities of the target system at pre-compile time.  
	
		\item We have created scripts to automate the visualization and interpretation of the results. For this we combine c/c++ with R. 
  
	\end{itemize}
\end{enumerate}

\section{Structure of the thesis}
\label{introductionstructurethesis}

The thesis is formed by 4 main parts (i) a background section about the involved elements of the thesis. (ii) An analysis of our problem with two possible solutions: compression and instruction overhead reduction. (iii) Finally in the last part, we discuss the result and see how much do they mach with the purposed solutions for our problem. 

\chapter{Background}
\label{background}

\begin{itshape}
As the background of optimizations made to a Graph 500 application, a short timeline of the history of Supercomputing will be listed to place the  cited concepts on the time. Some of the core optimizations technologies, used in both the compression and the main Graph 500 implementation will be reviewed. 

Regarding Supercomputing, it will be provided a background on some of its core concepts: High Performance Computing (HPC), General Processing Graphic Processing Units (GPGPU) and Message Passing Interface (MPI). 

Last, concepts about graphs (in a Linear Algebra context), and the datasets used often on graphs literature will also be described.
\end{itshape}

\section{Milestones in supercomputing}
\label{sectioncomputererahistory}

As an introduction to Supercomputing, some of the personal names and milestones of its history\footnote{\url{http://www.computerhistory.org}} are listed below. Many of these will be referenced along in this work.

\begin{itemize}
	\item \textbf{1930's} - Fist theoretical basis of computing. 
		\begin{itemize}
			\item \emph{\textbf{1936}: Alan Turing} develops the notion of a ``\textbf{\emph{Universal machine}}'' through his paper ``On Computable Numbers'' \cite{turing1936}. It is capable of computing anything that is computable.
		\end{itemize}
	\item \textbf{1940's} - The foundation of computing.
		\begin{itemize}

			\item \emph{\textbf{1945}: John von Neumann} introduces the concept of a stored program in the  draft report on the EDVAC design. This way he creates the design of a sequential computer \cite{vonNeumann} (known as the \textbf{\emph{von Neumann architecture}}).
			\item \emph{\textbf{1947}: Bell Laboratories} invents the \textbf{\emph{Transistor}}.
			\item \emph{\textbf{1948}: Claude Shannon} writes ``A Mathematical Theory of Communication'' (Sec. \ref{entrophy} ) \cite{shannon48} 
		\end{itemize}
	\item \textbf{1950's} - The first commercial general-purpose computers are born.  	
		\begin{itemize}
			\item \emph{\textbf{1951}: UNIVAC-1} is launched. This is the very first commercial \emph{general-purpose} computer. Even though it is general-purpose, it is focus on home use. 
			\item \emph{\textbf{1953}: IBM} announces Model 650. Stores the data on rotating tape. This is the first mass-produced computer.
			\item \emph{\textbf{1953}: IBM} announces Model 701. The first \emph{Mainframe}\footnote{These computers, also referred as \emph{``Big Iron''}, receive their name for the shape of their cabinet structure. They are often used in statistical computation, banking transactions, etc.} 
			\item \emph{\textbf{1956}: UNIVAC} announces an UNIVAC built with transistors. 
		\end{itemize}
	\item \textbf{1960's} - During this period the IBM Corporation dominated the general purpose early computing industry. This was the era of the \emph{Mainframes}. There are new additions to the foundation of computing.
		\begin{itemize}
			\item \emph{\textbf{1965}: Gordon Moore} coins the \textbf{\emph{Moore's Law}}. This predicts the number of in-chip transistors throughout time.
		
			\item \emph{\textbf{1966}: Michael Flynn} proposes a classification of computer architectures. This is known as \textbf{\emph{Flynn's Taxonomy}}. It is still used nowadays.
			\item \emph{\textbf{1968}: \textbf{CDC 7600}} is launched by Seymour Cray. It is considered by many the first true Supercomputer.
		\end{itemize}
	\item \textbf{1970's} - First Supercomputers enter in scene. They increase their performance through \emph{Vector Processors Units} among other methods. 
		\begin{itemize}
			\item \emph{\textbf{1971}: Intel Corporation} launches its first chip: Intel 4004
			\item \emph{\textbf{1972}: Cray Research Inc.} is created by Seymour Cray. He is consider as the father of Supercomputing.
			\item \emph{\textbf{1975}: Cray Research Inc} completes the development of \textbf{\emph{Cray 1}} - The first Vector-processor Supercomputer.
			\item \emph{\textbf{1978}: Intel Corporation} introduces the first 16-bit processor, the 8086.
		\end{itemize}
	\item \textbf{1980's} - Many technological improvements. First IBM/Intel personal computers (PC). First Connection Machine Supercomputers.
		\begin{itemize}
			\item \emph{\textbf{1982}: Cray Research Inc} introduces the \textbf{\emph{Cray X-MP}} Supercomputer. This version uses Shared Memory and a vector-processor. It is a \emph{'cleaned-up'} version of the CRAY-1. Successive versions of this Supercomputer increase the number of CPUs, raise the clock frequency and expand the instruction size from 24 bits to 32.
			\item \emph{\textbf{1983}: Thinking Machines} introduces \textbf{\emph{CM-1}}, the first \emph{Connection Machine} (CM). It is based upon the SIMD classification.	
			
			\item \emph{\textbf{1984}: IBM} introduces the IBM PC/AT based on the chip Intel 80286. The chip works at 16 bits.
			\item \emph{\textbf{1985}: Intel Corporation} introduces the 80386 chip with 32-bit processing and on-chip memory management.     

		\end{itemize}
	\item \textbf{1990's} - Next step in the Supercomputer era. First Massively Parallel architectures. The home computer industry is based on \emph{'clones'} of the original IBM PC.		
		\begin{itemize}
			\item \emph{\textbf{1991}: Thinking Machines} introduces \emph{CM-5}, a new version of their \emph{Connection Machine} running on an RISC SPARC, and replacing the connection network of the previous CM-2. The CM-5 changes, this way, to a MIMD design. This is the first NUMA architecture.
		
			\item \emph{\textbf{1992}: Touchstone Delta} Is an experiment carried out in \emph{Caltech} with 64x Intel 8086 microprocessors. It opens a door to a new era of parallelism. This is later referred as \emph{``The attack of the killer micros''}. Here, a large number of Commercial off-the-shelf (COTS) microprocessors, invaded a world dominated by ``Big Iron'' Vector computers.		
			\item \emph{\textbf{1993}: Intel Corporation} releases ``Pentium'' chip. Personal computers continue to grow.
			\item \emph{\textbf{1993}: \textbf{top500.org}} ranking is created. It uses a linear algebra \emph{LINPACK benchmark}.
			\item \emph{\textbf{1997}: The Intel ASCI Red Supercomputer} was developed based on the \emph{Touchstone Delta} experiment. This Massively parallel Supercomputer was the fastest in the world until the early 2010's. 
			\item \emph{\textbf{1999}: IBM PowerPC 440} microprocessor is launched. This 32-bit RISC high performance core will be the main processor of many future \emph{HPP} architectures like the Blue Gene/L or Cray XT3 
		\end{itemize}
	\item \textbf{2000's} - The fastest Supercomputers are \emph{Massively Parallel} architectures. 	
		\begin{itemize}
			\item \emph{\textbf{2001}: \textbf{General Purpose GPU processing} (GPGPU)} begins its development by the advent of the programmable shaders and floating point units on graphics processors.    
			\item \emph{\textbf{2004}: Intel Corporation} announces the cancelation of two of their processors \cite{endmoore}. This is known as the \textbf{\emph{End of Frequency Scaling}}. The new speedups are based upon parallel techniques developed in the previous Supercomputing era. With the Multicore the personal computer industry also go parallel. 
		\end{itemize}
	\item \textbf{2010's} - The only Supercomputer architectures in \emph{Top500}\footnote{\url{http://www.top500.org/statistics/overtime/}}	 are \emph{Massively Parallel}: \emph{MPPs} and \emph{Clusters}.
		\begin{itemize}
			\item \emph{\textbf{2010}: \textbf{graph500.org}} challenge is created. It uses graph-based data insensitive computation over a \textbf{\emph{Breadth-first Search}} (BFS) algorithm.
			\item \emph{\textbf{2012}: \textbf{green.graph500.org}} challenge is created. The current energy awareness leads to this energy focused variation of the \emph{Graph 500} Challenge. This new benchmark is also based upon the BFS algorithm but reflexes the power efficiency in the results.
		\end{itemize}

\end{itemize}  

\section{Global architectural concepts}
\label{sectionarchitecturalconcepts}

Continuing with the background about High Performance Computing (HPC) and related technologies like GPGPUs, we will introduce in this section some concepts which will be referenced in the document. This concepts encompass the internal methods used in the compression algorithm (\emph{SIMD} \slash \space Vectorization \cite{lemire}), other optimizations made to the selected compression algorithm (\emph{Super-Scalar} optimizations, improvements on the \emph{Pipelinig} \cite{heman}). These concept will also be referenced in this Background section for a better understanding of \emph{HPC, GPGPUs} and Supercomputing in general.

\subsubsection{The Von Neumann design }
\label{sectionflynns}

It was in 1945 when the physicist and mathematician John von Neumann and others, designed the first programmable architecture. It was called \emph{Von Neumann} architecture and had an intrinsically serial design. In this, there is only one processor executing a series of instructions and the execution flow occur in the same order as it appears in the original program \cite{vonNeumann}. 

\subsubsection{Moore's forecast }
\label{sectionmooreslaw}

\begin{figure}[hbt]
  \centering
  {\includegraphics[width=0.80\textwidth,viewport=0pt 0pt 480pt 450pt,clip]{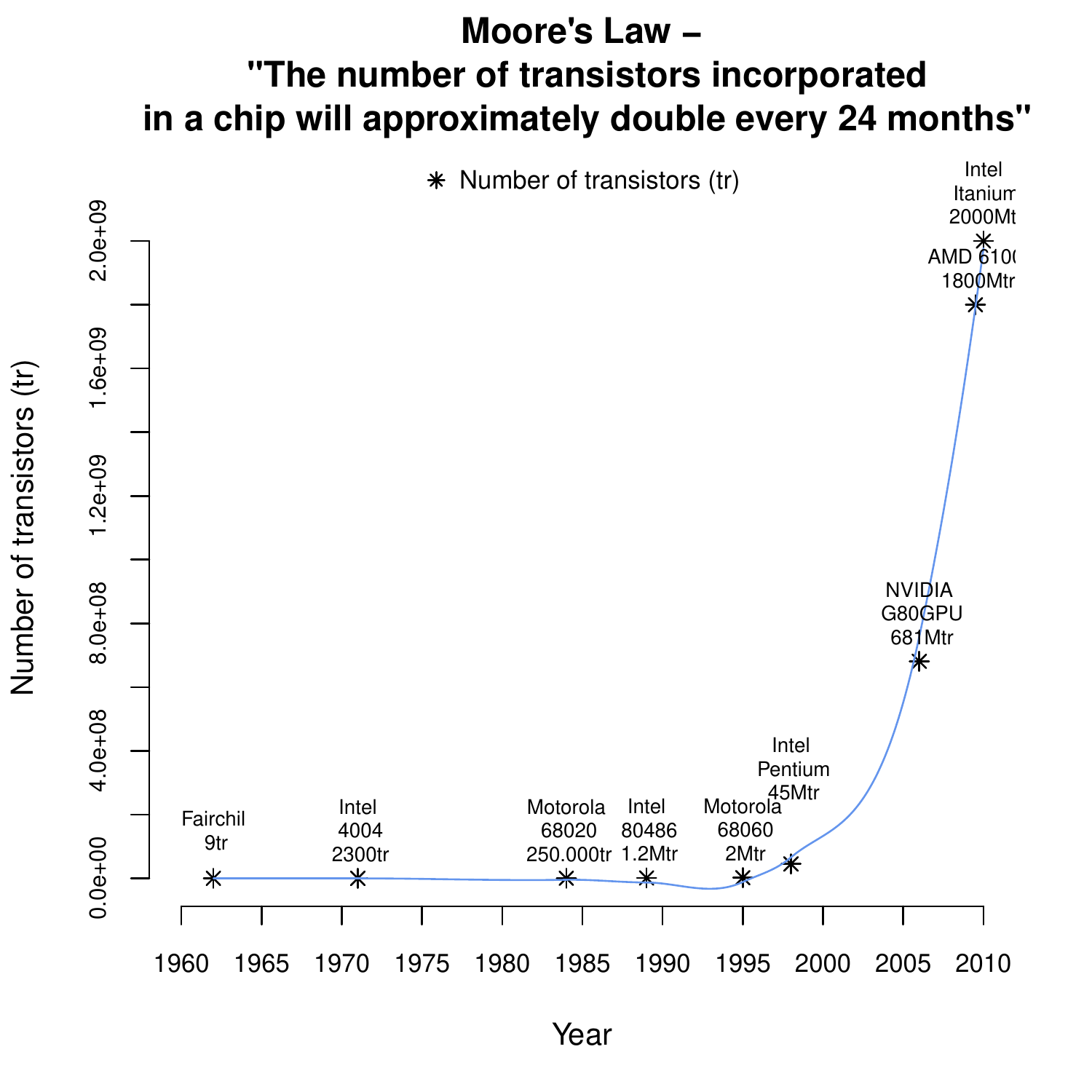}}%
  \caption{Moore's Law \cite{audio} }
  \label{fig:moorecurve}
\end{figure}

Gordon Moore, in 1965, created a forecast for the number of in-chip transistors on a silicon microprocessor \cite{moores65}. It was able to predicted the number of transistors on a chip troughtout the time:  ``In-chip transistror number would double every 2 years'' (Figure \ref{fig:moorecurve}).  \medskip 

This prediction was true for a long time and the clock frequency of  microprocessors was increasing due to deeper \emph{Pipelines} and more transistors. In the early 2000's the micro processors reached a temperature limit \cite{moores65}. Because of this, clock frequency has not been since 2005 (Sec. \ref{sectionhpchistorymulticorerevolution}). \medskip

This, has ultimately led to Multi-core technology (a chip with multiple cores) allowing a similar speed. As a result, nowadays software applications are required to have parallelism to benefit from this \cite{superscalarfuture}. \medskip

\subsubsection{Pipeline instruction parallelism  }
\label{pipelining}

The \emph{Instruction Pipelining} technique uses \emph{instruction-level} parallelism inside a single processor. The basic instruction cycle is chopped into a series which is called a \emph{Pipeline}. This allows a quicker execution throughput. Put simply, instead of processing each instruction sequentially (completing the instruction before the next one), each instruction is split into sub steps. These sub steps can be executed in parallel.

\subsubsection{Superscalar architectures }
\label{sectionscalarprocessor}

A \emph{Superscalar processor} is an architecture in which the processor is able to issue multiple instructions in a single clock. This is achieved using redundant facilities to execute an instruction. 

Each of the superscalar replicated execution units is a resource within a single CPU, such units may be an \emph{Arithmetic Logic Unit}, a \emph{Vector Processing Unit}, bitwise operators, or sets of CPU registers (Intel\texttrademark \space Multithreading). This excludes the case of the replicated unit being a separate processing unit (for example other \emph{core} in a Multi-core processor).

Note that, while a superscalar design is typically also pipelined, \emph{Pipelining} and \emph{Superscalar architectures} are different concepts.

\subsubsection{Flynn's taxonomy }
\label{sectionflynns}

In 1996, Michael Flynn creates the earliest classification systems for parallel and sequential computers \cite{flynn}. The categories in the taxonomy may be seen in Table \ref{flynnstaxonomy} \medskip

\begin{table}[htb]
\centering
\adjustbox{width=0.95\columnwidth}{%
\begin{tabular}{@{}lcc@{}}
\specialrule{1.5pt}{1pt}{1pt}
                      & \multicolumn{1}{l}{Single Instruction Stream} & \multicolumn{1}{l}{Multiple Instruction Streams} \\ 
                      \specialrule{1.5pt}{1pt}{1pt}
Single Data Stream    & SISD                                          & MISD                                             \\
Multiple Data Streams & SIMD                                          & MIMD                                             \\ 
\specialrule{1.5pt}{1pt}{1pt}
\end{tabular}
}
\caption{Flynn's Taxonomy}
\label{flynnstaxonomy}
\end{table}

He classified systems no matter if they were functioning using a single set or multiple sets of instructions. Also to whether (or not) the instructions were using a single set or multiple sets of data. \emph{Flynn's Taxonomy} is still in use today. \medskip

According to the authors Patterson and Hennessy, ``Some machines are hybrids of these categories, but this classic model has survived because it is simple, easy to understand, and gives a good approximation'' \cite{flynnisamix}. \medskip

Two terms in Flynn's Taxonomy relevant to this work: \emph{SIMD} and \emph{MIMD} are described below.

\begin{enumerate}
\setcounter{enumi}{0}

\item \textbf{Simple Instruction Multiple Data (SIMD)}

In a \emph{SIMD} architecture, the parallelism is in the data. There is only one program counter, and it moves through a set of instructions. One of these instructions may operate on multiple data elements in parallel, at a time. 

Examples in this category are a modern \emph{GPGPU}, or \emph{Intel SSE} and \emph{AltiVec} Instruction-sets.

\item \textbf{Multiple Instruction Multiple Data (MIMD)}

\emph{MIMD} are autonomous processors simultaneously executing different instructions on different data. \medskip

Distributed systems are generally categorized as \emph{MIMD} architectures. In these systems each processor has an individual memory. Also, each processor has no knowledge about the memory in other processors. In order to share data, this must be passed as a message from one processor to another. 

Examples of this category are a \emph{Multicore-superscalar} processor, a modern \emph{MPP}, or a \emph{Cluster}.

As of 2015, all the \emph{Top500} Supercomputers are within this category.
\end{enumerate}

\subsubsection{Types of parallelism according to their level of synchronization }
\label{sectiontypeparallfinegrain}

Other way to classify the parallelism is by the concept of \emph{Granularity}. In this context we relate the amount of computation with the amount of communication due to synchronization. This classification would be as follows:

\begin{itemize}
	\item \textbf{fine-grained parallelism} In this class of granularity the tasks have small computation and high amount of synchronization. As a  downside of this level of synchronization the overhead due to  communication is bigger.
	\item \textbf{coarse-grained parallelism} In this type of granularity, the tasks are bigger, with less communication between them. As a result the overhead due to communication is smaller. 
\end{itemize} 

In order to attain a good parallel performance, the best balance between load and communication overhead needs to be found. If the granularity is too fine, the application will suffer from the overhead due to  communication. On the other side, if the granularity is too coarse, the performance can suffer from load imbalance.

\section{Clusters and High Performance Computing}
\label{sectionclustersandhpc}

High Performance Computing (HPC), or supercomputing, is a kind of computation focused in high consuming computational operations where specific parallel paradigms and techniques are required. Some examples of these are predictive models, scientific computation, or simulations. The types of Supercomputers (and their features), and how they evolved in  history until today, are defined in this section.

\subsection{Hardware architectures }
\label{sectionclustershwarchs}

\begin{figure}[hbt]
  \centering
  {\includegraphics[width=0.80\textwidth,viewport=0pt 0pt 900pt 600pt,clip]{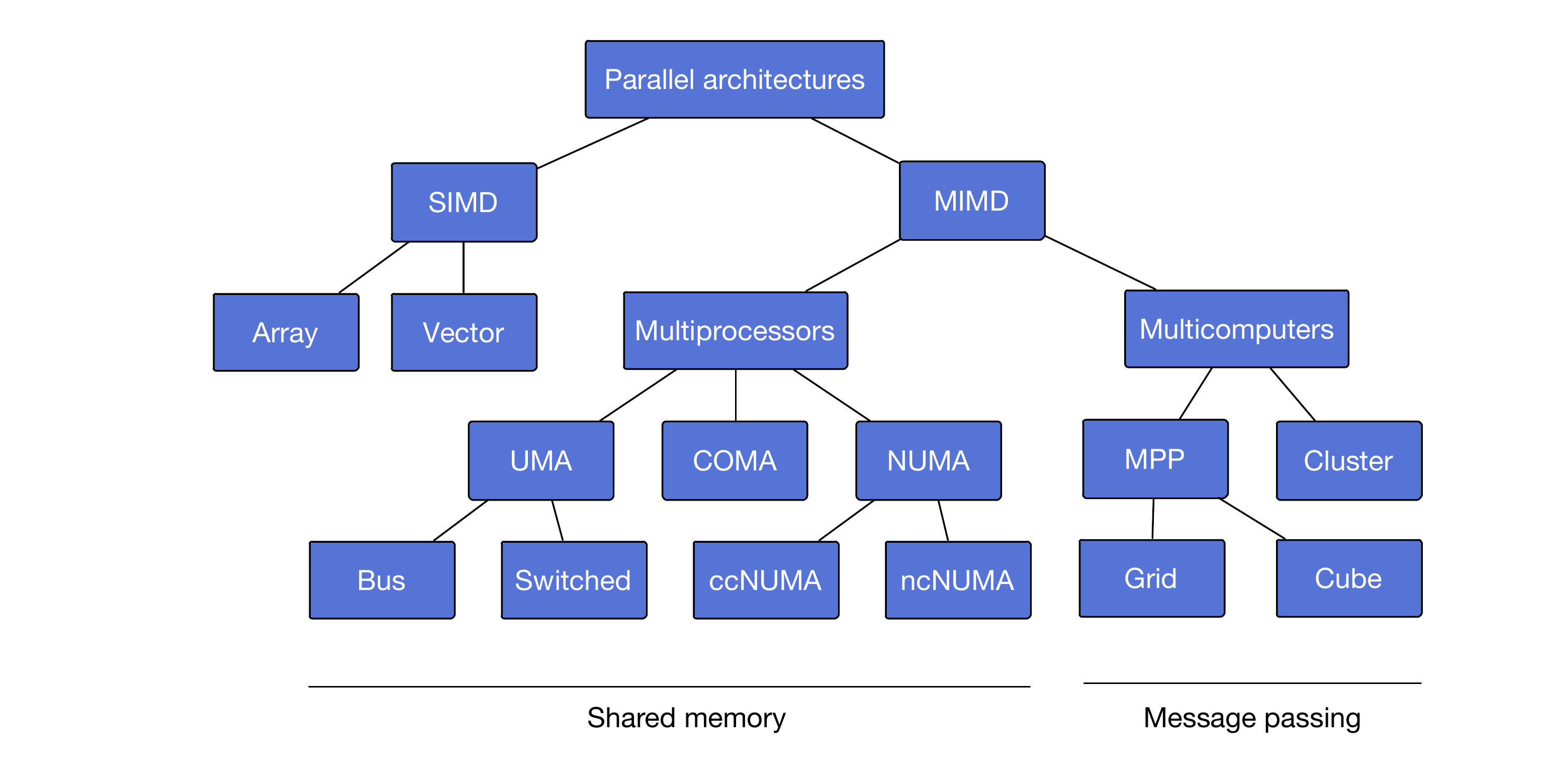}}%
  \caption{Architectural classification in HPC \cite{architecturesTaxonomies}.}
  \label{fig:archclassification}
\end{figure}

Based upon the taxonomy purposed by Flynn we show a general classification of the HPC architectures. This can be see in Figure \ref{fig:archclassification}.

\subsection{Interconnection networks in HPC }
\label{sectionclustershwarchs}

As this topic is very broad we will only review a concept related to networking, GPGPU, hardware architectures and in an extend, to this work. 

The \emph{Remote Direct Memory Access} (RDMA) is a technique created to solve a problem in networking applications. First, the problem is the performance penalty imposed when a system (sometimes, a processor unit within the main system) accesses the memory of another system (sometimes the original one) involving the Operating system (e.g. CPU, caches, memory, context switches) of the first one. The solution for this problem consists of the access via the network device to the peer system. 

Some implementations of this technique use hardware specific vendors as Infiniband\texttrademark \space , and others give support to already existing technologies as Ethernet, such as RoCE and iWarp.

The specific BFS implementation based on GPGPU, in which multiple GPUS might be allocated within the same system, would suffer a high penalty without this technique. In our tests, the RDMA support is enabled and managed by the Message Passing Interface \emph{MPI} implementation (Section \ref{messagepassinginterfacempi})

\subsection{The classification of supercomputers used in HPC  }
\label{sectionhpchistorysupercgoparallel}

The first computers in the market were mainly built for governmental, scientific or military purposes. It could be stated that computing started with Supercomputing. Here we list a classification of supercomputers according to similar design characteristics and sorted chronologically.

\subsubsection{Vector supercomputers}
\label{sectionhpchistorysupercgoparallelearlydays}

These were the first to appear \cite{audio}. In these architectures, the parallelism was achieved with the help of \emph{Vectorizing Compilers}. The applications needed to be adapted for their use with these compilers. 

In this scenario some loops could sometimes be vectorized through annotations in the language. Other loops needed to be arranged in order to remove the dependencies. 

The main benefits of these Supercomputers were for programs that would involve a lot of array processing. 

They enter in the Flynn's Taxonomy (Table \ref{flynnstaxonomy} page \pageref{flynnstaxonomy}) under the \emph{SIMDs} category and unlike other \emph{SIMD} architectures like the (described in the next section) \emph{Connection Machines}, the modern \emph{GPGPUs} or the \emph{Intel SSE / AltiVec instruction sets} these used larger and variable vector sizes. \medskip

\subsubsection{Data-parallel architectures}
\label{sectionhpchistorysupercgoparalleldparallelarchs}

This was a brief moment in time. In this category enter the initial \emph{Connection Machines}. These architectures also enter in the Flynn's \emph{SIMD} category. 

Here, the data-parallel language will distribute the work. The algorithms were very clever and involved data-parallel operations. Examples of those operations were  segmented scans, Sparse Matrix operations or reductions. 

Some parallel languages were \emph{C*}, \emph{CMF} (\emph{Connection Machine Fortran}) or \emph{*Lisp} \cite{audio}.

\subsubsection{Shared Memory}
\label{sectionhpchistorysupercgoparallelsharedmem}

In these systems, several processors share a common memory space. The programming model in these systems is based on \emph{Threads} and \emph{Locks}.

The algorithms were some variation of Parallel Random Access Machine (PRAM) algorithm, or hardware-based algorithms like cache coherence implementations. This latter were supported by hundreds of technical papers. In this classification enter the sub-taxonomy listed in Figure \ref{fig:archclassification}. Two of this types will be described below. 

\begin{enumerate}
	\item in a \textbf{UMA} model all the processors share the physical memory uniformly and access time to a memory location is independent of which processor makes the request.
	\item in a \textbf{NUMA} model the memory access time depends on the memory location relative to the processor: the local memory of a processor will be accessed faster than its remote memory (memory local to another processor or memory shared between other processors). This type of architectures are common in cluster programming and have relevance in this work due to the new advances made over the distributed BFS algorithm for big cluster supercomputers. A diagram of the local and remote banks of memory may be seen in figure \ref{fig:numa}. This architecture will be referenced in section \ref{relatedwork}.

\begin{figure}[hbt]
  \centering
  {\includegraphics[width=0.7\textwidth,viewport=0pt 0pt 580pt 420pt,clip]{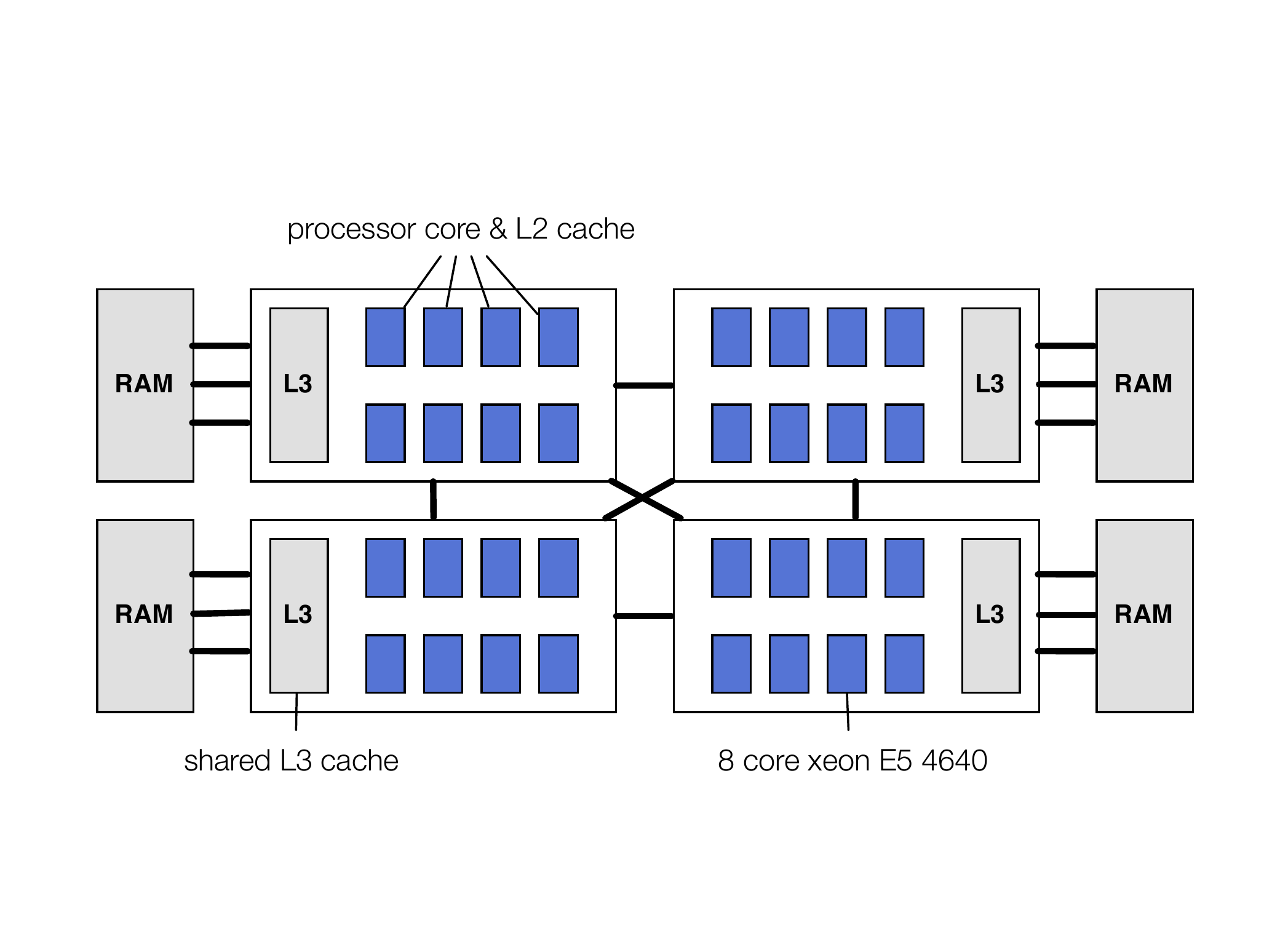}}%
  \caption{Detail of a NUMA architecture. Intel\texttrademark \space SandyBridge-EP system \cite{numaAwareFujisawa}.}
  \label{fig:numa}
\end{figure}

\end{enumerate}

\subsubsection{Multicomputers}
\label{sectionhpchistorysupercgoparallelmasiveparallelism}

The \emph{Multicomputers} replaced the \emph{Vector Supercomputers}. The Shared memory machines were still popular but in contrast with them, these new architectures provided thousands of processes rather than just tens. These two models are not incompatible and are often combined. The programming model was made using \emph{Message Passing}. In its early days each organization used its own message-passing implementation.  Later, the scientific community created an standard, \emph{Message Passing Interface} or simply \emph{MPI} \cite{audio}. From an algorithm point of view,  there had to be made complete changes to the code. This was called \emph{domain decomposition} and required a complete re-structuring of the program, which needed to be split.

These architecture designs enter in Flynn's \emph{MIMD} category. In this group there are two main classes: \emph{Massively Parallel Processor (MPP)} and \emph{Commodity Clusters} (or simple \emph{Clusters}). Both are described below. 

\begin{enumerate}

\item \textbf{MPP} \emph{Massively Parallel Processor (MPP)} are complex and expensive systems. An \emph{MPP} is a single computer system which sometimes is formed by many nodes. The nodes contain many connected processors which are tightly coupled by specialized interconnect networks (e.g: Hypercube, Mesh or Torus  \cite{torusinterconnect}) \cite{flynnisamix,dongarra}. Examples of MPPs are the IBM Blue Gene series or the Cray Supercomputers XT30 and XC40 among others. As it will be difined in the section Graph computations (Section \ref{graphcomputations}), these systems are referred as \emph{\textbf{Lightweight}} by the \emph{Graph 500 BoF}\footnote{\url{http://www.graph500.org/bof}}.  

	\item \textbf{Clusters} systems are usually cited as: ``a \emph{home-made} version of an MPP for a fraction of the price'' \cite{tanenbaum}. However this is not always true, as it can be the example of the Tianhe-2, in China. These, are based upon independent systems (Usually Shared Memory Systems) connected through high speed networks (e.g. Infiniband, Myrinet, Gigabit Ethernet, etc). 
\end{enumerate}

As of November 2015, the Top10 architectures in the \textbf{\emph{Top500}} \footnote{\url{http://top500.org/lists/2015/11/}} list are 2 Clusters and 8 MPPs (being the fastest, the Tianhe-2 Cluster in China) \medskip

In the Top10 of the \textbf{\emph{Graph500}} \footnote{\url{http://www.graph500.org/results_nov_2015}} list, as of November 2015, there is 1 Cluster and 9 MPPs (being the fastest, the K-computer MPP in Japan)

\subsection{The popularization of the multicore architectures}
\label{sectionhpchistorymulticorerevolution}

Due to the change in the manufacturing process of the silicon-chips, the rest of the world followed suit.

\begin{figure}[hbt]
  \centering
  {\includegraphics[width=0.65\textwidth,viewport=0pt 0pt 590pt 450pt,clip]{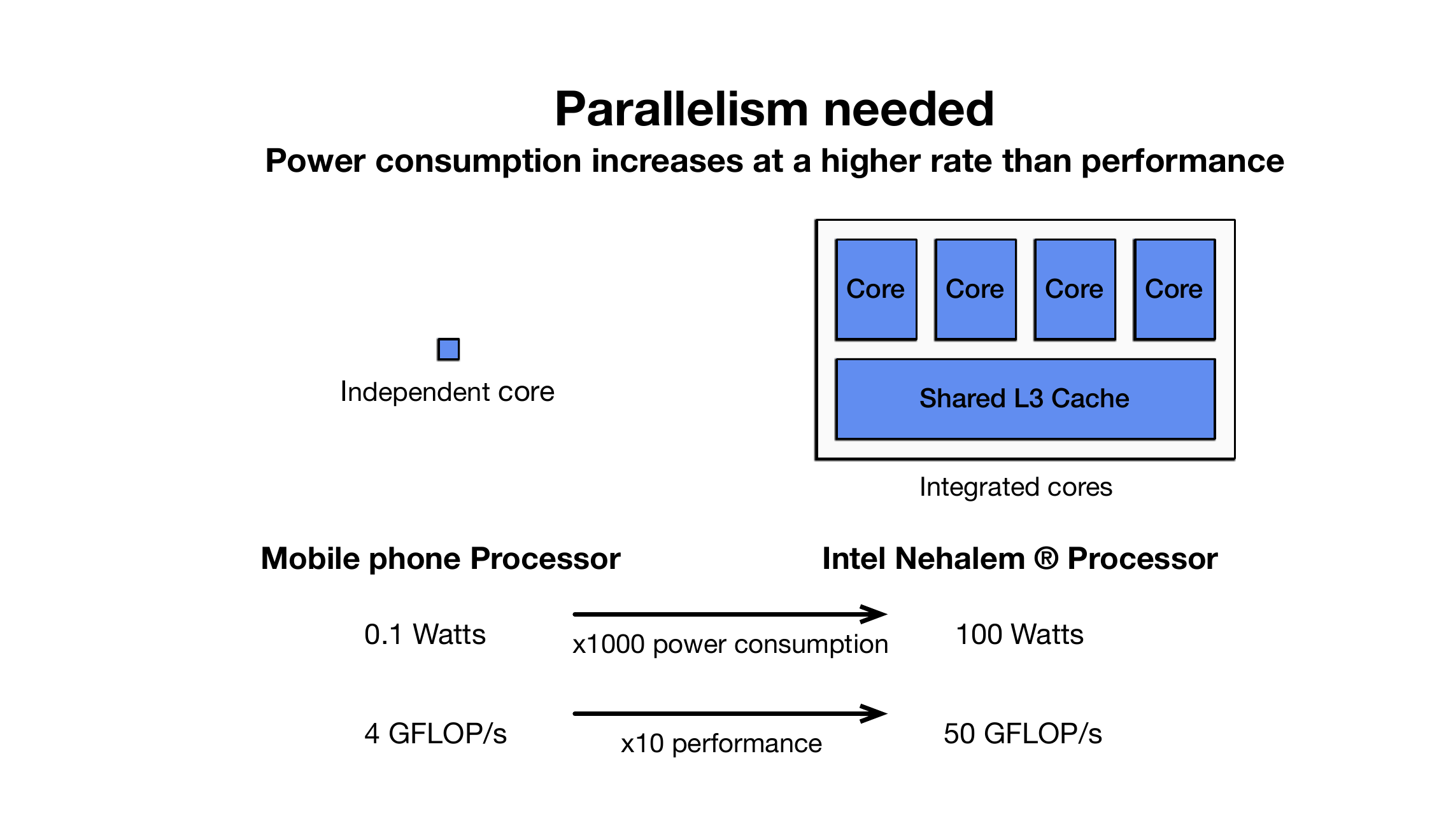}}%
  \caption{Power consumption increases at a higher rate than performance \cite{audio}.}
  \label{fig:moores2}
\end{figure}

The required power to enhance the performance (by increasing the area of the circuit block) varies at a higher rate than the performance does. (Figure \ref{fig:moores2}) \cite{audio}. 

One direct effect of the latter, is that the generated heat increases (proportional to the consumed power). As a result, alleviating the effects of this heat, and keeping the silicon chips within an operative range is difficult. The physical explanation of this is: 

The generated heat in a chip is proportional to its Power consumption \(P\) \cite{powerconsumption}, which is given by  \eqref{eq:powerconsumption}. On this, when the frequency increases also the power consumption and the heat rise. 

\(\textbf{C} \longrightarrow \) The \emph{Capacitance} being switched per clock cycle (proportional to the number of transistors)

\(\textbf{V} \longrightarrow \) Voltage 
 
\(\textbf{F} \longrightarrow \) The processor frequency
\[
P = C * V^{2} * F \tag{1}\label{eq:powerconsumption} 
\]

\begin{figure}[hbt]
  \centering
  {\includegraphics[width=0.50\textwidth,viewport=0pt 0pt 440pt 450pt,clip]{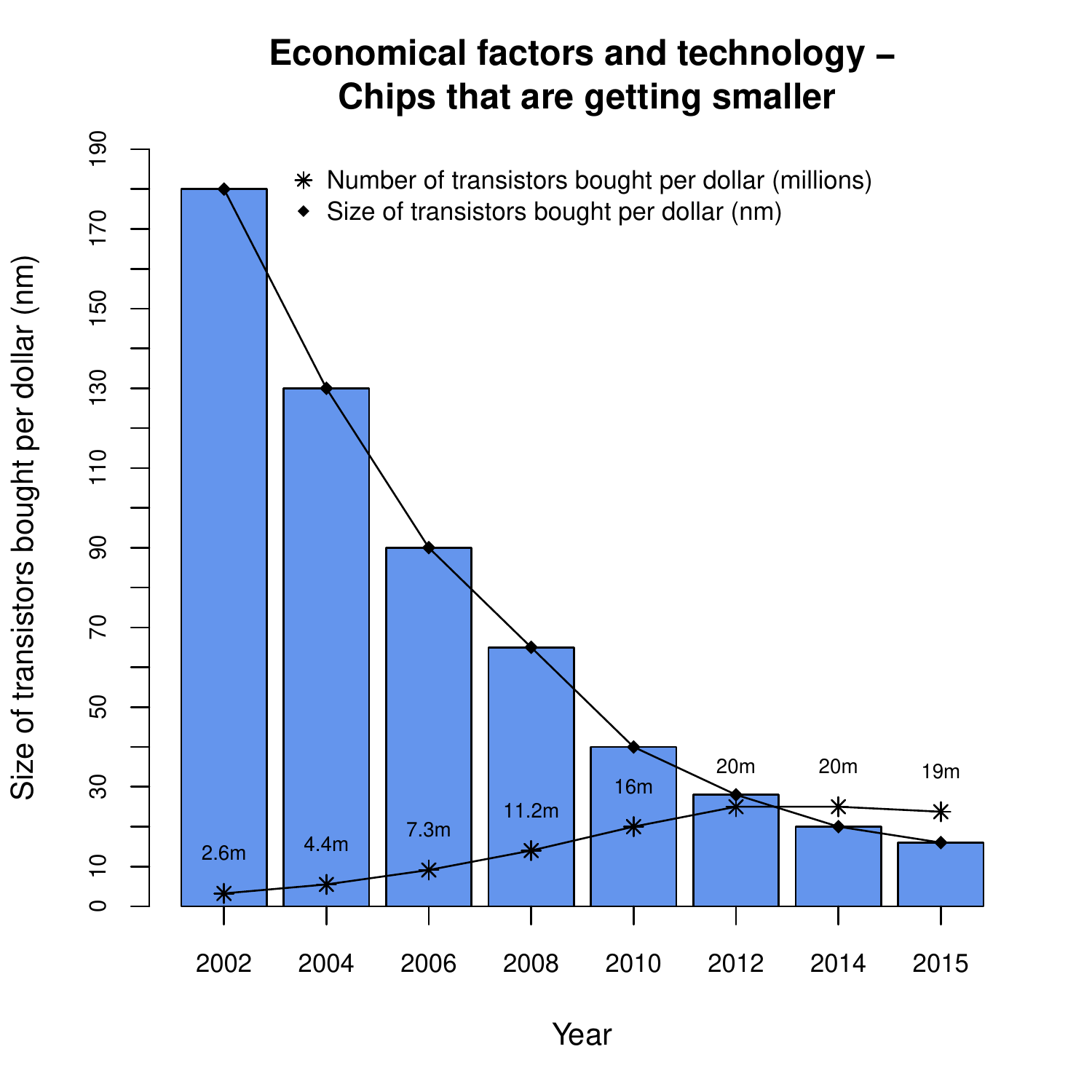}}%
  \caption{Manufacturing costs for smaller technologies become uneconomical \cite{audio}.}
  \label{fig:moores1}
\end{figure}

This is what ultimately led to Intel's cancellation of several microprocessors in 2004 \cite{endmoore}. Also, this moment in time is generally cited as the ``\emph{End of Frequency Scaling}''. As this stopped the increasement of the frequency, many cores started to be added into one single chip. There are also more reasons (economical) which affected the evolutions in this chip technology. For example, the relationship between the profits and the technology costs (smaller technologies are more expensive). This ultimately led to an slow down to new smaller technologies \cite{audio} (Fig. \ref{fig:moores1}). \medskip

From the point of view of the software, this evolution made \emph{threads} a popular programming model in the last years (as it had been learned a lot about of them with Shared Memory architectures, used in Supercomputing). Also, other programming models like the less restricted \emph{``fork-join'' threads} used in CILK, started to be more popular and practical.

\subsection{Energy efficiency}
\label{sectionhpchistorymulticorerevolution}

After going through the limitations that heat imposes to chips, and how the energy consumption rises with the in-chip number of transistors, it can be better understood why energy efficiency is becoming today one of the main design constrains in High Performance Computing. 

\begin{figure}[hbt]
  \centering
  {\includegraphics[width=0.50\textwidth,viewport=0pt 0pt 430pt 450pt,clip]{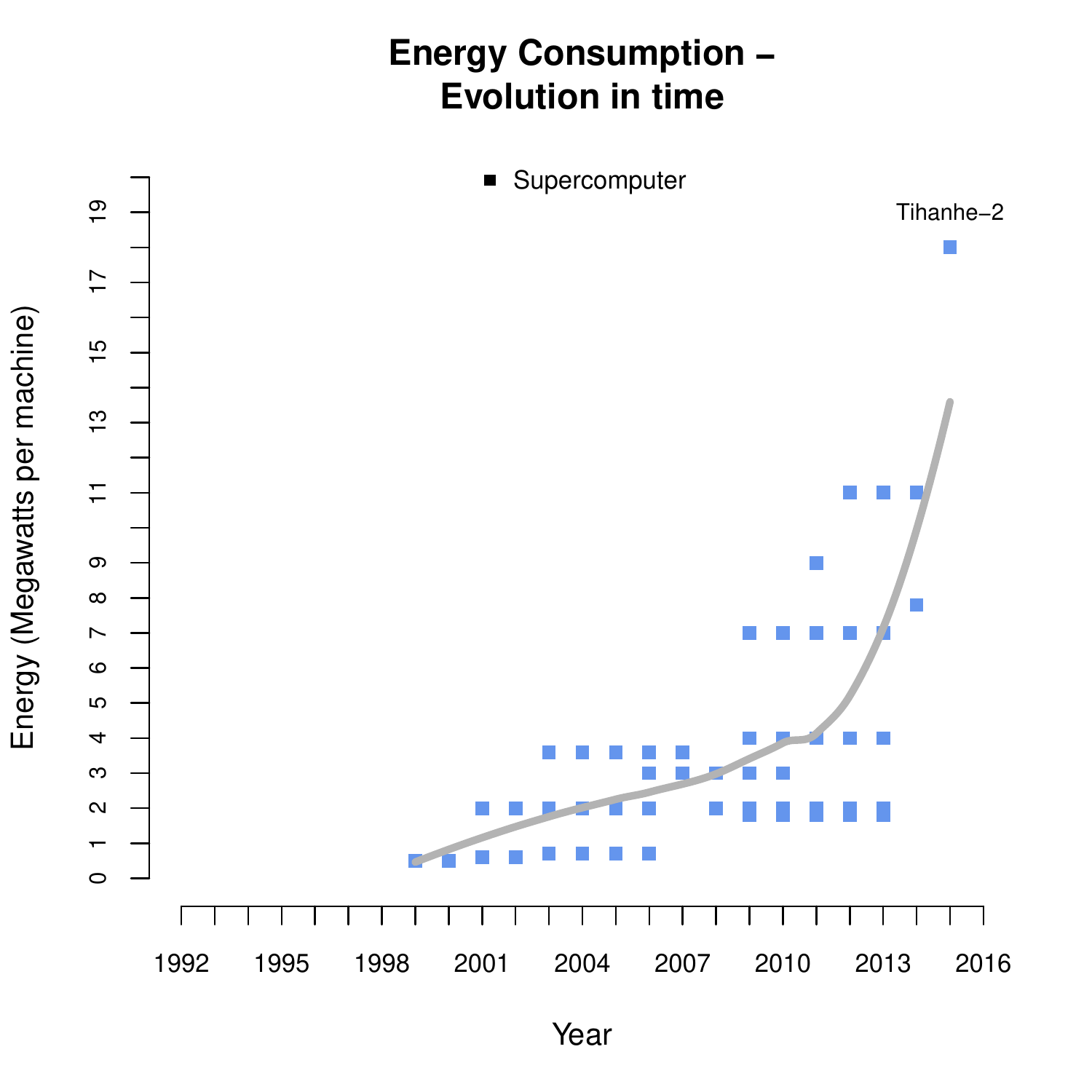}}%
  \caption{Power consumption increases at a higher rate than performance \cite{audio}.}
  \label{fig:energy}
\end{figure}

To illustrate the importance of this, a real example (using modern Supercomputers) will be is used. 

As of 2015, in the United States of America, the cost per megawatt is around 1 million dollars \cite{audio} and the consumed energy usually follows a \emph{chip scaling}. With this in mind, in the scenario that we are going to describe, in 2008 a 1 petaflop system used 3 megawatts to run. Following this scaling, in 2018 a 1 exaflop Supercomputer will require 20 megawatts. 

To contrast the importance of the energy saving plans, a 20 megawatts supercomputer would require an approximate budget of 20 Million dollars which was a quarter of the yearly budget destined to nursing in that country in 2015. \cite{audio}). 

The current energy cost in super computers is serious and not sustainable. In the last years researching on energy efficiency has moved also to algorithms and software design. As it will be illustrated in Section \ref{graphcomputations}, new benchmarks (like the \emph{\textbf{green.graph500}}\footnote{\url{http://green.graph500.org/}}) are being created to measure this \emph{Energy} impact of computation.  

\section{General Purpose Graphic Proccessor Units}
\label{graphicproccessorunitgpucomputing}

\begin{itshape}
Overview
\end{itshape}

\subsection{GPGPU architectures}
\label{secbackgroundgpuarchitectures}

\begin{figure}[hbt]
  \vspace{-10mm}
  {\includegraphics[width=0.8\textwidth,viewport=0pt 0pt 780pt 550pt,clip]{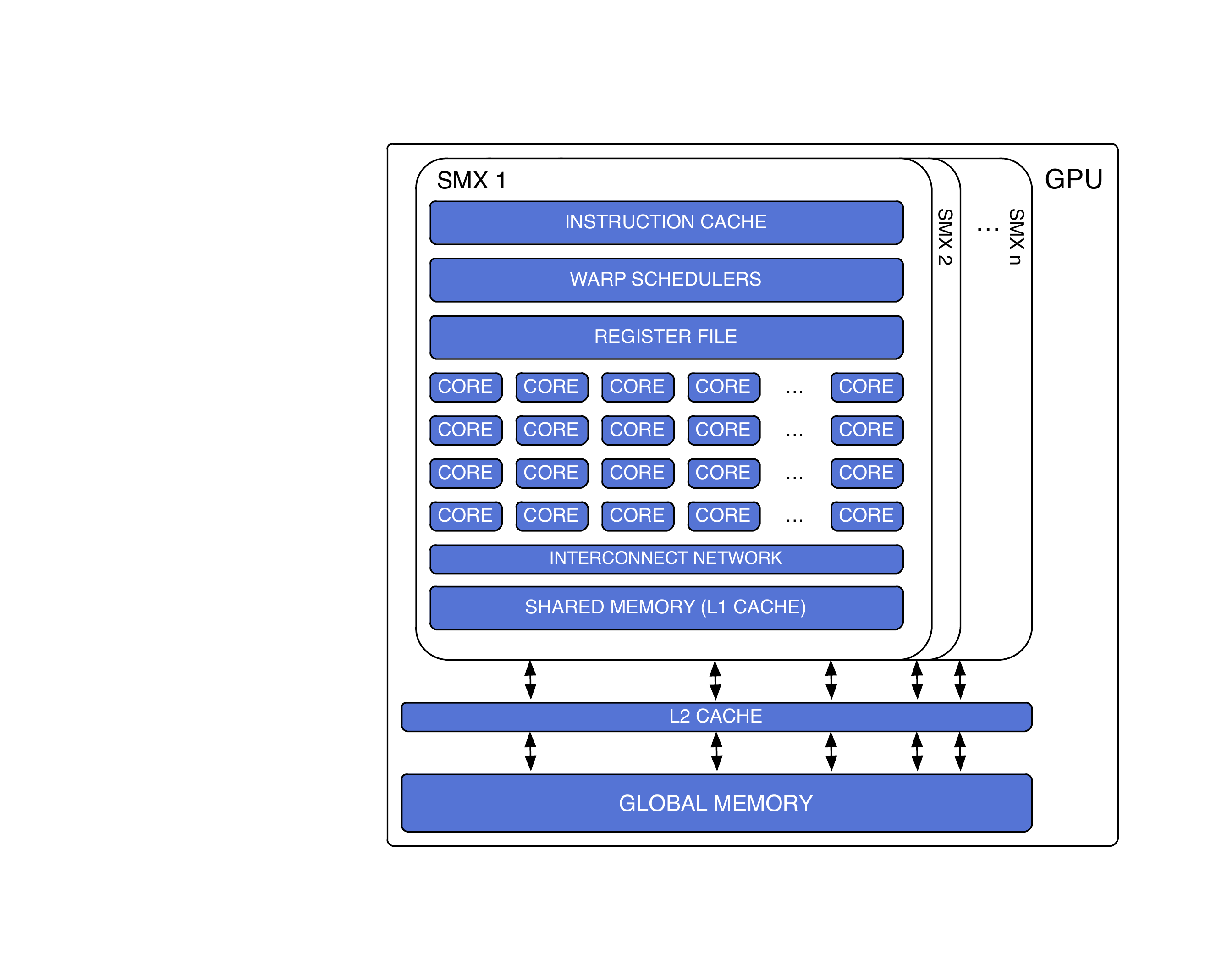}}%
  \vspace{-10mm}
  \caption{Simplified architecture of an NVIDIA\texttrademark \space Kepler GPU}
  \label{fig:nvidia}
\end{figure}

GPUS are widely used as commodity components in modern-day machines. In the case of NVIDIA\texttrademark \space architectures, a GPU consists of many individual \emph{execution units} (SMs/ SMXs), each of which executes in parallel with the others (Figure \ref{fig:nvidia}). During runtime, \emph{threads} on each execution unit are organized into thread blocks, and each block consists of multiple 32-thread groups, called \emph{Warp}. If threads within a Warp are set to execute different instructions, they are called \emph{``diverged''}. Computations in diverged threads are only partially parallel, thus reducing the overall performance significantly.

The GPU includes a large amount of device memory with high bandwidth and high access latency, called global memory. In addition, there is a small amount of shared memory on each execution unit, which is essentially a low latency, high bandwidth memory running at register speeds. Due to such massive amounts of parallelism, GPUs have been adopted to accelerate data and graph processing. However, this maximum potential parallelism is many times difficult to achieve.

\subsection{Regular and non-regular problems}
\label{secbackgroundgpuregular}


\begin{itemize}
	\item In \textbf{regular code}, control flow and memory references are not data dependent, for example SpMVM is a good example. Knowing only the source code, the input size, and the starting addresses of the matrix and vectors we can predict the program behavior on an in-order processor.

\item In \textbf{irregular code}, both control flow and memory addresses may be data dependent. The input values determine the program’s runtime behavior, which therefore cannot be statically predicted. For example, in a binary-search-tree implementation, the values and the order in which they are processed affect the control flow and memory references. Processing the values in sorted order will generate a tree with only right children whereas the reverse order will generate a tree with only left children.

\end{itemize}

Graph-based applications in particular tend to be irregular. Their memory-access patterns are generally data dependent because the connectivity of the graph and the values on nodes and edges determine which graph elements are computed next accessed, but values are unknown before the input graph is available and may change dynamically. 

\section{Message Passing Interface}
\label{messagepassinginterfacempi}

The Message Parsing Interface (MPI) resulted as a standardisation of the multiple implementations of communication protocols in early supercomputing.

MPI only defines an interface of calls, datatypes or restriction that are separately implemented by other organizations. Two popular implementations of MPI are OpenMPI and MPICH.

On MPI exist two types of communication calls:

\begin{itemize}
	\item \textbf{Point-to-Point calls}. Allow the direct data send and receive from / to other 	node. Some examples are MPI\_send, MPI\_isend (non-blocking) or MPI\_Recv.
	\item \textbf{Collective communications} allow us to send data from / to multiple nodes at a time. Some examples are MPI\_Reduce, MPI\_Allgatherv, MPI\_scatter. 
\end{itemize}
 
\section{Graph computations}
\label{graphcomputations}

\begin{itshape}

As examined in the \emph{Introduction section} (Section \ref{subintroduction1}) Graphs and Graphs computations are becoming more important each day.

In a first part of this section it will be reviewed the Graphs structures and their characteristics. \medskip

On a second part of the section it will be discussed the Graph 500 challenge. About the latter it will be described what is it main motivation, what parts conform a Graph 500 application, what is the basic algorithm in a non parallelized application. Lastly it will be discussed why the parallelization of this algorithm (BFS) is theoretically and computationally difficult.

Also other recent benchmark (Green Graph 500), which is very related with this one, will be briefly described.   

\end{itshape}

\subsection{Graphs}
\label{graphcomputationsgraphs}

The graph algorithms encompass two classes of algorithms: traversal algorithms and analytical iterative algorithms. A description of both follows:

\begin{itemize}
	\item \textbf{Traversal algorithms} involve iterating through vertices of the graph in a graph dependent ordering. Vertices can be traversed one time or multiple times. This class includes search algorithms (such as breadth-first search or depth-first search), single source shortest paths, minimum spanning tree algorithm, connectivity algorithms and so on and so forth.
	\item  \textbf{Analytically iterative algorithms} involve iterating over the entire graph multiple times until a convergence condition is reached. This class of algorithms can be efficiently implemented using the Bulk Synchronous Parallel (BSP) model. Algorithms in this class include page rank, community detection, triangle counting and so on.
\end{itemize}

We will focus in a parallel implementation of the traversal algorithm Breadth-first Search (BFS). This algorithms requires some form of data partition, concurrency control, and thread-level optimizations for optimal performance.

\begin{itemize}
	\item \textbf{Degree of a graph} is the number of edges connected to a vertex.
	\item \textbf{Distance between to vertices on a graph} is the value of the shortest path between to vertices.
\end{itemize}

\subsection{Graphs partitioning}
\label{graphcomputationsgraphspaart}

\subsubsection{Parallel SpMV}

\subsubsection{1D Partitioning-based BFS (Vertex)} In 1D data partitions each processor gets a row of vertices of the adjacency matrix. This kind of partitioning is the default one and has network communication complexity of \(\theta(n \times m)\), where \texttt{n} is the number of rows, and \texttt{m} is the number of columns.


 \[ A = \left[\begin{array}{cc}
	A_{1} \\ \hline
 	A_{2} \\ \hline
    \vdots \\ \hline
    A_{k}
	\end{array}\right] \]

\subsubsection{2D Partitioning-based BFS (Edge)} The 2D-partition based BFS is known to perform better than the 1D in graph with low degree. Unfortunately, for high degree graphs the situation becomes the opposite. First of all, as a pre-condition for doing this kind of partition that adjacency matrix must be symmetric in relation to the number of processors, i.e with 36 processors we need to be able to divide the matrix into \(6 \times 6\) blocks. 

 \[ A = \left[\begin{array}{ccc}
	A_{1,1}^{(1)} & \cdots & A_{1,C}^{(1)} 	\\ \hline
 	\vdots & \ddots &  \vdots  				\\ \hline
	A_{R,1}^{(1)} & \cdots & A_{R,C}^{(1)} 	\\ \hline
 	\vdots & \ddots &  \vdots  				\\ 
 	\vdots & \ddots &  \vdots  				\\ 
	A_{1,1}^{(C)} & \cdots & A_{1,C}^{(C)} 	\\ \hline
 	\vdots & \ddots &  \vdots  				\\ \hline
	A_{R,1}^{(C)} & \cdots & A_{R,C}^{(C)} 
	\end{array}\right] \]

\subsection{Graph 500 challenge}
\label{graphcomputations500}

Some concept attaining the Graph 500 challenge are described below.

\paragraph{Performance} the Traverse Edges per Second (TEPS) is used as the main performance metric of the Graph 500 applications. TEPS is proposed by the \emph{graph500} benchmark to report the throughput of supercomputers on graph processing.

\paragraph{Energy efficiency} is also a metric which measures the efficiency in terms of energy expressed as traversed edges per joule. Since platforms have heterogeneous hardware resources and different power consumptions, their performance (TEPS) is divided by their power to calculate the energy

\paragraph{Provided reference implementations} The \emph{Graph 500} provides five reference implementations with different characteristics. This implementations have the following characteristics 

\begin{enumerate}
	\item \textbf{Sequential} This reference implementation uses a sequential BFS algorithm and introduces no parallelism but the data vectorizations introduced by the compiler. It could fit for example a UMA architecture, using one only processor on its CPU, such as a mobile phone or a tablet.
	\item \textbf{OpenMP} The benefit of this implementation comes from the hand of the introduced thread parallelism which may be achieved with OpenMP and a multicore architecture. An example architecture which would fit in this category would be a laptop or a Personal computer.
	\item \textbf{Cray XMT} This implementation is specific for Cray\texttrademark \space MPPs. Takes advantage on the specific interconnectors of this HPC supercomputers.
	\item \textbf{MPI} On this implementation the parallelism is based on Flynn's MIMD category (Table \ref{flynnstaxonomy} and Figure \ref{fig:archclassification}). This is the problem is partitioned and divided into several processors through Message Passing Interface (Section \ref{messagepassinginterfacempi})
	\item \textbf{2D data partitioned MPI}.This implementation has been added with posteriority and is based on the 2D problem partitioning purposed in  \cite{generalAnalisysBFSpossibleCompressiom}. As the previous one distribute the load among processors using MPI. This reference implementation is the more structurally similar to our one.  .  
\end{enumerate}

All these implementations are based on CPU processors.

\subsubsection{Problem sizes}
\label{graphcomputations500sizes}

According to the Graph 500 consortium problem sizes enter into different categories depending on their size. The size is expressed in logarithmic scale and represents the number of total vertices inside the graph. An overview of the categories may be seen in table Table \ref{graph500sizes}.

As it will be described in section \ref{inputdata} and it is cited briefly here, The graph 500 challenge makes use of a Kronecker generator \cite{kroneker1} to create the graph. The generated graph (as required in this challenge) contains 16 edges per vertex (\emph{Edge factor}) . This latter is the reason why the Graph 500 applications works over sparse graphs.    

In the table \ref{graph500sizes} Real world graphs whit a higher number of edges per vertex (degree of the graph) have an scale greater than 30. As it was previously described in this same section a node with logarithmic scale of 30 would have \(2^{30}\) vertices and \(2^{34}\) Edges.

One note which will also be referenced next in the document is that the meaning of using an Sparse vector with elements of (SpMV) of 64-bit means that a Row (or column for symmetric 2d partitions) of the CSR matrix representing the whole graph, also has 64 elements. This would mean that the maximum represented graph size could have \(2^{64}\) Vertices. If we think about this in terms of compression. Is our compressor algorithm works over 32-bit integers (integer is other way of calling the bitmap SpMV) would mean that the maximum archivable graph size (in Table \ref{graph500sizes}) would be \texttt{small} size.

\begin{table}[hbt]
\centering
\begin{tabular}{llll}
\hlinewd{1.2pt}
Problem class     & Scale & Edge factor & Approx. storage size in TB \\ \hlinewd{1.2pt}
Toy (level 10)    & 26    & 16          & 0.0172                     \\
Mini (level 11)   & 29    & 16          & 0.1374                     \\
Small (level 12)  & 32    & 16          & 1.0995                     \\
Medium (level 13) & 36    & 16          & 17.5922                    \\
Large (level 14)  & 39    & 16          & 140.7375                   \\
Huge (level 15)   & 42    & 16          & 1125.8999                  \\ \hlinewd{1.2pt}
\end{tabular}
\caption{Problem sizes for a \emph{Graph 500} BFS Benchmark}
\label{graph500sizes}
\end{table}

\subsubsection{Structure and Kernels}
\label{graphcomputations500kernels}

\begin{algorithm}
\caption{A Graph 500 application in pseudocode}\label{g500globalalg}
\begin{algorithmic}[1]
\State \textbf{Graph generation} (not timed) ``Graph Generation as a list of Edges''
\State \textbf{Kernel 1} (timed) ``Graph Construction and conversion to any space-efficient format such as CSR or CSC''
\For {i in 1 to 64}
\State \textbf{Kernel 2} (timed) ``Breadth First Search (timed)''
\State \textbf{Validation} (timed) ``Validation for \emph{BFS tree} using 5 rules''
\EndFor
\end{algorithmic}
\end{algorithm}

A graph 500 application is formed by 4 different steps, some of which are timed and some of which not. Also, has two differentiated core parts called \textbf{kernels}, in the first one improvements over the structure containing the graph are allowed and timed. The second kernel contains the code of the BFS algorithm. This latter step is repeated 64 times and on each of them the results are verified. 

As a result of a benchmark are required some statistical parameters, such as the mean execution time or the Traverse Edges per Second metric (TEPS) calculated as an harmonic mean of all the 64 required iterations. The algorithm of the benchmark may be seen as pseudocode in table \ref{g500globalalg}.

\subsubsection{The \emph{Green Graph 500} Challenge}
\label{graphcomputations500green}

The \emph{Green Graph 500}\footnote{\url{http://green.graph500.org/}} challenge is an Energy-Aware Benchmark launched in the year 2012. It is based on the similar technical principles than the \emph{Graph 500}. Some aspects like the metrics (TEPS\slash Watt), the kernels structure, or the submission rules \cite{green500rules} differ from the previous benchmark. \medskip

As it was discussed in the previous section \emph{Looking for energy efficient technologies} (Section \ref{sectionhpchistorymulticorerevolution}), energy is becoming a concern between the HPC community due to the increasingly maintenance costs.  

Due to this fact new improved algorithms and implementations are starting to be focused also on a energy-efficient implementation. 

\section{Input data}
\label{inputdata}

\subsection{Synthetic data}
\label{syntheticdata}

In the graph 500 challenge's algorithm of table \ref{g500globalalg}, the step on line \texttt{1} generates a list of edges, that in next steps will be converted into other structures. 

The graph 500 consortium sets some specifications for the generated graph and for the used generator. The generator \cite{kroneker1} builds open scale graphs which have similar statistic distributions to the graphs observed in the real world. the specifications on the Graph 500 challenge require, apart from the usage of a Kronecker generator, a number of 16 edges per vertex.

\begin{figure}[H]
  \centering
  \vspace{-6mm}
  {\includegraphics[width=0.65\textwidth,viewport=0pt 0pt 460pt 450pt,clip]{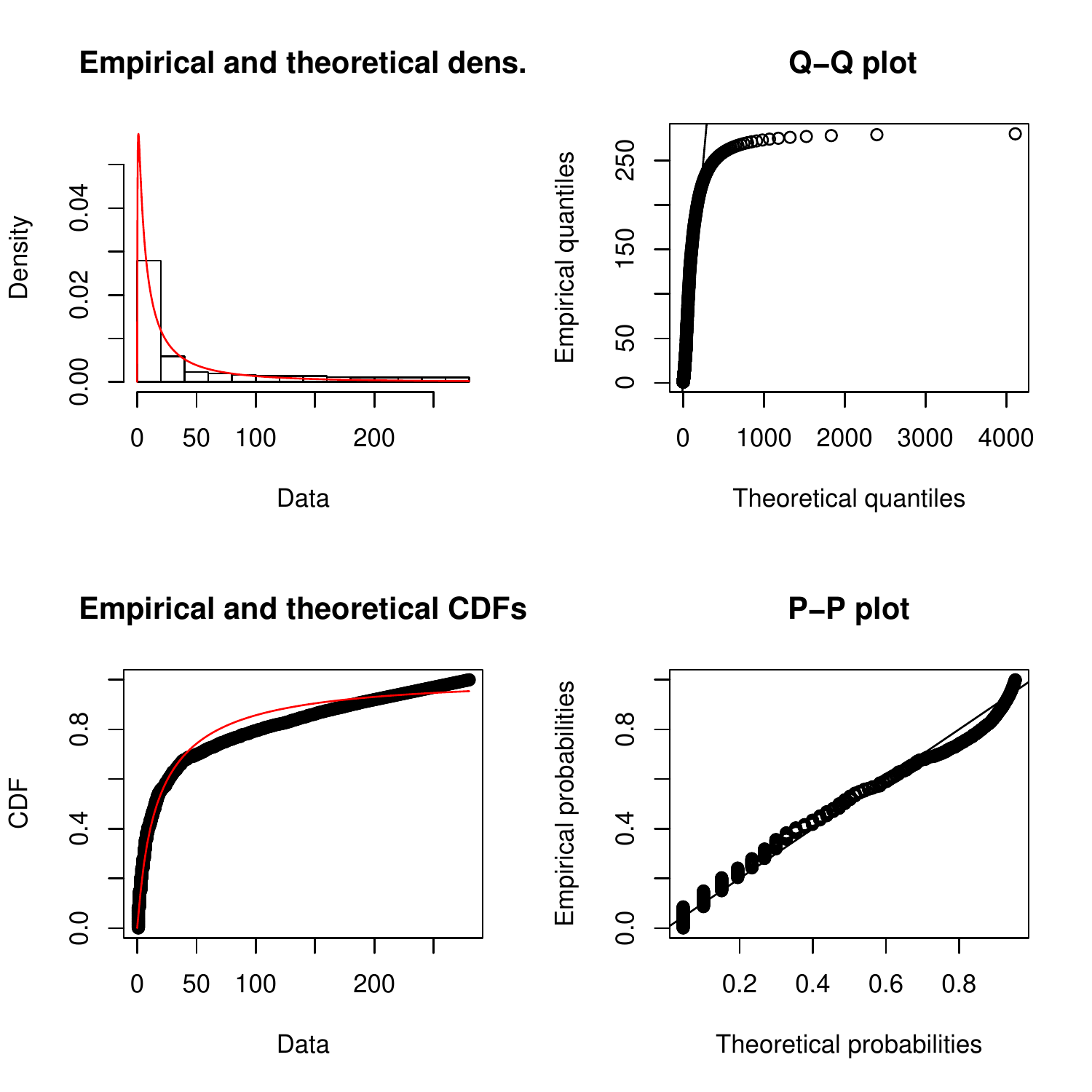}}%
  \vspace{-6mm}
  \caption{Mass distribution of a synthetically generated graph. The distribution has been generated using \cite{zipfR}}
  \label{fig:kronek}
\end{figure}

This graphs generated synthetically this way have been target of many statistical research. In the figure \ref{fig:kronek} it may be seen a typical graph 500 graph, following a Power law distribution. One reason for showing how a graph looks like statistically, is to see the comparison with that data that it is distributed by the processors and will be compressed later.   

Kroneker generated graphs meet the following relations:

\[
vertices =  2^{scale} \wedge edges = 2^{scale} \times edgefactor
\]

\subsection{Real world graphs (datasets)}
\label{syntheticdata}

A real world graph usually is greater than a scale 30 kronecker generated graph. Other differences with kronecker are the number of edges per vertex. real world graphs have a much larger edge-factor, and thus a higher diameter (maximum distance between two vertex in the graph).

\chapter{Related work}
\label{relatedwork}

\section{Optimizations}
\label{relatedworktechniques}

\begin{enumerate}
	\item \textbf{2D decomposition} This optimization is described in the section Graph computation.
	\item \textbf{Vertex sorting} Vertex sorting is a technique that helps to alleviate one of the major problems of the graph computations. Big size graphs are represented with structures with a very low spatial locality. This means that for example to traverse a graph with a distributed algorithm, the required data to compute the computation may be allocated in other processor. Unfortunately there is no known way to prevent this.  The Vertex sorting technique is combined with the previous one (2D decomposition) in such a way that with the decomposition each processor would have a subset of the Adjacency matrix and by adding an extra phase just at the begging to relabel the vertex, we add extra locality to the edges allocated in a same processor.   
	\item \textbf{Direction optimization} . Beamer et al. \cite{bottom-up, fighting-children} proposed two new techniques to deal with the unnecessary vertex computations when performing the distributed BFS algorithm. As an introduction, the usual way to transverse a graph is called Top-down: the adjacency matrix is used in such direction that provides us with the neighbour nodes to the ones in our frontier queue. One characteristic of the Top-bottom algorithm is that performs bad with high degree graphs (graphs with many neighbours in one vertex). to avoid unnecessary computations. the Adjacency matrix is duplicated to be able to use the opposite direction to traverse. This way when the number of neighbours is greater than a \(\delta\), the number of nodes in vertex in the opposite direction is computed and the minimum among them is chosen ase direction to follow.    
	\item \textbf{Data compression} Since the latency in the communication between the processors limits the maximum computational speed in the nodes by stalling them without new data, this method intend to alleviate this effect. This is the main goal of this work.  
	\item \textbf{Sparse vector with pop counting}. The pop instruction (\_\_popc in CUDA C) are instructions to deal with bitmaps with hardware support. Some new implementations of the graph500 use this optimization.   
	\item \textbf{Adaptive data representation} 
	\item \textbf{Overlapped communication} The technique of overlapping communication (also called software pipelining) allows to reduce the stall time due to communication. (Described in further detail in section \ref{sectionoptcommunication}). 
	\item \textbf{Shared memory}
	\item \textbf{GPGPU} General purpose GPUs are reviewed in section \ref{sectionlibrariesturbopfor}. 
	\item \textbf{NUMA aware} With NUMA aware specific libraries it is possible to ``pin'' a full data structure to an specific (core, processor, bank of memory) in systems with the possible penalty of accesses from one processor to a bank of memory belonging to other processor.   
\end{enumerate}

\section{Other implementations}
\label{relatedworkimplementations}

\begin{enumerate}
	\item \textbf{Hybrid (CPU and GPGPU)}
	\begin{itemize}
		\item \textbf{``B40C'' (Merrill et al.)} \cite{duane-merrill}
		\item \textbf{``LoneStar-Merrill''} \footnote{\url{http://iss.ices.utexas.edu/?p=projects/galois/lonestargpu}} \cite{lonestarMerrill}
		\item \textbf{``BFS-Dynamic Parallelism'' (Zhang et al.)} \cite{bfsDynamicParallelism}
		\item \textbf{``MapGraph''} \cite{mapgraph}
		\item \textbf{``GunRock''} (Wang et al.) \cite{gunrock}
		\item \textbf{``Enterprise'' (Liu et al.)} \cite{enterprise-topbottom}
		\item \textbf{SC11 (Ueno et al.)}
		\item \textbf{ISC12 (Ueno et al.)} \cite{ueno103teps}
		\item \textbf{SC12 (Ueno et al.)} \cite{ueno-et-al}
	\end{itemize}
	\item \textbf{CPU-based}
	\begin{itemize}
		\item \textbf{ISC14 (Fujisawa et al.)} \cite{numaAwareFujisawa2015}
	\end{itemize}
\end{enumerate}

\begin{table}[hbt]
\centering
\begin{tabular}{lllll}
\hlinewd{1.2pt}
Optimizations                              & SC11       & ISC12 & SC12 & ISC14 \\ \hlinewd{1.2pt}
2D decomposition                           & \checkmark & \checkmark     & \checkmark    & \checkmark     \\
Vertex sorting                             & \checkmark &      &     &      \\
Direction optimization                     &           &      &     &   \checkmark   \\
Data compression                           & \checkmark & \checkmark  & \checkmark    &      \\
Sparse vector with pop counting            &           &      &     &   \checkmark   \\
Adaptive data representation               &           &      &     &   \checkmark   \\
Overlapped communication                   & \checkmark & \checkmark     & \checkmark    & \checkmark     \\
Shared memory                              &           &      &     &   \checkmark   \\
GPGPU                                      &   & \checkmark     &  \checkmark   &      \\ 
NUMA-aware                                 &   &      &     & \checkmark     \\ \hlinewd{1.2pt}
\end{tabular}
\caption{Other implementations}
\label{relatedimplementations}
\end{table}

\section{Compression}
\label{relatedworkcompression}

In this subsection we present a list of other work on compression applied to Breadth-first Search implementations. We discuss their algorithm and compare this to ours.  
\begin{enumerate}
	\item \textbf{``Compression and Sieve: Reducing Communication in Parallel Breadth First Search on Distributed Memory Systems'' (Huiwei et al.)} \cite{compressingbitmap-sieve} \medskip
	
	In this work a 1D partition-Based BFS using SpMV multiplications is analyzed and with and without compression on its communication. The approach followed to compress the sparse vectors is to state that they have low cardinality as they are bitmaps. The compression codecs they analyze enter into the category of \emph{Bitmap Indexes}. The selected codec is WAH (Word Aligned Hybrid) a patented codec with an medium to low compression ratio but on the other hand an acceptable speed. The result of the study shows a reduction in the BFS time of over the 50\% thanks to the compression.  
	\item \textbf{``Highly Scalable Graph Search for the Graph500 Benchmark'' (Ueno et al.)} \cite{ueno-et-al}. \medskip
	
	The implementation of Ueno et al. had a good position in the graph 500 challenge during June 2012. It was it that year when they introduced compression. In this case the compression was implemented over a 2D Partition-based BFS executing on a GPFPU device. In contrast with Huiwei et al. Uneno treated the data as a sequence of integers instead of a sequence of bits. Also they chose a Varint-based encoder (VLQ) with neither a good compression ratio nor a good compression / decompression speed.  The main goal of this algorithm was its implementation simplicity. Once selected they implemented it on both the GPU device and the CPU. Also they reimplemented two more versions, a 32-bit based and a 64-bit.
	
	A not about GPGPU devices is their current lack of native arithmetic support for 64-bit integers. The choose of a full 64-bit compression on the device would have make the algorithm to incur on a performance penalty. For this reason and also to avoid performing compression for low amount of data they implement (as we have also done) a threshold mechanism.

\end{enumerate}



\chapter{Problem Analysis}
\label{problemanalysis}

\begin{itshape}
As aforementioned in the \emph{Introduction} the optimizations presented in this work have been implemented on a previously developed code \cite{matthias}. 

In this section, we will analyze the previous implementation and decide (based on the results of the analysis of the code and the instrumentation experiments)  what changes will be made to increase the overall performance. \medskip

Also, we will review the optimizations already applied to our code based on other state of-the-art research (discussed in Section \ref{relatedwork} - \emph{Relared work}). Next, we will explain how we have measured the data using internal (coded timers) and external instrumentation (profilers). In the case of the external instrumentation we will compare the available candidates and review the chosen one according to our needs. 

In addition, we will also include the measured data of our instrumentations. Regarding the communicated data, it will be highlighted what is its type, as it will be useful to understand the effectiveness of the chosen compression scheme. \medskip    

Finally, and based on the data of the instrumentation results, we will analyse and decide what approaches are considered to be feasible to alleviate the detected issues and optimize the overall performance.
\end{itshape}

\section{Our initial implementation}
\label{sectionanalysisourpreviousimplem}

A list of the optimizations implemented on the \texttt{Baseline} version of this work is listed in the table \ref{relatedimplementationsmade}.

\subsection{Optimizations table in ``\texttt{Baseline}'' implementation}
\label{sectionalgorithmbfsbfsalgo}

The list of optimizations may be seen on table \ref{relatedimplementationsmade}.

\begin{table}[hbt]
\begin{threeparttable}[b]
\centering
\begin{tabular}{lllll}
\hlinewd{1.2pt}
Optimizations                              & \texttt{Baseline} implementation \\ \hlinewd{1.2pt}
2D decomposition                           & \checkmark       \\
Vertex sorting                             &                  \\
Direction optimization                     &                  \\
Data compression                           & \checkmark \tnote{1}      \\
Sparse vector with pop counting            &                  \\
Adaptive data representation               &                  \\
Overlapped communication                   &                  \\
Shared memory                              &                  \\
GPGPU                                      &  \checkmark      \\ 
NUMA-aware                                 &                  \\ \hlinewd{1.2pt}
\end{tabular}
\caption{Other implementations}
\label{relatedimplementationsmade}
\begin{tablenotes}
\item [1] Feature to be implemented in this work
\end{tablenotes}
\end{threeparttable}
\end{table}

\subsection{Our ``\texttt{Baseline}'' algorithm}
\label{sectionalgorithmbfsbfsalgo}

\begin{algorithm}
\caption{Our BFS algorithm based on replicated-CSR and SpMV}\label{alggeneral}
\begin{algorithmic}[1]
\Statex Input : s: source vertex id
\Statex Output : $\Pi$: Predecesor list
\Statex f: Frontier queue / Next Queue
\Statex t: Current queue
\Statex A: Adjacency matrix
\Statex $\Pi$: Visited node /predecessor list 
\Statex $\otimes$: SpMV multiplication
\Statex $\odot$: element-wise multiplication
\State $f(s) \gets s$
\For{\textbf{each} processor $P_{i,j}$ in parallel}
\While{ f $\neq \varnothing$}
\State $TransposeVector(f_{i,j})$
\State $f_{i} \gets \textbf{ALLGATHERV}(f_{i}, P_{\ast ,j})$
\State $t_{i} \gets A_{i,j} \otimes f_{i}$
\State $t_{i,j} \gets \textbf{ALLTOALLV}(t_{i}, P_{i, \ast})$
\State $t_{i,j} \gets t_{i, j} \odot \overline{\Pi_{i}}$
\State $\Pi_{i,j} \gets \Pi_{i,j} + t_{i,j}$ 
\State $f_{i,j} \gets t_{i,j}$
\EndWhile
\EndFor
\end{algorithmic}
\end{algorithm}

In our initial algorithm, where the data is partitioned in 2D fashion, each of the processors operates over is own assigned block of the global Matrix. The  

As is was discussed in the section \emph{General Purpose Graphic Processor Units} (Section \ref{graphicproccessorunitgpucomputing}), some requirements are needed to build an optimal algorithm over a GPGPU device. 

In this section it will be listed the pseudo-code of the used parallel BFS algorithm implemented over CUDA. This used algorithm is based on the implementation of Merrill et Al. \cite{duane-merrill,matthias} and includes modifications to adapt the code to a 2D-partitioning. 

\subsection{General communication algorithm}
\label{sectioncommoverview}

The code presented in this work makes use of a 2-Dimensional partitioning of the CSR initial Graph Matrix.

The 2D partitioning were proposed as an alternative to the 1D to reduce the amount of data transmitted through the network \cite{1dpartitioning}.

Being \(G(V,E)\) a \emph{Graph} with \(\abs{V}\) \emph{Vertexes} and \(\abs{E}\) \emph{Edges}, on an original \texttt{1D} strategy the distribution unit across the network would be the \texttt{Vertexes} performing a complexity of \(O(P)\) where \texttt{P} is the total number of Processors. \medskip

On the other hand, on a 2D strategy the transferred unit are the \texttt{Edges}. This leads to a complexity reduction of \(O(\sqrt{P})\). 

Described briefly, on a 2D partitioning (the used one) there are two phases where each processo operates and transmit part of their Matrixes. 

On the first sub-phase the BFS operates over rows and as result performs an \emph{``AllReduce''} Operation that transmits the Column. This may be viewed on 2 and 3 of \texttt{Algorithm \ref{alg2d}}.

The next sub-phase Multicasts rows \cite{matthias} Line 4 of the algorithm.    

\begin{algorithm}
\caption{Simplified communication algorithm for 2D-partitioning}\label{alg2d}
\begin{algorithmic}[1]
\While{``there are still Vertexes to discover''}
\State $updateDataStructures \gets \Call{BFS\_iteration}{}$
\State \Call{column\_communication}{}
\State \Call{row\_communication}{}
\EndWhile\label{while1}
\State \Call{combine\_predecessorList}{}
\end{algorithmic}
\end{algorithm}

\subsection{Data and communications}
\label{sectionanalysisdata}

This section details the types of data that we will use in our communications. These are based on two structures, the first one consists of a sequence of long integers used to distribute vertexes between the nodes \footnote{Update: In our implementation, for the used graph generator and the maximum reachable \emph{Scale Factor} , these integers are in the range \([1..2^{24}]\) and are sorted in ascending order. Being the smallest, the first one in the sequence.}  \cite{matthias}.

A second structure is used to prevent nodes of re-processing already visited vertexes.

\paragraph{The Frontier Queue}

The frontier Queue is the main data to be compressed. Each of the elements on its sequence is a sparse vector of bits (SpMV). The analysis that we have performed to multiple big buffers of this data shows an uniform distribution, slightly skewed (Section \ref{compression}). 

\paragraph{The Predecessor list bitmap}

The bitmap is used in the last call of the BFS iteration is a 32-bit sequence of integers that is used to convert the frontier queue in the predecessor list.

\section{Instrumentation}
\label{sectioninstrumentation}

In this section It will be covered the selection of the \emph{Performance analysis tool}. It will be discussed the used criteria, the needed requirements in our application and some of the available the options. \medskip

As mayor requirements it will be set a low execution overhead, and the ability to wrap specific code zones for measurement. 

Minor requirements will be the availability of data display through console. This feature will enable the data to be tabbed more easily into diagrams. 
 
\begin{itemize}
	\item \textbf{TAU} \footnote{\url{https://www.cs.uoregon.edu/research/tau/home.php}} Originally implemented in the University of Oregon. Offers console interface. Narrow code wrapping is not implemented.  
	\item \textbf{Vampir} \footnote{\url{https://www.vampir.eu/}}  Offers graphical interface. We have not tested this option.
	\item \textbf{Score-P} \footnote{\url{http://www.vi-hps.org/projects/score-p/}} Offers both graphical (using Cube \footnote{\url{http://www.scalasca.org/}}) and console interface. Allow direct code zone wrapping. 
\end{itemize}

\subsection{Instrumented zones}
\label{sectioninstrumentationzones}

Listed below are the instrumented zones corresponding to the second Kernel (timed), which involves the communication, the BFS iteration.   
 
\begin{itemize}
	\item \textbf{BFSRUN\_region\_vertexBroadcast} initial distribution of the Start vertex. Performed only once. 
	\item \textbf{BFSRUN\_region\_localExpansion} Execution of the BFS algorithm on the GPU device.
	\item \textbf{BFSRUN\_region\_testSomethingHasBeenDone} Checks if the queue is empty (completion). To do so performs an MPI\_Allreduce(isFinished, Logic-AND) between all nodes.
	\item \textbf{BFSRUN\_region\_columnCommunication}. Performs a communication of the Frontier Queue vectors between ranks in the same column. This steps contains the next one.
	\item \textbf{BFSRUN\_region\_rowCommunication}. Performs a communication of the Frontier Queue vectors between ranks in the same row.
	\item \textbf{BFSRUN\_region\_allReduceBitmap}. Merges the predecessors in all nodes and check performs the validation step of the results. Run only once at completion of the BFS iteration.
\end{itemize}



\section{Analysis}
\label{sectionanalysismhaucksresults}

In this section it will be analyzed the previous instrumentation and instruction complexity data. 

As a result, a solution will be given for each of the issues found. The implementation, results and final conclusions, follow in the next sections of this document.

\subsection{Communication overhead}
\label{sectionanalysiscommunication}

As is has been mentioned at the beginning of this chapter, our \texttt{Baseline} implementation is mainly based upon a GPGPU BFS kernel, based in the implementation of Merrill et al. \cite{matthias}, with a 2D data partition, which enables the implementation to perform within a multi-GPU platform and in a multiple node environment through Message Passing Interface (MPI).     

In the results of the previous work, it was noticed a loss of performance in comparison to the original Merrill et al. \cite{duane-merrill} implementation. This was despite the application was running on multiple distributed multiGPU nodes within a fast cluster.

As a result of analyzing the reasons of the performance loss, it was decided to mitigate latency generated in the communications through compressing the transfers.

That new improvement is the work presented in this document.

\subsection{Instruction overhead}
\label{sectionanalysisinstructions}

As we were going to add a compression system which would be enabled at compile time, we still had space to add and test new improvements. The analysis of the code showed for example that the execution using thread parallelism (OpenMP) was slower than the one where the application was running on a single thread. Even though the experiment platforms has capabilities for multiple thread execution.

In a more on detail analysis we saw that some of the code entering in the timed parts of the application performed matrix intensive operations and allowed parallelism by the compilers. We also saw that many scalar improvements could be applied without much complexity.

In the last part we ran the compiler in inspection mode to retrieve the successfully applied optimizations in our \texttt{Baseline} implementation. Also, and to set a reference point with other distributed BFS implementation, we also performed this test on those. The other implementations are: (i) a Graph500 2D partition-based with MPI reference implementation. (ii) and the latest state-of-the-art Graph 500 implementation using GPGPU \cite{ueno-et-al}. The comparison may be seen in table .    

\chapter{Optimizing data movement by compression}
\label{compression}

\begin{itshape}

By compressing the data movements in the \emph{graph500} execution, we intend to solve the issues detected in the previous section (Section \ref{sectionanalysiscommunication}).  \medskip 

In this section we will first start describing some compression-related concepts for a better understanding of the optimizations. We will describe the available compression techniques. For each of them, we will expose how they match the Graph 500 data types. We will also discuss the related and most state-of-the-art compression algorithms (also known as \textbf{codecs} or \textbf{schemes}). After describing the algorithms we will focus on the available open-sourced library choices. Last, we will focus on the integration and discuss the decisions taken here.
\end{itshape}

\section{Concepts}
\label{compressionsectionconcepts}

We list below related compression terms that are used in this work and its related literature \cite{heman,lemire,glasgow,varint,simd-pfor,simple16, simple8b}.

\subsubsection{Information Theory}
\label{sectioninformationtheory}

The \emph{Information Theory} was the result of applying  statistical analysis to measure the maximum compression achievable in a communication channel. It was created by Claude Shannon in 1948, in his paper \emph{A Mathematical Theory of Communication} \cite{shannon48}.  \medskip

In the successive years, the \emph{Information theory} has applied statistics and optimal codes to assign the shorter codes to the most probable symbols on sequence to increase that way the \emph{compression ratio} (defined in this section). This \emph{ratio} depends on some parameters of the data, like its \emph{Entropy} (also defined in this section).  \medskip

\subsubsection{Information Retrieval Systems (IRSs)}
\label{sectioninformationretrievalsystems}

An \emph{Information Retrieval Systems} (IRSs) is a system which searches and retrieves data in environments of large data-volumes (for example Database or the World Wide Web).

To achieve this, modern techniques like \emph{Indexing} are used. These techniques make use of specific compression schemes which have required to be adapted to numerical data (\emph{indexes}). This benefits our \emph{Graph 500} application in two ways: 
\begin{enumerate}
	\item Since our application also uses numerical data - on each communication it transmits a \emph{``Frontier queue''} (which is a sequence of integers), we are able to use an integer-specific encoding for our purpose. Also, because of the current importance of this IRS systems, this algorithms improves quickly.  
	\item These techniques, due to the requirements of this IRS systems, have evolved to offer high speeds with low-latencies. This factor is important since one requirement in our application is the performance. 
\end{enumerate}

\subsubsection{Indexer}
\label{sectionindexer}

An \emph{Indexer} is a type of the \emph{IRS}. As defined before they manage the access to the stored indexes. A part of this \emph{Indexer} systems is usually compressed due to the high amount of data that they need to store.  

Common web search engines like Google\footnote{\url{https://www.google.com/}} are examples of \emph{Indexers}. In the case of Google, the indexing system is based in the \emph{Variable Byte} algorithm (an specific variant called \emph{varint-GB})  \cite{google} which will be discussed in further detail below.

\subsubsection{Inverted Indexes}
\label{sectioninvertedindexes}

The \emph{Inverted indexes} are the most commonly used data-structures (among other options) to implement \emph{indexer}  systems \cite{invertedindexes}.  They consist of two parts: 

\begin{enumerate}
	\item \textbf{A lexicon:} it is a dictionary-like structure that contains the \emph{terms} that appear in the documents plus a \emph{``document frequency''}. The latter is somehow similar to a ``ranking'', and indicates in how many documents those terms appear.
	\item \textbf{Inverted list} (also called \emph{Posting list}), is an structure which contains (for each term) the \emph{document identifiers}, its \emph{frequency}, and a \emph{list of placements} within the documents. \medskip

	This latter part usually becomes very large when the volume of the data is high, so it is often compressed to reduce the issue. The used  schemes are required to meet this criteria:
	
\begin{itemize}
	\item they need to considerably reduce the size of the \emph{Inverted list} by achieving a high \emph{compression ratio}.
	\item the need to allow a very fast decompression to not affect the search speed.      
\end{itemize}	
	 
\end{enumerate}

\subsubsection{Shannon's Entropy }
\label{entrophy}

As aforementioned Shannon's \emph{Information theory} settled the foundation of the data compression. One of the main concepts of this theory: the ``Entropy'' enables us to know the maximum theoretical limit of an optimum compression.   The definition of the concept is given below. \medskip

Put simply, \emph{Shannon's Entropy} can be defined as the amount of unpredictability of an event in terms of information. In \emph{Shannon's Entropy} events are binary.  

According to Shannon, in a distribution where a binary event happens with probability \(p_{i}\), and the event is sampled \(N\) times, the total amount of information we receive is given by \eqref{eq:entropy} \footnote{This expression applies to discrete distributions.}. 

\begin{align*}
H(X) = - \sum_{i=0}^{N-1} p_{i} \log_{2}{p_{i}} \tag{2}\label{eq:entropy}	
\end{align*}

\begin{figure}[hbt]
  \centering
  {\includegraphics[width=0.5\textwidth,viewport=0pt 0pt 460pt 450pt,clip]{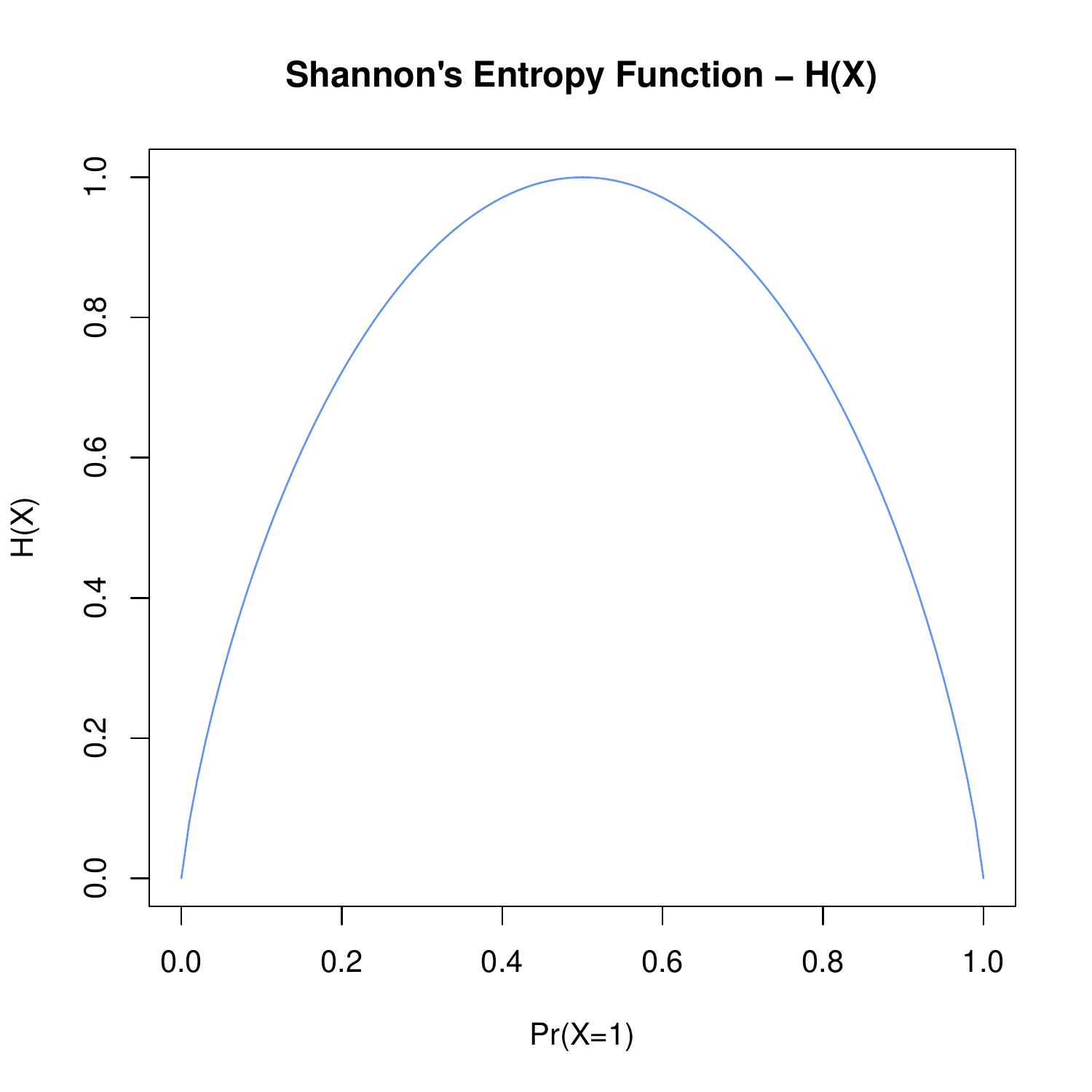}}%
  \caption{Shannon's Binary Entropy Function }
  \label{fig:entropy}
\end{figure}

To illustrate the concept we will use the example of tossing a coin with equal result probability. In this case, because we have no information about the possible result (Each side of the coin has \(p(x)=0.5\)), the \emph{Entropy} (Unpredictability) will be maximum  according to Equation \eqref{eq:entropy} (\(H(X)=1\)). Visually, this can be seen\footnote{To display Entropy as a continuous function it has been approximated to \( H(p) = -p log{(p)} - (1-p) \log{(1-p)}\) \cite{continuosentropyestimation}. }  in the Figure \ref{fig:entropy}. \medskip

\subsubsection{Limits of compressibility}
\label{limits}

The minimum number of bits per symbol will be given by the \emph{Entropy} of the symbols in the sequence. Expression \eqref{eq:numBits}.

\[
bits = \big[ H(X) \big] \tag{3}\label{eq:numBits}
\]  

\subsubsection{Compression Ratio }
\label{ratio}

According to the \emph{Information Theory} \cite{shannon48}, the maximum compression resulting from a data sequence, depends on its probability density function. Accordingly, the bigger the predictability in the sequence, the greater the achieved compression.\medskip

The resulting compression is proportional to the \emph{compression ratio} (greater is better). This \emph{ratio} is shown in the expression \eqref{eq:ratio}. 

This concept will be useful to measure the impact of the optimization of our communications in the final results.

\begin{align*}
ratio &= \frac{original\_ size }{compressed\_ size} \tag{4}\label{eq:ratio}	
\end{align*}

\subsubsection{\emph{``Lossless''} and \emph{``Lossy''} compressions}
\label{lossless}

These terms are related with compression and thus defined below. First, the compression of our data needs to meet these requirements:

\begin{enumerate}
	\item the size of the encoded data needs to be lower than the size of the original size.
	\item the encoding algorithm must allow recovering the original sequence from the compressed sequence, without any data loss.  
\end{enumerate}

When both previous criteria are met we are talking about \emph{``Lossless codecs''}. These are the ones used in this work.  

In other literature, only the first criteria above is required, and some tolerance is allowed in the second one. These are called  \emph{``lossy codecs''}. They are often used to \emph{reduce the size} of \emph{media} data (images, sound and video). These are not being discussed in this work. 
    
\subsubsection{Compression algorithms, Schemes, Codecs and Encodings}
\label{sectioncompressionscheme}

In this work the \emph{compression algorithms} will also be named  \emph{schemes},   \emph{codecs} or \emph{encodings}. These four terms are equivalent.

\subsubsection{Dependency}
\label{sectioncompressiondep}

To perform the compression of a full integer set some of the codecs operate over each integer independently. Others act over a group of integers and return a sequence in which each compressed \emph{block} have dependencies to the others. These dependencies are called \emph{oblivious} (independent) and \emph{list-adaptive} (dependant) in the literature \cite{obliviousname,glasgow}. They also appear as \emph{Integer encoders} and \emph{Integer sequence encoders} in the literature \cite{indianscomp}.  \medskip

Usually because independent (\emph{oblivious}) codecs compress each value by its own and discard the information about the rest may miss global characteristics which could benefit the global compression. Codecs belonging to this method are reviewed on this section even though the main focus will be on the dependent (\emph{list-adaptive}) schemes.      

\subsubsection{Compression techniques}
\label{compressiontechniques}

The compression techniques discussed below are the foundation blocks of the algorithms reviewed in this document. 

Next to these techniques are listed others (\emph{``Novel techniques and optimizations''}) that are usually combined with the former.\medskip

As a result, very optimized algorithms, operating over a very specific data domain (integers and integer sequences) are built. These encodings are, as of the moment of writing this document, the start-of-the-art integer compression algorithms.      

\begin{enumerate}
\item \textbf{Dictionary Compression} This is the first of the main compression technique exposed in this document and encompass multiple algorithms families. \medskip

The \emph{dictionary} technique is the most general compression method, and can be applied to any data domain. Usually this technique is made by using \emph{Huffman or Arithmetical} optimal codes. The symbols to be compressed (\emph{dictionary elements}) are assigned a dense code according to their relative probability of appearance in the document. The higher the probability of the symbol the shorter the assigned code. This technique has the downside of requiring a full recoding if new symbols are added. A way to check the efficacy of this method for our data, is by generating the empirical \emph{Entropy} (defined next, in this section) of our distribution of numbers. 

\item \textbf{Prefix Suppression \emph{PS} (or length encoding)} This is the second global compression mechanism described in the document. This technique is applied to data know \emph{a-priory} to be a integer. Prefix suppression (PS) suppresses the 0's at the beginning of each element in the sequence. As a direct result, we reduce the sequence's domain. This technique generates a \emph{bit-aligned} variable-sized sequence.

\item \textbf{Frame of Reference (\emph{FOR})} The next global compression mechanism described this document is Frame-of-reference. This method was first described in \cite{originalfor}. This compression technique is applied to a sequence of integers. The way it works is by splitting the sequence of elements in \(n\) blocks, so that for each block \(B\)  

 \[
  \exists \ min \in B \cdot \forall \ i \in B \cdot (i^{`} = i - min)  \tag{5}\label{eq:forblock}
 \]

This process is repeated for all blocks in the sequence. The size of each of the \(n\) blocks \(B\)  is:   
\[ 
size = \big[ \log_{2}{(max + 1 - min)} \big]   \tag{6}\label{eq:forsize}
\] 
This technique contrasts with PS in that the generated sequence has a fixed size. This resulted sequence has a   byte-level alignment.\medskip

This family of algorithms will be the main focus of this work (discussed below) as they give a good compression ratio and low compression / decompression times.

\end{enumerate}

\subsubsection{New techniques focused on integer/ integer sequences compression}
\label{compressionnoveltechniquesoptimizations}

These optimizations are applied to the previous techniques to improve the algorithm. The technique \emph{Binary Packing} described below is a full technique \emph{per se}, and has been used as the core compression algorithm in this work (\emph{S4-BP128} codec). Any of these techniques below may be combined. 

\begin{enumerate}
	\item \textbf{Delta compression, differential coding or delta-gaps technique } The concept is firstly described in \cite{origindelta}. The delta compression, may also appear with the names \emph{Differential coding}, \emph{delta-gaps}, \emph{d-gaps} or \emph{\(\delta\)-gaps} in the literature. \cite{heman,lemire,glasgow,varint}.  

The technique works by storing the distances between elements instead of the elements themselves (which have lower size than the main integers).\medskip

Regarding the research tendency over this technique, it is changing in the last few years:

\begin{itemize}
	\item the initial work was focused on the gap probability distributions to achieve the optimal \emph{compression ratio}. For example, depending on the frequency of each term we could adjust the selected Huffman or Golomb's codings \cite{bernoulli}.

	\item the most recent work emphasizes in a better \emph{compression ratio} and lower decompression latency \cite{trotman}.   
\end{itemize}

\item \textbf{Binary packing} This technique implements both the concepts of \emph{Frame-of-Reference} (FOR) and \emph{differential coding}. If in FOR each integer in a range is coded and then represented in reference to that range. For example, for a block of size 128, and integers in the range [2000, 2127] the can be stored using 7 \emph{bits} as offsets of the \emph{minimum} value of the range (i.e. 2000). Using the expression \eqref{eq:forsize} the resulting size can be seen in \eqref{eq:binpacksize}

\[ 
\big[ \log_{2}(2127+1 - 2000) \big] = 7 \tag{7}\label{eq:binpacksize}
\] 

With the technique of \emph{differential coding}, now the \emph{minimum} value is coded and the rest of elements are represented as increments \(\delta\) to that value.

With this technique there are two main factors which give theresulting size of the compressed sequence. These are:
\begin{itemize}
	\item the number of bits \texttt{\emph{b}} used to encode the blocks (in the worst case block)
	\item the block length \texttt{\emph{B}}
\end{itemize}
  
Initial schemes of this technique, refered in the literature as \emph{PackedBinary} \cite{binpack} or AFOR-1 \cite{afor1}, use a fixed block size of 128 or 32 integers. Future variations of the algorithm make use of variable size blocks (AFOR-2, AFOR-3) \cite{afor2}.

\item \textbf{Patched coding and Exceptions} Binary coding may lead at some cases to very bad efficiency, resulting sometimes in compressions needing more bits than the original representation. This is for the example the case of compressing sequences like the following:

\[
5 , 120, 300, 530, 990, 25, 980, 799273656454 \tag{8}\label{eq:badforsample}
\] 

In the sequence given by \eqref{eq:badforsample} in case of only needing to compress the 7th first numbers the required bits would be  

\[ 
\big[ \log_{2}(990+1 - 5) \big] = 9,9 \tag{9}\label{eq:badforvalue}
\] 

Unfortunately, if the 8th number would need to be included in the block, the compression would even result in a size bigger than 32-bit per integer.

To alleviate this problem Zukowski-et-al \cite{heman} proposed the concept of exceptions, where the the 8th number in this sample sequence would be stored in a separate space. This is the concept of \emph{exceptions}, and will encompass all values greater than \(2^{b}\), being \(\texttt{b}\) the value defined in the \emph{Binary packing} concept.    

\end{enumerate}

\section{Compression algorithms (codecs or schemes)}
\label{compressionalgorithms}

\begin{table}[hbt]
\centering
\adjustbox{width=0.90\columnwidth}{%
\begin{tabular}{lccccr}
\hlinewd{1.2pt}
Names/Families  & Alignment & Dependency  & Technique & Target data                     \\ \hlinewd{1.2pt}
Unary             & bit       & Independent & PS & Integers                            \\ 
Gamma             & bit       & Independent & PS & Integers      \\ 
Huffman-int       & bit       & Independent & PS & Integers      
\\ 
Golomb            & bit       & Group       & PS & Integers 
\\ 
Rice (Rice\(_{k}\))              & bit       & Group       & PS  & Integers                     \\ 
Variable Byte     & byte      & Independent & PS  & Ints. Seq.   \\ 
Simple family     & word      & Group       & PS & Ints. Seq.                            \\ 
FOR family       & byte      & Group       & FOR & Ints. Seq.                           \\ 
PDICT             & -         & -           & Dict.  & Ints. Seq.                                  \\ 
Lempel-Ziv family & -         & -           & Dict.  & low H(x) Seq.                                   \\ \hlinewd{1.2pt}
\end{tabular}
}
\caption{Compression algorithms (also called \emph{codecs} or \emph{schemes})}
\label{table:compressionclasification}
\end{table}

To describe the compression algorithms used in this document, we will group them first, according to their resulting alignment in the compressed sequence. These groups and their features are shown in Table \ref{table:compressionclasification} . In addition to this, some terms used here are described in the introduction of the current section.    

The main compression groups to be described are \emph{Bit aligned}, \emph{Byte aligned}, \emph{Word aligned} and \emph{Dictionary based}.  

\begin{enumerate}[label=\textbf{\Alph*}.]
	\item \textbf{Bit aligned algorithms}
	\begin{enumerate}
		\item \textbf{Unary} This \emph{oblivious} (group independent) codec represents an integer as N \emph{one bits} followed by a \emph{zero bit} which acts as a terminator. 
		\item \textbf{Gamma} This \emph{oblivious} codec consists of two parts. In the first it represents an integer as \(log_2{N}\) \emph{bits} in \emph{Unary} scheme. In the second part the \(floor(log_2{N})\) bits of the integer are also represented with \emph{Unary}. This scheme is usually inefficient as integers become larger. To avoid this sometimes \emph{Delta compression} is used in the second half of the encoding.     
		\item \textbf{Huffman-int} This scheme is also an independent integer encoder. This is a variation of \emph{Gamma} in which the first half (the first \(log_2{N}\) \emph{bits}) are encoded whith Huffman optimal codes instead of \emph{Unary}.
		\item \textbf{Rice} This \emph{group dependent} (List-adaptative) integer sequence encoder  consists of two parts. In the first part the  \(floor( \frac{N}{2*k})\) \emph{bits} are encoded using in \emph{Unary}. The secon part uses (\(N* mod 2*k\)) \emph{bits} for a given \(k\). This schema is also known as \textbf{Rice\(_k\)} in literature. 
		\item \textbf{Golomb} This \emph{group dependent} integer sequence encoder is  similar to \emph{Rice\(_k\)}. In this scheme \(k\) is a power of 2.  
	\end{enumerate}
	\item \textbf{Byte aligned algorithms}
		\begin{enumerate}
		\item \textbf{Frame Of Reference Family} This codecs are group dependent sequence integer encoders. As described above the FOR technique has multiple scheme implementations. Some of the differences of these schemes are explained here.
			\begin{enumerate}
				\item \textbf{PFOR} \cite{heman} Original Patched coding Frame-of-Reference algorithm. In this scheme the \emph{exceptions} where  stored without being compressed including only its relative order in the linked list.
				\item \textbf{PFOR2008} \cite{pfor2008} as a compression optimization it was proposed the compression of the exception storage, using 8, 16 or 32 bits.
				\item \textbf{NewPFOR} \cite{newpfor-optpfor} The exceptions are stored in a dedicated part of the output. This area is divided in two subareas. Simple-16 is used as \emph{exception} encoder in one of the areas. 
				\item \textbf{OptPFOR} \cite{newpfor-optpfor} This scheme is very similar to the previous \emph{NewPFOR} but the size of the exceptions subareas is selected accordingly to achieve better compression ration an lower decompression latency. Simple-16 is also used as encoder scheme for the \emph{exceptions}.
				\item \textbf{ParaPFOR} \cite{parapfor-gpufor} This is a variation of \emph{NewPFOR} optimized for the execution on GPGPUs. As an improvement, the compression of the \emph{exceptions} using \emph{Single-16} is moved from the  exception part to the original segment. This modifications worsens the compression ratio but leads to a much faster decompression. 
				\item \textbf{Simple-PFOR} \textbf{Fast-PFOR} \textbf{SIMD-FastPFOR} \cite{simd-pfor}  In this codec, \emph{exceptions} are placed in 32 different reserved areas of the output. These are encoded using Vectorized (with \emph{SIMD instructions}) Binary packing. The main differences in Simple and FastPFOR schemes is in how they compress high bits of the \emph{exceptions} values. In all other aspects Simple and FastPFOR are identical. Regarding the SIMD-PFOR codec, it is identical to FastPFOR except that its packs relies on vectorised bit packing for some of the exception values.
				\item \textbf{S4-BP128\footnote{This is the codec that has been used in this work.}}  This codec was also introduced by Lemire et al. \cite{lemire}. It uses the \emph{Bit packing} technique described above with blocks  (\texttt{\emph{B}})	of 128 integers.	 For each block, the bit-width (\texttt{\emph{b}}) ranges fron 0 to 32 bits. The ``S4'' stands for 4-integer \emph{SIMD}.	 This algorithm uses the concept of \emph{Metablock} to refer to 4 blocks of 128 integers (Partitioned that way to perform better the \emph{SIMD} instructions). Before \emph{Bit packing} is performed, the \emph{metablocks} use some bits to contain two \texttt{\emph{b}} values as a header. \emph{VByte} compression is added to the last blocks that can not be grouped in a \emph{metapackage}. This is the fastest Lemire et al. family of codecs \cite{lemire} and is the one used in this work.  
				\item \textbf{TurboPFOR} A fixed size Vectorized Binary packing is also used to encode the \emph{exceptions}. As novelty, a  flag bitmap is included to map the placement of the \emph{exceptions}\footnote{\url{https://github.com/powturbo/TurboPFor/issues/7}}.
    		\end{enumerate}
    	\item \textbf{Variable Byte Code} \hfill \\
			This popular codec is group independent (\emph{Oblivious}), as it encodes each value on its own. As a difference with the aforementioned \emph{Oblivious} codecs (Unary, Gamma and Huffman-int), this one uses byte alignment instead of bit. A benefit of this schemes over the previous ones is that even though the compression ratio is usually worse, the decoding speed is faster \cite{vintOvergamma}. 
			The main algorithm is known under many names: v-byte, Variable Byte, variable-byte, var-byte, VByte, varint, VInt, or VB. An  algorithm that enter this category is the Group Varint Encoding (varint-GB) \cite{google}, used by Google. Some improved codecs over the initial \emph{Variable Byte} algorithm will be listed below.
			 \begin{enumerate}
			 \item \textbf{varint-G8IU} \cite{varint} This scheme is based on Intel \texttrademark \space SSE3 \emph{SIMD} instructions. By the usage of vectorisation Stephanoph-et-al outperformed by a 300\% the original VByte. They also showed that by using vectorisation in the CPU it is possible to improve the decompression performance more than 50\%. The varint-G8IU is an \emph{SIMD} optimized variation of the varint-GB used by Google \cite{simd-pfor}. It was patented by its authors.
			     
			 \item \textbf{maskedvbyte} \cite{maskedvbyte} This version also uses a \emph{SIMD instruction set} based on Intel \texttrademark \space SSE2 to achieve the performance improvements over the original VByte algorithm. The codec adds a Mask (\emph{Bitmap}) to be able to perform the decoding faster with the vectorisation. 
			 \item \textbf{maskedvbyte} Variable-Length Quantity (VLQ). Only provide in this work to reference the compression added to their  Graph 500 implementation by \cite{ueno-et-al}.
			 \end{enumerate}
			 
    	\end{enumerate}
	\item \textbf{Word aligned algorithms}
		\begin{enumerate}
			\item \textbf{\emph{Simple} family } This family of algorithms was firstly described in \cite{originsimple}. It is made by storing as many possible integers within a single \texttt{n}-bit sized word. Within the word, some bits are used to describe the organization of the data. Usually this schemes give a good compression ratio but are generally slow.
			\begin{enumerate}
			 	\item \textbf{VSimple} \footnote{\url{https://github.com/powturbo/TurboPFor}}  This codec is a variant of  Simple-8b. It can use word sizes of 32, 40 and 64 bits. As other improvements it uses a \emph{``Run Length Encoding''}\footnote{Note from the author inside the source code files.}. 
			 	\item \textbf{Simple-8b} \cite{simple8b} Whereas the Variable-byte family of codecs use a fixed length input to produce a variable sized output, these schemes may be seen as the opposite process, with multiple variable integers they produce a fixed size output. In the case of this particular scheme a 64-bit word is produced. It uses a 4-bit fixed size header to organize the data, and the remaining 60 bits of the word for the data itself. It uses a 14-way encoding approach to compress positive integers. The encoded positive integers may be in a range of \([0, 2^{28})\). It is considered to be the most competitive codec in the Simple family for 64-bit CPUs \cite{simple8b}. 
			 	\item \textbf{Simple-9}, \textbf{Simple-16} \cite{originsimple,simple16} These two formats (also named S9 and S16 respectively in the literature) use a word size of 32 bits. The former codec uses a 9-way encoding approach to compress positive integers, but in some cases bits are wasted depending on the sequence. To alleviate the issue the later codec uses a 16-way encoding approach. Both of them use a 4-bit fixed header for describing the organization in the data and 28 bits for the data itself. Due to the limitation that this implies: they are restricted to integers in the range \([0, 2^{28})\) they will not be  reviewed in further detail as do not meet  the \emph{Graph 500} requirements for a vertex size (48-bit integers).    
			\end{enumerate}
		\end{enumerate}
	\item \textbf{Dictionary based algorithms}
		\begin{enumerate}
			\item \textbf{Lempel Ziv Family and PDICT} Dictionary algorithms are based in the \emph{Dictionary technique} described above.  Some algorithms that use this technique are the Lempel Ziv algorithms family (commonly named using a \emph{LZxx} prefix) and PDICT \cite{heman}. Their performance compressing / decompressing integer sequences is outperformed by the integer domain-specific algorithms aforementioned. For this reason they will stay of the scope of this work. 
		\end{enumerate}
\end{enumerate}

\section{Compression libraries}
\label{implementationsandlibraries}

For the usage of each of described algorithms in our \emph{Graph500} implementation we could have either implemented from scratch each of the above algorithms or either use a state-of-the-art library containing some of those algorithms. As time is a limitation in this work and the compression algorithms are beyond of the scope of our work, we have used several open sourced libraries based on the work of other authors. The selection criteria of each of those libraries is described below.\medskip

Regarding these libraries, and because we noted that all of them presented  different benefits, we opted to implement a modular library integration, so we were able to easily include (or remove) any of them so we could and test and demonstrate the positive impact of each. The library integration has been made with the following design parameters in mind:

\begin{itemize}
	\item The design is simple and modular. This allows new maintainers to add new desired compression libraries with the minimum effort. This has been done by the usage of a software design pattern called \emph{``Factory''} \cite{gof}. In this way we  have managed to improve of three of the \emph{Graph 500}'s evaluation criteria:  
	\begin{enumerate}
		\item Minimal development time: addition of new libraries by new maintainers in an small amount of time.
		\item Maximal maintainability: modification and debugging of the current code.
		\item Maximal extensibility: addition of new libraries by new maintainers with a minimal impact on the already developed code.
	\end{enumerate}
	\item Minimal possible impact in the performance of the BFS kernel. We have managed to achieve this by creating the \emph{Factory} call prior to the initialization of the second Kernel (BFS cycles and validations) and passing a memory reference to the compression / decompression routine. 
\end{itemize}

\begin{table}[hbt]
\centering
\adjustbox{width=0.95\columnwidth}{%
\begin{tabular}{p{0.25\linewidth}p{0.2\linewidth}p{0.2\linewidth}p{0.205\linewidth}}
\hlinewd{1.2pt}
\textbf{Features}                & \textbf{Lemire et al. }                                                             & \textbf{TurboPFOR}                                                                                                  & \textbf{Alenka Project}                                  \\ \hlinewd{1.2pt}
Languages               & C++,Java                                                                            & C,C++,Java                                                                                                                     & CUDA C                                                         \\ 
Input types\slash \space sizes & Unsigned Integer 32-bit                                                                    & Unsigned Integer 32\slash \space 64-bit                                                                                                   & Unsigned Integer 32\slash \space 64-bit                                        \\ 
Compressed type & Unsigned Integer 32-bit (vector)                                                                   & Unsigned Integer 32\slash \space 64-bit (vector)                                                                                                   & 8-bit aligned Structure.                                        \\ 
License   & Apache 2.0                                                                          & GPL 2.0                                                                                                                   & Apache 2.0                                                     \\ 
Available codecs  & S4-BP128, FastPFOR, SimplePFOR, SIMDFastPFor, v-byte, maskedvint, varintG8IU                              & \emph{``Turbo-PFOR''}, Lemire et al.'s , Simple8b/9/16, LZ4, v-byte, maskedvbyte, varintG8IU                                                               & Para-PFOR                                                   \\
Required hardware   & SSE4.1+                                                                             & SSE2+                                                                                                                     & Nvidia GPGPU                                                 \\ 
Technology            & SIMD                                                                                & SIMD                                                                                                                      & GPGPU                                                          \\ 
Pros              & Clean code                                                                  & Very good compression ratio                                                                                                    & Clean code. Very good latency \\ 
Cons           & Integer size adds extra latency because conversion is required .                                                                                   & Source code is difficult to understand.                                                                              & Compression less optimized than other schemes                                      \\ \hlinewd{1.2pt}
Integrated in code              & Yes                                                                  & Partially                                                                                                    & Partially \\ \hlinewd{1.2pt}
\end{tabular}
}
\caption{Main feature comparison in three used compression libraries}
\label{table:implementationscomparison}
\end{table}

\subsection{Lemire et al.}
\label{sectionlibrarieslemire}

This library\footnote{\url{https://github.com/lemire/FastPFor/}}, has been created as proof-of-concept of various  academical works \cite{heman,lemire,originalfor}. One of the main contributions of this work is the achievement of a very high performance and compresion-ratio. This has very high value in the compression of \emph{Inverted indexes} (Section \ref{sectioninvertedindexes}), which are very used nowadays (as it has been exposed before). 
		 
Lemire et al. achieved this high performance by implementing  a version of the algorithm using \emph{SIMD} instruction sets in the CPUs.   

The base of this work is the Frame-of-Reference algorithm first purposed by Goldstein-et-al \cite{originalfor}, and later improved by Heman-et-al who among other additions to the algorithm, also enhanced the performance by using \emph{Superscalar} techniques \cite{heman}.

Some characteristics about the Lemire's implementation are listed below.

	\begin{description}

		\item[Compression input types and sizes] This library allows only unsigned 32-bit integer input type.

		\item[Compression output] The output type is the same as the input:  \emph{unsigned 32-bit integer}.
		
		\item[Included codecs] FastPFOR, SIMD-FastPFOR, S4-BP128(Used), Vbyte, maskedvbyte, Varint, Varint-G8IU.

		\item[Pros and Cons] Good compression ratio for the best case scenario. A 32-bit integer is reduced to near 4-bit (value near to the Entropy) (\( ratio = \frac{32}{4} = 8  \)). Also, the library has been implemented in a clean way and the source code is easy to understand.

		As disadvantages to fit this library in a \emph{Graph 500} application, we have found the next: \medskip
		
		The implementation restricts the size of the compressed 32-bit integers.  This limitations implies:
		
		\begin{itemize}
			\item for each compression-decompression call, a size and type pre-conversion must be carried out carried out to adjust the type to the required by the \emph{Graph 500}. 
			\item to avoid integer overflows in the previous conversions, it is required to perform a pre-check over all the integers in the Graph just after the graph generation. 
		\end{itemize}

	\end{description}

\subsection{Turbo-PFOR}
\label{sectionlibrariesturbopfor}

The Turbo-PFOR\footnote{\url{https://github.com/powturbo/TurboPFor}} library is an open sourced project containing a full set of state-of-the-art integer compression algorithms from various authors. The library main contribution is an improved version of the \emph{Lemire et al.} \cite{lemire} implementation, mixed with the ideas and code from the other authors. The Lemire's code has been fully optimized\footnote{\url{https://github.com/powturbo/TurboPFor/issues/7}} and the library itself makes a massive use of the CPU's \emph{SIMD} instructions.\medskip

The package also includes an statistical \emph{Zipf} Synthetic Generator which allows to pass the \emph{skewness} of the distribution as a parameter. The statistical generator allows to include into the compared terms the value of the empirical \emph{Entropy} of the generated integer sequence.

The package also compares the new optimized implementation with other state-of-the-art from other libraries. As the result of the optimizations, the author claims a better \emph{compression ratio} and a higher decompression speed for integer sequences more similar to phenomenological \emph{Pareto} distributions.  


	\begin{description}
		\item[Compression input types and sizes] The library allows sizes of 32 and 64 \emph{bits} integers without sign.

		\item[Compression output] The implementation relays on the C Standard Template Library (STL) and the output type is the same than the input. 
				
		\item[Included codecs] \emph{``Turbo-PFOR''}, Opt-PFOR, Lemire et al.'s set of codecs, Simple-8b, Simple-9, Simple-16, LZ4, Vbyte, maskedvbyte, varint-G8IU.

		\item[Pros and Cons] Similar latency than the analogous codec in Lemire et al. library, with an  2\(\texttt{X}\) \emph{compression ratio} improvement. 
		
		Also, the integer input size matches the \emph{``64-bit unsigned integer''} required by \emph{Graph 500}. This prevents our code from the pre-conversion step in \textbf{lines 2, 4, 6, 8 of algorithm \ref{algcompintegration}}. 
		
		As a possible downside, the code is difficult to understand since it heavily relays on C macros.
		
	\end{description}

\subsection{Alenka CUDA}
\label{sectionlibrariesalenka}

The Alenka\footnote{\url{https://github.com/antonmks/Alenka}} project is aimed to be a fast database implementation with an inbuilt compressed \emph{inverted indexed} based on CUDA and GPGPU technology. The query implementation is also based in this latter technology. 

The main compression algorithm is an state-of-the-art Patched Frame-Of-Reference \emph{Para-PFOR} (Section \ref{compressionalgorithms}). Its main features are:  

	\begin{description}
		\item[Compression input types and sizes] One of the main benefits of this implementation is that implements a compression for \emph{unsigned} integers. These size and sign suit the \emph{Graph 500} requirements which are \emph{``unsigned 64-bit integers''}. Therefore it is not needed the pre conversion step in \textbf{lines 2, 4, 6, 8 of algorithm \ref{algcompintegration} }. The possible integer input types are 32 and 64 \emph{bits} without sign.
		
		\item[Compression output] the algorithm generates a data structure containing information to re-ensemble the data in the other peer.  

		\item[Included codecs] \emph{Para-PFOR} implementation over CUDA (Section \ref{compressionalgorithms}).

		\item[Pros and Cons] As an advantage, the code is clean (easy to read) and short. As a possible downside may be pointed that the transformation of the output structure to an equivalent MPI structure to perform the communication may require some extra programming effort. However, this coding is mostly complete.

	\end{description}

\section{Integration of the compression}
\label{sectioncompressionintegration}

As described in further detail in other parts of this document, the integration of these compression libraries (which combine a broad spectrum of HPC technologies) has been designed flexible (``pluglable'') so that it is simple to add a new ``library'' from other authors. This library addition would involve some steps like:

\begin{enumerate}
	\item The 3rd-party compression library need to be copied to the \emph{Graph 500}'s \texttt{compression/} folder.
	\item The ``Factory object'' must become aware of that  a new library. This is done by selecting a name for the new class within that object and  adjusting the  input and output types. For example, it would be correct to copy an already existing one, choosing a new name, and setting the types. 
	\item A compression-specific \texttt{Object} for that new library must exist. This Class contains a wrapper of the new library's compression \& decompression calls so they match our application's calls. An specific file, that can be copied and modified exists for this purpose. In this new created \emph{object} it is required to adjust the \emph{Class}'s name and the  new types.  
	\item Finally, the file containing a mapping between that \emph{compression} type and its MPI equivalent needs to be updated. 
\end{enumerate}

Other aspects regarding the integration are the low impact of the integration of the integration on our code. The code performs only one call to create the compression object, which remains in memory. This is performed outside the \emph{Second kernel} (BFS + Communication) to minimize its effect on the  the overall performance. The compression object is passed as a memory position to the BFS calls.\medskip

Other benefit of this implementation is that the compression library may be updated regardless from the \emph{Graph 500} implementation.

\subsection{Experimental performance comparison}
\label{sectioncompressionexpperf}

In order to select the compression scheme that best suits our purposes, a comparison table (Table \ref{table:compressionperformance} and Table  \ref{table:parapforcomparison}) containing the results of our compression experiments is presented below. In these tables, the parameters used in the headers are described below.

\begin{itemize}
	\item \emph{\textbf{C. ratio \%}} shows the achieved compression-ratio represented as percentile. The lower the better. The formulat can be seen in expression \ref{eq:compratiopercent}
	
	\[ratio \% = \frac{1}{compression\_ratio} * 100 \tag{10}\label{eq:compratiopercent}
	\]
	\item \emph{\textbf{Bits/Integer}} Resulting bits required to represent the 32-bit integer. This value has as lower limit the value of entropy (Section \ref{compressionsectionconcepts}) of the sequence distribution. The lower the better. 
	\item  \emph{\textbf{C Speed MI/s}} represents the compression speed measured in Millions (\(10^{6}\)) of 32-bit integers processed per second. The higher the better.
	\item \emph{\textbf{D Speed MI/s}} analogous to the previous parameter. For decompression speed. The higher the better.
\end{itemize}

The aim of generating the data is to be able to gather compression statistics from each codec and see the one that best suits our needs: we require good decompression speed (like an IRSs), but also a good compression speed, to not penalize the compression calls. A good compression-ratio is as important as the two previous aspects. For building the table we will be using a synthetical Zipf integer sequence generator. The generator is specific for generating sequences of indexes (integers) similar to the ones appearing in web pages, search engines indexes and databases indexes (Further discussion about those topics in the next sections of this chapter).

Regarding graphs datasets, they have been study to follow Power law distributions (like  Zipf, pareto or lognormal) \cite{gov2zipf}. So, that could be a clue of that using a Zipf generator would give a real insight about the data we intend to compress. Unfortunately, regarding our data, we do not compress the vertices. Instead, we do compress sparse vectors (SpMV) represented as integers. These integer sequence have some own numerical characteristicswhich are going to help us to choose the correct codec for of Frontier Queue compression.

If we could do a summary of what family of codecs to choose depending of the properties of the data to compress, it would be as follows:

\begin{itemize}
	\item For compressing an unique integer instead of a sequence, our option would be a Prefix Suppression (PS) type codec. Whiting the available codecs in this broad family we would choose the one which better adjust to or latency , speed and compression-ratio requirements.
	\item If we had a sequence of integers where the distance from one number to the other is a-priori known to  not be very high, the Frame Of Reference (FOR) type with a possible delta compression on top would be a good option.
	\item If the data is an integer sequence, where the numbers are known to be low (applying this family over high numbers could result counterproductive), then the Varint family could be a good option,
	\item If the cardinality of the data to compress is low, for example bitmaps with '0's and '1's, a Bitmap Index family would be a good option. Downsides of bitmap indexes are their low speed and their heavy memory footprint.
	\item for any other case where we have no knowledge about properties of the data to be compressed a general Purpose dictionary compression (for example Lempel Ziv family) would be a good choice.
\end{itemize}   

\begin{figure}[htb]
  \centering
  {\includegraphics[width=0.70\textwidth,viewport=0pt 0pt 460pt 450pt,clip]{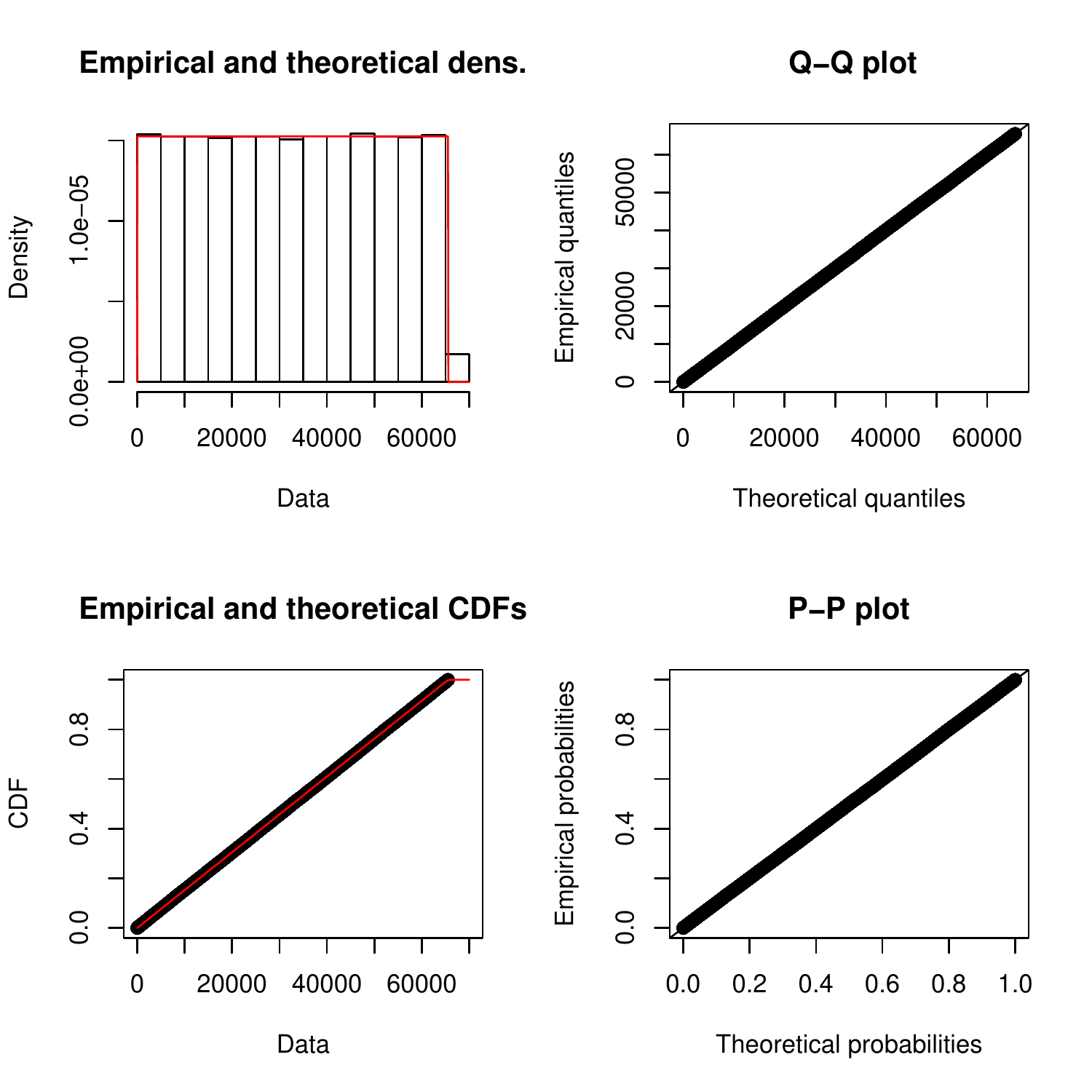}}%
  \caption{Mass distribution of a long SpMVM vector in our Frontier Queue. The distribution detection has been performed using \cite{fitdistR} }
  \label{fig:fquniform}
\end{figure}

To gather information about the sequence of integers to be compressed (the SpMVs from our Frontier queue) we have manually extracted several big buffers (bigger than 20M \footnote{M stands for Millions} integers) into files and  analysed them afterward with an statistical package \cite{fitdistR}. The result of the tests showed an slightly left-skewed Uniform distribution (Figure\ref{fig:fquniform}). In terms of compressibility the term Uniform distribution is similar to randomness. Because random sequences have a high entropy (Section \ref{entrophy}), the compression would most likely be bad. Even though the entropy of those sequences was high: 15-bit out of 32-bit of an integer, we have been able to perform a very high compression (over 90\% reduction) very close to the entropy value. This has been thanks to the fact of (1) the data was sorted due to our BFS implementation (2) the gaps in the sorted sequence where small. (3) this kind of sequences have raised interest due to the Big data + Indexer phenomena and have the support of efficient algorithms nowadays.




\begin{table}[hbt]
\centering
\begin{tabular}{lr}
\hlinewd{1.2pt}
                         & \multicolumn{1}{c}{Extracted Frontier Queue from our BFS } \\ \hlinewd{1.2pt}
Experiment platform      & ``Creek''                      \\
Prob. Distribution       & Uniform (slightly skewed)                        \\
Empirical Entropy        & 14.58845 bit                            \\
Skewness                 & -0.003576766                                \\
Integer Range            & [1-65532]                                  \\
Total integers in sample & 29899                                    \\ 
\hlinewd{1.2pt}
                         & TREC GOV2 Corpus                             \\ \hlinewd{1.2pt}
Experiment platform      & ``Creek''                       \\
Prob. Distribution       & Power Law 			\\                       
Total integers in sample & 23918M                                  \\ \hlinewd{1.2pt}
\end{tabular}
\caption{Configuration used in both the real sample of a Frontier Queue transfer and the TREC GOV2 Corpus. Both tests have been performed in the ``Creek'' experiment platform (Table \ref{table:creekplatform}). }
\label{table:configurationcomparisons}
\end{table}

Also, as aforementioned we are including in this work a compression based on GPU (which we believe can provide interesting performance benefits in comparison to the two ones presented in this work - based on CPU's \emph{SIMD instruction-sets}). Regarding compression implemented on GPUs, it has to be noted than currently, due to the limitation of a maximum integer size of 32-bit in the GPUs cores's ALUs, the use of GPU compression on a graph 500 application would limit the maximum achievable scale of the BFS to 32 (\(2^{32} Vertices\)). The option of using a 64-bit integers with non native hardware support would penalize the compression. \cite{ueno-et-al}.

In order to test compression our chosen Frame of Reference codec in GPGPU we are configuring our benchmarking tool to use a data corpus \footnote{TREC GOV2 corpus without sensible data, and valid licence for this work. \url{http://ir.dcs.gla.ac.uk/test_collections/access_to_data.html} } used in papers from other authors, where the algorithm for our CUDA compression (\emph{Para-PFOR}) is tested with the same dataset and on a similar platform. Our used system is a node of our \emph{``Creek''} cluster (Table \ref{table:creekplatform}).\medskip

The result of this comparison can be seen in Table \ref{table:parapforcomparison}. As a first result it can be noted that the decompression speeds of the \emph{Para-PFOR} algorithm on a Fermi architecture are faster than any other \emph{SIMD} techniques implemented on CPU. Regarding the \emph{compression ratio} we can observe that it is lower than other more optimized algorithms. The reasoning behind this is that the \emph{Para-PFOR} scheme sacrifices its \emph{exception} management   to benefit the overall decompression speed \cite{parapfor-gpufor}.\medskip 

Our results of this theoretical simulation\footnote{Based on a comparison with other existent research \cite{parapfor-gpufor} on a comparable environment. } of the CPU and CUDA compressions are the following.

\begin{itemize}
	\item the performance of the \emph{Para-PFOR} using GPGPUs performs better than other \emph{SIMD} based algorithms, but lacks of a good compression ratio. The latter disadvantage may be alleviated by using the \emph{exception} management of other PFOR codecs.
  
	\item the usage of the compression directly in the GPU device may result interesting because it may minimize the memory footprint of the used data structures of our BFS (in case of used encoded buffers directly in the card). Unfortunately, even though we would be able to use more data of the graph in the limited GPU devices' memory, we would still be limited to an scale of 32, if we want to keep the performance of the compression algorithm low (no native 32-bit integer support on GPU devices \cite{ueno-et-al}).   
\end{itemize}

\subsection{Observations about experimental performance}
\label{sectionexpperfobservations}

As it can be seen in the figures of the previous results, the described above algorithms are focused for their usage in IRSs and therefore designed to achieve a high \emph{compression ratio} with high decompression speed (Table \ref{table:compressionperformance}).\medskip

The higher importance of the decompression speed over the compression one is founded in that these systems need to access (decompress) \emph{inverted indexes} at near real-time with optimally almost zero latency. 

In our integration of these integer sequence compression schemes for a \emph{graph 500} application we should focus on the schemes offering a  good \emph{compression ratio} and balanced good \(\texttt{compression} \longleftrightarrow \texttt{decompression}\) speeds.\medskip

In this work three compression libraries have been integrated  (two of them partially). Each of them features a main optimised codec. Based on Table \ref{table:compressionperformance} and Table \ref{table:parapforcomparison} (resulted from comparing results from similar test environments), we will give an insight on how each codec fits in the \emph{Graph 500} application.

\begin{enumerate}
	\item \textbf{Turbo-PFOR ``Turbo-PFOR''} This library has only been partially integrated. The included codec \emph{Turbo-PFOR} offers a near 50\% better compression ratio than any other in this work. The decompression speed performs as well as \emph{SIMD-FastPFOR} (Lemire et al. library). As downside the compression speed is lower. We believe this scheme may offer promising results. 
	\item \textbf{Lemire et al. ``S4-BP128''} This is the library used and integrated in this work. Its most competitive codec S4-BP128-D4 (tested in this work) offers both a good compression ratio and good (and balanced) compression and decompression speeds. 
	\item \textbf{Alenka CUDA ``Para-PFOR''} The partially integrated Alenka library, includes as main algorithm \emph{Para-PFOR} implemented over CUDA. This codec has been described, and also theoretically compared here. The comparison has been made by adapting the data from other works (a dataset\footnote{TREC GOV2 corpus without sensible data, and valid licence for this work. \url{http://ir.dcs.gla.ac.uk/test_collections/access_to_data.html} } has been executed in a system with similar capabilities than the one in the paper). As a result this codec seems promising because with a close \emph{compression ratio} to the one used in this work (\emph{S4-BP128}) it outperforms all codecs in terms of decompression speed \footnote{This result has been calculated using theoretical data.}. The compression speeds are not listed in the reviewed technical document and therefore unknown.\medskip

	 
\end{enumerate}

\begin{table}[hbt]
\centering
\adjustbox{width=0.95\columnwidth}{%
\begin{tabular}{cccccc}
\hlinewd{1.2pt}
C. ratio (\%) & Bits/Integer & C. speed MI/s & D. speed MI/s & Codec           \\ \hline
\emph{45,56}      & \emph{14.58}        & -             & -             & \emph{\(H(x)_{empirical}\)} \\
47.15      & \textbf{15.09}        & 3094.58       & \textbf{5068.97}       & VB-TurboBP128      \\
47.15      & 15.09        & 428.43        & 4694.21       & TurboPFOR       \\
47.15      & 15.09        & 1762.95       & 2850.89       & TurboBP128      \\
47.15      & 15.09        & 1762.75       & 2503.91       & TurboFOR        \\
47.21      & 15.11        & \textbf{3207.85}       & 4701.06       & \textbf{S4-BP128 (Used)} \\
56.14      & 17.96        & 182.82        & 1763.01       & varint-G8IU     \\
57.86      & 18.52        & 320.88        & 1259.19       & VSEncoding      \\
68.69      & 21.98        & 427.30        & 579.62        & TurboVbyte      \\
68.69      & 21.98        & 587.20        & 381.70        & Variable Byte   \\
68.69      & 21.98        & 629.42        & 282.95        & MaskedVByte     \\
100.00     & 32.00        & 3399.26       & 4535.60       & \emph{Copy (No C / D)}           \\ \hlinewd{1.2pt}
-    & \space  -        & \space -       & -     & CUDA ParaPFOR \\ \hlinewd{1.2pt}
\end{tabular}
}
\\ \caption{\textbf{Comparison of the compression codecs on the ``Creek'' experiment platform (Table \ref{table:creekplatform}). The test uses 64-bit unsigned integers as data (even though the benchmark implementation also allows 32-bit tests and uses the latter as reference size). The empathized figures on the table represent the more optimum values on each column. The use integer size meets criteria of \emph{Graph 500} for the vertexes. }}
\label{table:compressionperformance}
\end{table}

\begin{table}[hbt]
\centering
\adjustbox{width=0.95\columnwidth}{%
\begin{tabular}{lrrrrr}
\hlinewd{1.2pt}
C. ratio (\%) & Bits/Integer & C Speed MI/s & D Speed MI/s & Function        \\ \hlinewd{1.2pt}
14.04    & 4.49         & 224.83      & 742.90      & TurboPFOR       \\
14.64    & 4.68         & 190.15      & 548.45      & VSimple         \\
16.56    & 5.30         & 242.81      & 396.39      & LZ4             \\
19.58    & 6.27         & 511.25      & 641.76      & TurboVbyte      \\
21.75    & 6.96         & 996.56      &  1244.29     & S4-BP128 (Used)  \\
26.00     & 8.32         & 1197.39     & 1240.31     & TurboFOR        \\
28.01    & 8.96         & 407.32      & 358.39      & Vbyte FPF \\
31.87    & 10.20        & 135.34      & 879.72      & varint-G8IU      \\
100.00   & 32.00        & 1005.10     & 1005.62     & \emph{Copy (No C/D)}       \\ \hlinewd{1.2pt}
\textbf{27.54}    & \space   N/A       & \space  N/A      & \textbf{1300.00}     & \textbf{CUDA ParaPFOR}   \\ \hlinewd{1.2pt}
\end{tabular}
}
\caption{S4-BP128 and Turbo-PFOR compared to a similar CUDA Para-PFOR setup in an external paper. The data used is the TREC GOV2 Corpus.}
\label{table:parapforcomparison}
\end{table}

\subsection{Usage of compression thresholds}
\label{sectioncolumnandrowcompressionthresholds}

By \emph{compression thresholds} we refer to the minimal size of the integer sequence, required to trigger the compression call. The motivation behind this concept is that the compression routines add an extra overhead to the Processor Unit (CPU or GPU, depending of the used library). This overhead is sometimes not justified in terms of benefit from the compression. 

This is the case of very short sequence of integers where the time of performing  \(\texttt{Compression} \longrightarrow \texttt{Transmission} \longrightarrow  \texttt{Decompression}\) is greater than the one achieved by just \(\texttt{Transmission}\).\medskip

Regarding this optimization, we have not noticed significant benefit in its use. However, it will be discussed in section Future work (Section \ref{futurework}) this technique may help other improvements.

\subsection{Communication before and after compression}
\label{sectioncolumnandrowcompressionalgorithm}

In this section it will be described the algorithms of the communication with the integrated compression call. As it was described in the section \emph{Graphs 500 Challenge} (Section \ref{graphcomputations500}), the specifications of the \emph{Graph 500} benchmark specify Vertex labels of 48-bit unsigned integers. To be compressed, our integer sequences (defined as 64-bit integer) we have proposed three codecs (i) S4-BP128 (used), (ii) Turbo-PFOR and (iii) Para-PFOR (over CUDA). The first one operates over 32-bit integers. The last two codecs operate over 64-bit integers.\medskip

To be able to fit the 64-bit integer sequence, on the used compressor (S4-BP128, 32-bit), we have favourably used the fact that our Graph generator did not overpassed the 32-bit range \([0, 2^{32})\), due to the Scale used in the experiments. As a result of this the 64-bit integers could be pre-converted to 32-bit and post-converted again to 64-bit.\medskip

The synthesis of the compression-integration is detailed in the Algorithm

\begin{algorithm}
\caption{Our BFS algorithm with SpVM and compression}
\label{alggeneral}
\begin{algorithmic}[1]
\Statex Input : s: source vertex id
\Statex Output : $\Pi$: Predecesor list
\Statex f: Frontier queue / Next Queue
\Statex t: Current queue
\Statex A: Adjacency matrix
\Statex $\Pi$: Visited node /predecessor list 
\Statex $\otimes$: SpMV multiplication
\Statex $\odot$: element-wise multiplication
\State $f(s) \gets s$
\For{\textbf{each} processor $P_{i,j}$ in parallel}
\While{ f $\neq \varnothing$}
\State $TransposeVector(f_{i,j})$
\State $f_{i}^{'} \gets compress(f_{i})$ 
\State $f_{i}^{''} \gets \textbf{ALLGATHERV}(f_{i}^{'}, P_{\ast ,j})$
\State $f_{i} \gets decompress(f_{i}^{''})$ 
\State $t_{i} \gets A_{i,j} \otimes f_{i}$
\State $t_{i}^{'} \gets compress(t_{i})$ 
\State $t_{i,j}^{''} \gets \textbf{ALLTOALLV}(t_{i}^{'}, P_{i, \ast})$
\State $t_{i,j} \gets decompress(t_{i,j}^{''})$ 
\State $t_{i,j} \gets t_{i, j} \odot \overline{\Pi_{i}}$
\State $\Pi_{i,j} \gets \Pi_{i,j} + t_{i,j}$ 
\State $f_{i,j} \gets t_{i,j}$
\EndWhile
\EndFor
\end{algorithmic}
\end{algorithm}
  
\begin{algorithm}
\caption{Integration of compression. Detailed}
\label{algcompintegration}
\begin{algorithmic}[1]
\Statex \space
\Procedure{Compression}{input, output\_compressed64}
	\State $input32 \gets \Call{transform32}{input}$
	\State $output32 \gets \Call{compress\_library\_call}{input32}$
	\State $output\_compressed64 \gets \Call{transform64}{output32}$	
\EndProcedure 
\Statex \space
\Procedure{Decompression}{input\_compressed64, output}
	\State $input32 \gets \Call{transform32}{input\_compressed64}$
	\State $output32 \gets \Call{decompress\_library\_call}{input32}$
	\State $output \gets \Call{transform64}{output32}$	
\EndProcedure
\Statex \space
\end{algorithmic}
\end{algorithm}

\chapter{Optimizing instruction overhead}
\label{sectionoptimizinginstructions}

\begin{itshape}
As a second group of optimizations made to our BFS base implementation we will be testing some optimizations based upon the efficient programming of our algorithm. In these optimizations we will be covering five areas: (i) \texttt{scalar optimizations} (ii) \texttt{vectorization} (iii) \texttt{thread parallelism} (iv) \texttt{memory access}, and (v) \texttt{communication patterns}. For each of these subsections, small code snippets illustrating the optimizations will be given. At the end of the section, it will be included a checkbox table showing which optimizations have been added. The results are listed in the Final Ressults section (Section \ref{sectionfinalresults}).      
\end{itshape}

\section{Scalar optimizations}
\label{sectionoptinstructionscalars}

Scalar operations \slash \space data are those which involve common data types or functions where the data consist of a single value which can not be operated in parallel. Examples of these in C language are an integer, a char or a data struct, and also any operation involving the use of any on them.   

In this section, some methods to improve the performance with scalar operations will be discussed. On each technique it will be annotated whether (or not) that technique has been used in the \emph{Graph 500} implementation, and where.

\subsection{Strength reduction}
\label{sectionoptinstructionscalarsstrengthred}

This technique has been intensively used in our code. By \emph{Strength reduction} we cover transformations in the data and functions that avoid either unnecessary value conversions (or floating point operations), and expensive arithmetical functions by the usage of its reciprocal. Some sample transformations are listed below using short code snippets.\medskip

\begin{enumerate}
	\item Usage of the correct value for its specific data type. In C language this would encompass the usage of the suffixes \emph{``L'',``U''} or \emph{``LU''} for \emph{Longs}, \emph{Unsigned} values or both the previous. 
 
This will also include the case of the suffix \emph{``f''} for a float value. 
By doing this and keeping a coherence in the arithmetic or floating point operations, unnecessary conversions are avoided.   
	\item The replacement of arithmetical operations (with big overhead) by a reciprocal (with less overhead) will also improve the results in the case of operations in critical loops. This is the case of replacements of Divisions (slower) by multiplications (faser).    
\end{enumerate}

\begin{figure}
\centering
\begin{BVerbatim}[commandchars=\\\{\}]
 
// Bad: 2 is "int"
long b=a*2;

// Bad: overflow
long n=100000*100000;

// Bad: excessive
float p=6.283185307179586;\space \space 

// Bad: 2 is "int"
float q=2*p;

// Bad: 1e9 is "double"
float r=1e9*p;

// Bad: 1 is "int"
double t=s+1;
\end{BVerbatim}
\begin{BVerbatim}[commandchars=\\\{\}]

// Good: 2L is "long"
long b=a*2L;

// Good: correct
long n=100000L*100000L;

// Good: accurate
float p=6.283185f;

// Good: 2.0f is "float"
float q=2.0f*p;

// Good: 1e9f is "float"
float r=1e9f*p;

// Good: 1.0 is "double"
double t=s+1.0;
\end{BVerbatim}
\caption{Strength reduction in variables and constants.}
\end{figure}

\begin{figure}
\centering
\begin{BVerbatim}[commandchars=\\\{\}]
 
// Bad: 3.14 is a double
float x = 3.14;

// Bad: sin() is a
// double precision function
float s = sin(x)

// Bad: round() takes double
// and returns double
long v = round(X);

// Bad: abs() is not from IML
// it takes int and returns int
int V = abs(x);
\end{BVerbatim}
\begin{BVerbatim}[commandchars=\\\{\}]

// Good: 3.14f is a float
float X = 3.14f;

// Good: sin() is a
// single precision function
float s = sinf(x)

// Good: lroundf() takes float
// and returns long
long V = lroundf(x);

// Good: fabsf() is from IML
// It takes and returns a float
float v = fabsf(X);
\end{BVerbatim}
\caption{Strength reduction with funtions}
\end{figure}

\subsubsection{Replacement of expressions with much overhead with cheaper ones}
\label{sectionoptinstructionscalarbitwise}

By expensive expressions we mean long operations involving some of the previous  division/multiplication or transcendental functions which can be partially pre-computed and transformed manually outside a critical loop, and substituted with an equivalent expression.

\begin{figure}
\centering
\begin{BVerbatim}[commandchars=\\\{\}]
 
// Common Subexpression Elimination \space \space 
// change division 
for (i = O; i < n; i++) \space \space 
	A[i] /= B;

// Algebraic transformations
// old:
for (i = O; i < n; i++)
	P[i] = (Q[i]/R[i])/S[i];

\end{BVerbatim}
\begin{BVerbatim}[commandchars=\\\{\}]

// to multiplication
const float Br = 1.0f/B;
for (i = 0; i < n; i++)
	A[i] *= Br;

// new:
for (i = 0; i < n; i++)
	P[i] = Q[i]/(R[i]*S[i]); \space \space

\end{BVerbatim}
\caption{The operations on the right have less overhead, than their left counterparts.}
\end{figure}

\subsubsection{Algebraic transformations. Usage of equivalent functions with less overhead}
\label{sectionoptinstructionscalarhwsupport}

The usage of transcendental functions with hardware support instead of their analogous. replacement of exp(), log() or pow(x, -0.5) with exp2(), pow2() or 1.0/sqrt(). This kind of strength reduction can also accelerate the global result when the operation is performed inside a critical loop.

\begin{figure}[hbt]
\centering
\begin{BVerbatim}[commandchars=\\\{\}]

// No hardware support 
double r = pow(r2, -0.5);
double v = exp(x);
double y = y0*exp(log(x/x0)*
	1og(y1/y0)/1og(x1/x0));
 
\end{BVerbatim}
\begin{BVerbatim}[commandchars=\\\{\}]

// Equivalent functions with hardware support
double r = 1.0/sqrt(r2);
double v = exp2(x*1.44269504089);
double y = y0*exp2(1og2(x/xO)*
	log2(y1/y0)/1og2(x1/xO));

\end{BVerbatim}
\caption{Usage of transcendental functions with hardware support}
\end{figure}

\subsubsection{Complexity reduction in loops by the use of caching }
\label{sectionoptinstructionscalarlowercomplexity}

This technique, also called \emph{loop hoisting} or \emph{loop invariant code-motion} in the literature can be placed also in the next subsection memory access optimizations. 

This technique involves loops consisting of several nested loops. This kind of transformations are usually detected and performed by the compiler. In some cases where the invariant variable is a vector, the compiler may skip it.\medskip

In this optimization, a memory position (e.g. a vector) indexed with the outer loop index does not need to be re-accessed in the inner loop. This variable can be copied to a scalar variable (``cached'') outside the inner loop and therefore prevent the positioning within the vector.       

\subsubsection{Efficient bitwise operations}
\label{sectionoptinstructionscalarbitwise}

The optimal bitwise transformations have been used intensively in this implementation. Bit-level operations are faster than their higher level analogous. Also, the usage of optimal methods which prevent branching whilst the bitwise operation is performed makes the result more optimal. 

\subsubsection{Control of floating point semantics}
\label{sectionoptinstructionscalarfpsemantics}

These optimizations (also called \emph{Precision control}) are usually activated in the compiler and therefore would also fit in the subsection about compiler optimizations. However, because they enable a balance between precision and speed of floating point scalars are placed in this section.

The control is done at compile time and allows the selection of multiple levels of precision. The lower the level of precision in the \emph{Arithmetic Logic Unit}, the faster the resulting code will be. \medskip

Because the \emph{Graph 500} test is not a floating point intensive benchmark, this implementation, as well as others like Titech's  \cite{ueno-et-al} use the lowest possible precision mode to gain performance.

\section{Vectorization}
\label{sectionvectorization}

As previously defined in the section \emph{Scalar optimizations} a variable based upon multiple data values allows the possibility of being vectorized. This is for example the case of an array in C. 

By saying that the variable can be vectorized we mean that in case some requirements are met, the variable will be processed by a \emph{Vector Processing Unit} (VPU) within the CPU. 

This will be done with the help of specific Instruction-sets (\emph{SSE, AVX}, etc in x86 architectures or \emph{AltiVec} in PowerPC ones). These instruction-sets are an example of \emph{SIMD} designs described in Section \ref{sectionflynns}.\medskip

When a vector is processed in \emph{SIMD} fashion (by the \emph{VPU}), as opposed to when it is processed as an scalar by the \emph{Arithmetic Processing Unit} (ALU), the calculations are accelerated by a factor of the number of values fitting the register. As it can be seen in the table \ref{intelinstructionset}, an Intel\texttrademark \space AVX instruction (256-bit width register) operating over 32-bit integers will enable an assembly instruction to operate over 8 integers at once.

A way to look at this is, if the code is not using VPUs where it could use them, it is performing only \emph{\( \frac{1}{8}\)th} of the possible achievable arithmetic performance.\medskip    

To manage the usage of the \emph{SIMD} instruction-set there are two possible approaches. 

\begin{enumerate}
	\item The usage of in-line assembly or special functions in higher level languages called intrinsics.
	\item Regular code in a way that allows the compiler to use vector instructions for those loops. This approach is more portable between architectures (requires a re-compilation). Also, it is more portable for future hardware with different instruction-sets.   
\end{enumerate}

The vectorization made in this work focuses on the second approach: \emph{automatic-vectorization} by the compiler. 

As the concept of aligned buffer is used multiple times in this sections we define this below. 
\[
\texttt{T *p is n-T aligned if  p \% n = 0}
\]

Where \texttt{T} is the type of the pointer and \texttt{p} contains its memory address.

\begin{figure}[hbt]
\centering
\begin{BVerbatim}

void main()
{ 
const int n=8; 
int i;
// intel compiler buffer allocation using 256-bit alignment (AVX) 
int A[n] __attribute__((a1igned(32))); 
int B[n] __attribute__((aligned(32))); 

// Initialization. There is scalar, thus not vectorized 
for (i=0; i<n; i++) 
	A[i]=B[i]=i; 
		
// This loop will be auto-vectorized 
for (i=0; i<n; i++) 
	A[i]+=B[i]; 
}
\end{BVerbatim}
\caption{Automatic vectorization of a for-loop on the main() section. Code for Intel compilers.}
\end{figure}

\begin{figure}[hbt]
\centering
\begin{BVerbatim}
// compiler is hinted to remove data aliasing with the ``restrict'' keyword.
void vectorizable_function(double * restrict a, double * restrict b)
{
	size_t i;

	// hint to the compiler indicating that buffers
	// has been created with 128-bit alignment (SSE+ in Intel)
	double *x = __builtin_assume_aligned(a, 16);
	double *y = __builtin_assume_aligned(b, 16);

	// for-loop size, is known
	for (i = 0; i < SIZE; i++)
	{
		// vector-only operations
		x[i] = y[i + 1];
	}
}
\end{BVerbatim}
\caption{Automatic vectorization of a buffer containing data dependencies (aliasing). The buffers are used at a different block level (function call) than they have been declared.}
\end{figure}

\begin{table}[hbt]
\centering
\adjustbox{width=0.98\columnwidth}{%
\begin{tabular}{cccc}
\hline
Instruction Set & Year and Intel Processor & Vector registers & Packed Data Types                                                             \\ \hline
SSE             & 1999, Pentium III        & 128-bit          & 32-bit single precision FP                                                    \\
SSE2            & 2001, Pentium 4          & 128-bit          & 8- to 64-bit integers; SP \& DP FP                                            \\
SSE3-SSE4.2     & 2004 - 2009              & 128-bit          & (additional instructions)                                                     \\
AVX             & 2011, Sandy Bridge       & 256-bit          & single and double precision FP                                                \\
AVX2            & 2013, Haswell            & 256-bit          & integers, additional instructions                                             \\
IMCI            & 2012, Knights Corner     & 512-bit          & 32- and 64-bit ints. Intel Xeon Phi \\
AVX-512         & (future) Knights Landing & 512-bit          & 32- and 64-bit ints. SP \& DP FP                        \\ \hline
\end{tabular}
}
\caption{SIMD instruction sets in Intel\texttrademark \space x86 architectures. The column \emph{\texttt{Vector registers}} shows the width of the Vector processing unit in \emph{bits}. This value is the optimum alignment for buffers/ arrays / vectors which are prone to be vectorized.}
\label{intelinstructionset}
\end{table}

\subsubsection{Optimization of critical loops}
\label{sectionoptvectorloops}

This is one of the main focus on instruction overhead reduction in this work. An introduction to this technique has done in the \emph{Vectorisation} section above. 

As it has been described, the usage of this technique over code with many loop iterations and array operations, may get a higher benefit from this. However, some requirements need to be met for the automatic-vectorisation being performed. These will be explained later in this section.

The main benefit of the usage of this technique on our application is in the second Kernel (BFS + Validation), and more specifically on the validation code.\medskip

Our validation code is an adaptation of the reference code provided by the \emph{Graph 500}\footnote{\url{http://www.graph500.org/referencecode}} community, with the addition of a 2D partitioning \cite{ueno-et-al}. The matrixes operations are performed by the CPU and iterated with non optimized loops which miss the chance of any possible vectorization.

These optimizations based on array vectorisations have been proved to be  successful on other state-of-the-art implementations like Ueno et al. \cite{ueno-et-al}. Also, in other works not directly related with the \emph{Graph 500}, which also makes intensive usage of arrays (the  compression library \cite{lemire}) this optimizations have resulted successful.\medskip

In order to successfully manage automatic-vectorisation the requirements listed below must be met. 

\begin{itemize}
	\item The loop must be a \emph{for-loop}. Other kinds like \emph{while} will not be vectorised.
	\item The number of iterations of the loop must be known either in compile or run time.
	\item The loop cannot contain complex memory accesses, i.e. memory accesses must have a regular pattern (ideally by the use a unit stride. e.g. V[n+STRIDE]).  

Note that some compilers may manage these issues, however this is not the general rule.
	\item The loop must not contain branching (for example \emph{if-then-else} statements).
	\item The data to be vectorised must not depend on other index pointing to itself (for example, a compiler will not vectorise data in the form  V[i]= V[i-1]. Sometimes even though a dependency may exist in the code, the compiler can be made aware to ignore it with the use of language keywords, like for example \emph{restrict} in C)     
\end{itemize}

In addition to these restrictions which applies to the loop, the compiler must be aware of this action by the use of a -O flag. Also, the internal structure of the data vector (this will be explained in further detail later in this chapter) must be declared ``aligned'' in memory. Only when the code is programmed accordingly, the compiler will detect and automatically vectorize the loops.  Even though these improvements may have impact in overall results , due to time constrains, they have been implemented partially. This  is reflected in the section \emph{Future work} (Section \ref{futurework}). \medskip

In other implementations of \emph{Graph 500}, \cite{ueno-et-al}, the vectorization of array structures executing on the CPU has been performed on detail. For that matter a fixed alignment of 128-bit (would allow \emph{Intel \texttrademark \space SSE*} instructions on x86 architectures) is used for the general memory data buffers. Also, other data structures which require temporal locality use a different alignment constants(``\emph{CACHE\_LINE\_ALIGNMENT}'', set to 32-bit for 32-bit integers). These tasks are performed on compile-time by the use of \emph{wrappers} over the functions that allocate the memory.

\subsubsection{Regularization of vectorization patterns}
\label{sectionoptvectorpatterns}

In auto-vectorisation, even though the compiler detects loops with incorrect alignment, it will try to vectorize them by adding scalar iterations at the beginning (called \emph{``peel''}), and also at the end  (called \emph{``tail''}). The rest of the vector will be vectorized normally. To perform a regularization of the vector a padding is introduced through a second container. Also, the compiler is made aware of this not required operation by the usage of a \emph{``pragma''}. \medskip

This optimization could be done in our \emph{Graph 500} application to increase the amount of vectorised loops. These transformations have not been implemented and are listed here as reference.

\subsubsection{Usage of data structures with stride memory access}
\label{sectionoptvectorstridemem}

When a data structure must be designed to solve a problem, sometimes two approaches are possible. Both are illustrated with code snippets below.

\begin{figure}[hbt]
\centering
\begin{BVerbatim}

// Example of an Array of Structures (AoS) - To avoid
struct AoS { // Elegant, but ineffective data layout
float X, y, Z, q;
};
AoS irregularMemoryAccess[m];

// Example of Struct of Array (SoA). Preferred data structure
struct SoA {
// Data layout permits effective vectorization due to the 
// uniform (n-strided) memory access

const int m; 
float * X; // Array of m-coordinates of charges
float * y; // ...y-coordinates...
float * Z; // ...etc.
float * q; // These arrays are allocated in the constructor
};

\end{BVerbatim}
\caption{Usage of data structures with stride memory access. Array of Structures (AoS) over Structure of Arrays (SoA)}
\end{figure}  

\subsection{Compiler optimizations}
\label{sectionoptinstructioncompiler}

To perform the optimizations, one possible strategy is the use of a preprocessor tool that prepares the build based on selected parameters, or detected features in the host machine's hardware. \medskip

The selection of build tool was difficult as currently exist multiple, and very similar options (e.g. \emph{cmake}, \emph{Automake}, etc). In our case we selected our decision on the ease of porting our current Built scripts to other tool. For this matter we chose \emph{Automake}, a mature project with active support developed by GNU.

Apart from being able to select options in our build or automatically detect optimal hardware features, it is possible to select good build parameters based on good practices. Those are exposed below.      

\subsubsection{Usage of an unique and optimal instruction-set}
\label{sectionoptinstructioncompileruniqueiset}

As it can be extracted from white papers of commercial or open source compilers\cite {intelgcc-whitepaper}, a good practise to follow when creating the binary code for an application is to specify the target architecture on which that code will run. This will enable the compile to select that known applicable software support for the specified hardware. Also the avoidance of including several instruction-sets which will increase the size of the generated binary code may impact on the overall performance. \medskip

As it has been specified in the introduction of this subsection, by the usage of a preprocessor tool, several tests are done before the build to detect the host platform. The assigned architecture will depend on this result. \medskip

The Table \ref{table:compilercomparison} compares the used compiler parameters for different Graph 500 implementations. The explicit usage of a pre checked target architecture may be seen under the column \emph{march} in that table.

\subsubsection{Inclusion of performance oriented features}
\label{sectionoptinstructioncompileaddflags}

Some research has been done on the benefits on applying \emph{Inter-procedural optimizations} (IPO) to array-intensive applications to improve the general  cache hit rate \cite{ipographs}. By contrast with other implementations like \cite{ueno-et-al} this implementation applies this technique. \medskip

Other performance oriented parameters added to the compiler are the -O and -pipe flags. In Table \ref{table:compilercomparison} can be seen a comparison of the different configurations of common \emph{Graph 500} implementations.

\subsubsection{Removal of non performance-oriented features}
\label{sectionoptinstructioncompilerremoveflags}

In contrast with the previous compiler parameters, other parameters are known to reduce the overall performance \cite{pie}. The removal of the -fPie flag from our previous implementation may have influenced in the performance increasement.

\subsubsection{Ensure auto-vectorisation is enabled in the compiler}
\label{sectionoptinstructioncompilerautovect}

All the requirements listed in the section \emph{Vector optimisations} are enabled by making the compiler aware of this. Usually vector optimisations are performed when the -O level is greater then a value. The value depends on the compiler. For the specific case of GNU's gcc compiler this value equals ``2'' (-O2).  

\begin{table}[hbt]
\centering
\adjustbox{width=0.95\columnwidth}{%
\begin{tabular}{lcccccccc}
\hlinewd{1.2pt}
Implementation       & Maketool & march & CUDA caps. & -O  & fmath & pipe & ipo & other \\ \hlinewd{1.2pt}
\emph{Graph500}'s ref. & No        & No             & No                & 4        & Yes    & No         & No  & -      \\
Ours (previous)      & No        & No             & No                & 3        & Yes    & Yes        & No  & -fPie  \\
Ours (new)           & automake  & Yes            & compile-time      & 3        & Yes    & Yes        & Yes & -      \\
Titechs              & No        & No             & No                & 3        & Yes    & No         & Yes & -g     \\
Lonestar-merril      & cmake     & No             & run-time          & 3        & No     & No         & No  & -g     \\ \hlinewd{1.2pt}
\end{tabular}
}
\caption{Comparisom of several build parameter for several open graph500 implementations. \emph{\texttt{Maketool}}=Build framework (e.g. automake, cmake, none). \emph{\texttt{march}}=whether or not the target architecture is detected at pre-process time. \emph{\texttt{CUDA caps}}=whether or not CUDA capabilities are detected (and when). \emph{\texttt{-0}}=default optimization level. \emph{\texttt{fmath}}=Control of floating point semantics by default. \emph{\texttt{pipe}}=pipe optimization enabled by default. \emph{\texttt{ipo}}=Linker-level optimization enabled by default (Inter-procedural Optimizations). \emph{\texttt{other}}=other compile flags. }
\label{table:compilercomparison}
\end{table}

\subsubsection{Ensure auto-vectorisation has been successfully enabled}
\label{sectionoptinstructioncompilerautovectenabled}

The \emph{Compiler reporter} is a reporting mechanism that outputs all the successfully performed optimizations automatically. It is usually activated by a flag.

By checking that the code transformations that have been performed we can confirm the changes in the generated code. \medskip

Due to time restrictions this testing was not fully performed in our \emph{Graph 500} application, and the part of Vectorisation is only partially verified.

\begin{figure}[hbt]
\centering
\begin{BVerbatim}

host% icpc -c code.cc -qopt-report -qopt-report-phase:vec
host% cat vdep.optrpt

LOOP BEGIN at code.cc(4,1)
<Multiversioned v1>
remark #25228: LOOP WAS VECTORIZED
LOOP END

\end{BVerbatim}
\caption{Ensure auto-vectorisation has been successfully enabled. Example for an Intel\texttrademark \space compiler}
\end{figure}

\section{Thread parallelism}
\label{sectionoptinstructionthreads}

The previusly described optimizations act over the achievable parallelism in the \emph{data layer}. In this new layer it is possible to generate serial codes executed concurrently by the creation of ``\emph{threads}''.
In HPC, this threads make use of the parallelism that introduces the multiple cores within the processor.\medskip

These threads must be used with a parallel Framework. Some example implementations of parallel threads Frameworks are Pthreads, OpenMP, Cilk+, Boost threads.

We will focus this work on \textbf{OpenMP} as it is the most popular and standardised framework, with the most active support. 
It implements one of the most common parallel patterns in HPC, the parallel-for or \emph{Map-reduce model}. On it, when a \emph{for} needs to iterate \emph{n} items and uses this model, the framework distributes the items by \emph{X}.\medskip

Concurrent paradigms may increase the overall performance. However, they have the downside of an added programming complexity. The writes and reads in shared memory by multiple codes may create \emph{``race conditions''}.

Put simple a \emph{race condition} is a non-deterministic state where the multiple processes have own copies of a variable and operate over it. When at some point the variable is copied to a global position, different threads will change the value and it will be the value of the slowest process, the one remaining in memory. 

In our \emph{Graph 500} implementation we were required to constantly test the application when the OpenMP threads was re-adjusted.\medskip

For a better understanding of the section \emph{Thread contention} below, the concept of synchronisation of threads, and solutions for the \emph{race condition} issue will be defined next. The problem of race conditions is usually solved ``\emph{locking}'' one or more threads. This is called \emph{synchronization}.

\subsection{Usage of fork-join model when suitable}
\label{sectionoptinstructionthreadsforkjoin}

In contrast with the parallel-for (or \emph{map-reduce} model), the OpenMP framework allows also the usage of a very mature model based on \emph{Parent} and \emph{Child} processes. 

At some cases when it is not possible the usage of a \emph{Map-reduce} model, it is recommended to study the possibility of including this one.

\subsection{Optimal thread scheduling}
\label{sectionoptinstructionthreadsscheduling}

The threads framework OpenMP, in the \emph{Map-reduce} model (parallel-for) allows the implicit specification on how the load of tasks per thread will be balanced. This is done by specifying a scheduling policy. The three main policies available are listed below. Also, it is specified the one that we consider the most suitable. 

\begin{description}
	\item 
	\item[static(n)] this is the default policy. Each thread has a fixed number \emph{n} of tasks to perform. Even though other threads have already finished their tasks, if one task in one thread takes longer to run, the other threads must wait.
	\item[dynamic(n)] this policy assigns dynamically the tasks according to the load in other threads. One downside of this policy is that threads may not have enough temporal locality resulting on bad cache hit rates.
	\item[guided(n, min)] this is the policy chose in our implementation. It mixes both previous scheduling policies. It is basically a dynamic policy with a minimum number of fixed tasks in the thread. The temporal and spatial locality of the accessed data may result in a better cache hit rate.
\end{description}

A note about the scheduling policies is that the correct selection is better done when the execution has been analysed with a profiler. 

\subsection{Thread contention prevention}
\label{sectionoptinstructionthreadscontention}

The threads contention may have two main causes. The first one is an excess of synchronisation (defined above). The second one is called \emph{false sharing} and is defined next.

As it was explained before an address is \emph{n}-aligned when it is multiple of \emph{n}. In the case of \emph{false sharing} the retrieved data of some threads (even though is aligned) is partially divided between two cache lines. This can translate into overhead due to complex memory accesses. In the case of a big-sized critical loop, this alignment issue may affect the maximum performance benefit resulting from the use of threads.

The issue is solved by adding a padding which corrects the alignment of the vector. \medskip

This optimization would be recommended in the critical loops of the validation code.

\begin{figure}[hbt]
\centering
\begin{BVerbatim}

// Padding to avoid sharing a cache line between threads
const int paddingBytes = 32; // Intel AVX+
const int paddingElements = paddingBytes / sizeof(int);
const int mPadded = m + (paddingElements-m%paddingElements);
int hist_containers[nThreads][mPadded]; // New container

\end{BVerbatim}
\caption{Prevent thread contention by preventing \emph{false sharing}: Use of a new padded container.}
\end{figure}     

\subsection{NUMA control / NUMA aware}
\label{sectionoptinstructionthreadsnumacontrol}

In \emph{Shared memory} systems, platforms with multiple processors have access to different banks of physical memory. The banks which physically belong to other processor are called ``remote memory'' for a given processor. The remote memory access have a greater latency penalty compare to local memory access (This can be seen in the Figure \ref{fig:numa}). This same effect happens in multicore systems and each core's first level cache. 

During the initial assignation of threads in this NUMA model, a thread running in one processor may be using the memory of a different one. All of this results in the \emph{threads} adding overhead to the application. \medskip
This issue affecting \emph{Non Uniform Memory Access} (NUMA) architectures can be alleviated with several approaches. 

\begin{itemize}
	\item Because memory allocation occurs not during initial buffer declaration, but upon the first write to the buffer (``first touch''), for better performance in NUMA systems, data can be initialized with the same parallel pattern as during data usage. This approach can be seen on Figure \ref{firsttouch}. By using this technique an initial assignation is done using the \emph{threads} framework - this prevents memory-access issues in the threads. \medskip
	\item Other approach is the one used by the current \texttt{Top 1} implementation in the Graph 500. They have implemented their own library \footnote{ULIBC - \url{https://bitbucket.org/yuichiro_yasui/ulibc}} which allows to \emph{``pin''} any data structure from memory to any specific CPU, memory bank o core.
\end{itemize}

In our work, after the analysis of the results we have included in the section \emph{Future work} (Section \ref{futurework}) a revision of our thread scheduling policy, as well as the introduction of this technique  to alleviate the bad performance when the thread execution is active.

\begin{figure}[hbt]
\centering
\begin{BVerbatim}
// Original initialization loop
for i < M
  for j < N 
    A[i * M + j] = 0.0f

// Loop initialization with ``First Touch''
#pragma omp parallel for
for i < M
  for j < N 
    A[i * M + j] = 0.0f
    
\end{BVerbatim}
\caption{Pseudo-code of allocation on \emph{First Touch}.}
\label{firsttouch}
\end{figure}

\section{Memory access}
\label{sectionoptinstructionmemaccess}

The memory access optimizations performed by alignments and more sequential structures, (as it was stated previously in the document), will help in the vectorisation of many loops. In this section a different approach is defined. This technique helps also with critical loops and may be an advantage in some zones of our second kernel (BFS + verification). \medskip 

\subsection{Locality of data access}
\label{sectionoptinstructionmemaccesslocality}

The ``Rule of Thumb'' for memory optimization is the locality of data access in
space and in time.

\begin{itemize}
	\item By \textbf{spatial locality} we mean a correlative access to the data in memory. This is managed by a proper election of the used data structures. The use of Struct of Arrays (SoA) over Arrays of Structs (AoS) improves this situation. The proper packing of the data to avoid  having to iterate in 1-stride fashion at the beginning and end of the structure (padding the structure) also improves the locality. 
	\item By \textbf{temporal locality} we mean that the required data at one point is as close as possible to the one in the next point. We manage this by changing the order of the operations (e.g., loop tiling, loop fusion or loop hoisting).
\end{itemize}

\begin{figure}[hbt]
\centering
\adjustbox{width=0.90\columnwidth}{%
\begin{BVerbatim}
// Original code
for (i = 0, i < m, i++)
	for (j = 0; ; j < n; j++)
		compute(a[i], b[j]); // Memory access is unit-stride in j 

// Step 1: strip-mine 
for (i = O; i < m; i++) 
	for (jj = 0; jj < n; j += TILE) 
		for (j = jj; j < jj + TILE; j++) 
			compute(a[i], b[j]); // Same order of operation
					     // as original 
					
// Step 2: permute 
for (jj = 0; jj < n; j += TILE) 
	for (i = O; i < m; i++) 
		for (j = jj; j < jj + TILE; j++) 
			compute(a[i], b[j]); // Re-use to j=jj sooner	
\end{BVerbatim}
}
\caption{Improve temporal locality by using loop-tiling (Cache blocking)}
\end{figure}

\subsection{Merge subsequent loops}
\label{sectionoptinstructionmemaccessfussion}

In this operation in different but subsequent loops (with same iterator distance) may be groped if the flow of the program allows this. This produces that the retrieved memory block to the CPU cache can be reused more close in time. This transformation can be seen in Figure \ref{fig:fusion}.

\begin{figure}[hbt]
\centering
\begin{BVerbatim}

// Original code
for (i = 0, i < m, i++)
	compute(a[i], b[j]); 
for (i = 0, i < m, i++)
	secondcompute(a[i], b[j]); 

// Loop fusion
for (i = 0, i < m, i++)
	compute(a[i], b[j]); 
	secondcompute(a[i], b[j]); 

\end{BVerbatim}
\caption{loop fusion}
\label{fig:fusion}
\end{figure}

\subsection{Variable caching in the outer loop}
\label{sectionoptinstructionmemaccesshoisting}

This optimization (\emph{loop hoisting}) has been described in the previous section \emph{Scalar Optimizations}. It is listed here as it is also a memory-access based optimization. As previously stated, this optimization has been included in many parts of the code (both kernels of the \emph{Graph 500} application). This transformation can be seen in Figure \ref{fig:hoisting}.  

\begin{figure}[hbt]
\centering
\adjustbox{width=0.80\columnwidth}{%
\begin{BVerbatim}

// Original code
for (i = 0, i < m, i++)
	for (j = 0, j < n, j++)
		for (k = 0, k < p, k++)
			// unnecessary and multiple access 
			// to a memory block with complexity 
			// O(m*n*k)
			compute1(a[i][i]);
			compute2(a[i][i]); 

// New code
for (i = 0, i < m, i++)
    const int z = a[i][i];
	for (j = 0, j < n, j++)
		for (k = 0, k < p, k++)
			compute1(z);
			compute2(z); 
			
\end{BVerbatim}
}
\caption{variable caching to provide better temporal and spatial locality  (\texttt{loop hoisting})}
\label{fig:hoisting}
\end{figure}

\section{Communication patterns}
\label{sectionoptcommunication}

Previously in this section, the optimizations over the multiple data in one  instruction was defined as the \emph{first layer of parallelism}. Next, in the case of the \emph{threads} which are generally used in HPC to control the cores inside a processor, it was defined the \emph{second layer of parallelism}.

In this section it will be defined a \emph{third layer}. On it, the ``parallel processors'' are each one of the nodes communicating by message passing (MPI).  

\subsection{Data transfer grouping}
\label{sectionoptcommunicationgrouping}

Small and subsequence data transfers can be prevented by grouping them in a MPI custom datatype. This improves the communication in two ways: (i) a longer usage of the network, and hence a better throughput is achieved (ii) the  memory access improves when relaying on the internal MPI packing system (Zero copy). 

We have applied these optimizations several times to alleviate the overhead of successive MPI\_broadcasts and MPI\_allgatherv for the pre distribution of communication metadata (i.e. buffer sizes in the next transfer) 

\subsection{Reduce communication overhead by overlapping}
\label{sectionoptcommunicationasynchronous}

The usage of this technique implies the continuous execution of tasks during the delays produced by the communication delay. To perform this optimization an asynchronous communication pattern, and a double (Send and Receive) buffers must be used. \medskip

This optimization has been proposed as other possible improvement in the communications and is listed un the the section \emph{Future work} (Section \ref{futurework}).

\begin{figure}[hbt]
\centering
\adjustbox{width=0.80\columnwidth}{%
\begin{BVerbatim}

// Non optimized communication

if (r == 0) {
// Overlapping comp/comm 
for(int i=0; i<size; ++i) {
    arr[i] = compute(arr, size);
}
MPI_Send(arr, size, MPI_DOUBLE, 1, 99, comm); } else {
MPI_Recv(arr, size, MPI_DOUBLE, 0, 99, comm, &stat); }

\end{BVerbatim}
}
\caption{No communication overlapped with computation}
\label{fig:hoisting}
\end{figure}

\begin{figure}[hbt]
\centering
\adjustbox{width=0.85\columnwidth}{%
\begin{BVerbatim}

// Optimized communication. Software pipeline

if (r == 0) {

  MPI_Request req=MPI_REQUEST_NULL:
    for(int b=0; b < nblocks; ++b) {
      if(b) {
        // pipelining
        if(req != MPI_REQUEST_NULL) MPI_Wait(&req, &stat);
          MPI_Isend(&arr[(b-1)*bs], bs, MPI_DOUBLE, 1, 99, comm, &req); }

          for(int i=b*bs; i<(b+1)*bs; ++i) {
            arr[i] = compute(arr, size); 
          }
    }   
      MPI_Send(&arr[(nblocks-1)*bs], bs, MPI_DOUBLE, 1, 99, comm); 
  } 
  else {

      for(int b=0; b<nblocks; ++b) {
         MPI_Recv(&arr[b*bs], bs, MPI_DOUBLE, 0, 99, comm, &stat);
      }
  }

\end{BVerbatim}
}
\caption{No communication overlapped with computation}
\label{fig:swpipelinbadg}
\end{figure}

\section{Summary of implemented optimizations}
\label{sectionsummaryimplemented}

In this section we have covered many optimizations, from which only a subset have been added into our work. Mainly, the added optimizations have been as follows:

\begin{enumerate}
	\item \textbf{Scalar optimizations}. From this group we have added all of the listed optimizations and covered most of out code. 
	\item \textbf{Vectorization}. The optimizations covered in this section been the main aim of our instruction overhead reduction. These optimizations have been added with more or less success (this will be discussed in the section Final results, Section \ref{sectionfinalresults})
	\item \textbf{Thread parallelism}. As we have mentioned above, to deal with threads we have used the OpenMP framework. When first analyzing our \texttt{Baseline} implementation we noticed a paradoxical performance loss when using threads compared to the 1 single thread case. This problem has been constant during all the development of our new 2 versions. In the last version we have achieved to slightly increase the performance of the OpenMP version over the 1-threaded one.
	\item \textbf{Memory access}. The memory access optimizations usually require the refactoring of large portions of code. For this reason sometimes it becomes more difficult the optimization of an already created code the its new reimplementation. Hence, we have only applied a few of this optimizations to the new structures that have been required to implement.
	\item \textbf{Communication patterns}. Among these optimizations we have only covered the regrouping of individual transfers performed serially into a custom MPI\_Datatype which would achieve in one transmission a bigger amount of data and a bigger use of the network resources. This modifications have been made majorly for the broadcasted metadata prior to other communications. Regarding the communication overlapping it has successfully proved to be a successful addition to a Graph 500 application \cite{numaAwareFujisawa2015, ueno-et-al} (Section \ref{relatedwork}) Despite of this, we have not implemented it in our code.    
\end{enumerate}

\chapter{Final Results}
\label{sectionfinalresults}

\begin{itshape}

In this section we present the results from our experiments and try to solve the problem detected in the previous section Analysis (Section \ref{sectionanalysismhaucksresults}). In this same section we will discuss the results and set the initial thought for the next sections, Conclusions and Future work. 

The chapter starts with data related with the systems used for testing. One is allocated in the Georgia Institute of Technology; other, is in the Rupretcht-Karls Universität Heidelberg. 








The results will be grouped depending on the type of performed optimization: (i) communication compression or (ii) instruction overhead optimizations.

\end{itshape}

\section{Experiment platforms}
\label{sectionresultsintroductiontestenvironments}






The Tables \ref{table:creekplatform} and \ref{table:kidsplatform} show the main details of the used architectures. The big cluster (Keeneland) has been used to perform main tests. The development and analysis of the compression has been performed in the smaller one. 

A third table \ref{table:victoriaplatform}  shows the specifications of the platform (also allocated in University of Heidelberg) used to test the overhead added by the compression. The experiments performed on this third platform have not been included as they have only been concluded partially. Even though, some conclusions of the executions on this machine are are added in this chapter. The table is therefore added, as a reference if needed.

\begin{table}[hbt]
\centering
\adjustbox{width=0.65\columnwidth}{%
\begin{tabular}{lc}
\hlinewd{1.2pt}
\textbf{Operating System} & 64-bit Ubuntu 12 with kernel 3.8.0     \\ \hlinewd{1.2pt}
CUDA Version              & 7.0                                    \\
MPI Version               & 1.8.3 (OpenMPI)                        \\ \hlinewd{1.2pt}
\textbf{System}           & SMP Cluster                            \\ \hlinewd{1.2pt}
Number of Nodes           & 8                                      \\
Number of CPUs / node     & 1                                      \\
Number of GPUs / node     & 2                                      \\ \hlinewd{1.2pt}
\textbf{Processor}        & Intel E5-1620 @ 3.6 GHz                \\ \hlinewd{1.2pt}
Number of cores           & 4                                      \\
Number of threads         & 2                                      \\
L1 cache size             & 4 x 64 KB                              \\
L2 cache size             & 4 x 256 KB                             \\
L3 cache size             & 1 x 10 MB                              \\
Memory                    & 4 GB (DDR3-1333), ECC disabled                       \\
Max. vector register width& 256-bit, Intel AVX                     \\
PCIe Speed (CPU \( \longleftrightarrow \) GPU)    & 5 GT/s         \\
QPI Speed (CPU \( \longleftrightarrow \) CPU)       & none 			\\  
\hlinewd{1.2pt}
\textbf{GPU Device}       & NVIDIA GTX 480 (Fermi architecture)    \\ \hlinewd{1.2pt}
Cores                     & 15 (SMs) x 32 (Cores/SM) = 480 (Cores) \\
Total system Cores        & 7680 (Cores) \\
Compute capability     	  & 2.0 \\
Memory                    & 1.5 GB, ECC disabled                                 \\
Memory Bandwidth          & 177.4 GB/s                             \\ \hlinewd{1.2pt}
\textbf{Interconnect}     & GigabitEthernet (Slow network fabric)           \\ \hlinewd{1.2pt}
Rate                      & 1 Gbit/sec                             \\ \hlinewd{1.2pt}
\end{tabular}
}
\caption{Experiment platform \textbf{``Creek''}.}
\label{table:creekplatform}
\end{table}

\begin{table}[hbt]
\centering
\adjustbox{width=0.65\columnwidth}{%
\begin{tabular}{lc}
\hlinewd{1.2pt}
\textbf{Operating System} & 64-bit Ubuntu 14 with kernel 3.19.0     \\ \hlinewd{1.2pt}
CUDA Version              & 7.0                                    \\
MPI Version(s)            & 1.8.3 (OpenMPI)                        \\ \hlinewd{1.2pt}
\textbf{System}           & SMP System                             \\ \hlinewd{1.2pt}
Number of Nodes           & 1                                      \\
Number of CPUs / node     & 2                                      \\
Number of GPUs / node     & 16                                      \\ \hlinewd{1.2pt}
\textbf{Processor}        & Intel E5-2667 @ 3.2 GHz                \\ \hlinewd{1.2pt}
Number of cores           & 8                                      \\
Number of threads         & 1                                      \\
L1 cache size             & 8 x 64 KB                              \\
L2 cache size             & 8 x 256 KB                             \\
L3 cache size             & 1 x 20 MB                              \\
Memory                    & 251 GB (DDR3-1333), ECC disabled                      \\
Max. vector register width& 256-bit, Intel AVX                     \\
PCIe Speed (CPU \( \longleftrightarrow \) GPU)      & 8 GT/s      \\
QPI Speed (CPU \( \longleftrightarrow \) CPU)       & 8 GT/s                               \\ \hlinewd{1.2pt}
\textbf{GPU Device}       & NVIDIA K80  (Kepler architecture)      \\ \hlinewd{1.2pt}
Cores                     & 13 (SM\emph{X}s) x 192 (Cores/SM\emph{X}) = 2496 (Cores) \\
Total system Cores        & 39936 (Cores) \\
Compute capability     	  & 3.7 \\
Memory                    & 12 GB, ECC disabled                                 \\
Memory Bandwidth          & 480 GB/s                              \\ \hlinewd{1.2pt}
\textbf{Interconnect}     & Internal PCIe 3.0 bus
           \\ \hlinewd{1.2pt}
Rate                      & PCIe's rate                             \\ \hlinewd{1.2pt}
\end{tabular}
}
\caption{Experiments platform \textbf{``Victoria''}.}
\label{table:victoriaplatform}
\end{table}

\begin{table}[hbt]
\begin{threeparttable}[b]
\centering
\adjustbox{width=0.65\columnwidth}{%
\begin{tabular}{lc}
\hlinewd{1.2pt}
\textbf{Operating System} & 64-bit Ubuntu 14      \\ \hlinewd{1.2pt}
CUDA Version              & 7.0                                    \\
MPI Version(s)            & 1.6.4 (OpenMPI)                        \\ \hlinewd{1.2pt}
\textbf{System}           & SMP Cluster                             \\ \hlinewd{1.2pt}
Number of Nodes           & 120 (used 64)                                      \\
Number of CPUs / node     & 2                                      \\
Number of GPUs / node     & 3 (used 1)                                    \\ \hlinewd{1.2pt}
\textbf{Processor}        & Intel E5-2667 @ 2.8 GHz                \\ \hlinewd{1.2pt}
Number of cores           & 6                                      \\
Number of threads         & 2                                      \\
L1 cache size             & 6 x 64 KB                              \\
L2 cache size             & 6 x 256 KB                             \\
L3 cache size             & 1 x 12 MB                              \\
Memory                    & 251 GB (DDR3), ECC disabled                      \\
Max. vector register width& 128-bit, Intel SSE4                     \\
PCIe Speed (CPU \( \longleftrightarrow \) GPU)      & 5 GT/s      \\
QPI Speed (CPU \( \longleftrightarrow \) CPU)       & 6.4 GT/s                               \\ \hlinewd{1.2pt}
\textbf{GPU Device}       & NVIDIA M2090  (Fermi architecture)      \\ \hlinewd{1.2pt}
Cores                     & 16 (SMs) x 32 (Cores/SM) = 512 (Cores) \\
Total system Cores        & 32768 (Cores) \\
Compute capability     	  & 2.0 \\
Memory                    & 6 GB, ECC disabled                                 \\
Memory Bandwidth          & 178 GB/s                              \\ \hlinewd{1.2pt}
\textbf{Interconnect}     & Infiniband (Fast network fabric)
            \\ \hlinewd{1.2pt}
Rate                      & 40 Gb/sec (4X QDR)                       \\ \hlinewd{1.2pt}
\end{tabular}
}
\caption{Experiments platform \textbf{``Keeneland''}\space \tnote{1}}
\label{table:kidsplatform}
\begin{tablenotes}
\item [1] \url{http://keeneland.gatech.edu/}
\end{tablenotes}
\end{threeparttable}
\end{table}

\section{Results}
\label{resultsinstructionsresults}

For this work we are going to focus in three main points of our development. This will help us to identify our two main goals: compression and instruction overhead. They are described below.

\begin{enumerate}
	\item \textbf{\emph{Baseline}} was our first BFS implementation. It its studied in the section Analysis. The Baseline implementation does not compress data and is a good scenario to study traditional overhead reduction on a graph500 application  
	\item \textbf{\emph{No-compression}} is our latest version without compression. The later has been disabled at compile time and introduces zero overhead. Measuring the difference in performance between this and the previous one (none of them have compression capabilities) is a good chance to measure the instruction-oriented improvements.
	\item \textbf{\emph{Compression}} same code the prior one. It has its compression capabilities enabled. 
\end{enumerate}

Here follows a description of the diagrams listed below. The diagrams have been generated from the trace files resulting from the executions. The three main types of graphs shown here are (i) strong scaling diagrams (ii) weak scaling (iii) time breakdowns. To know what conclusions we are getting from each type of graph they will be described.

\begin{enumerate}
	\item \textbf{\emph{Strong scaling}} keeps some value fixed and shows evolution over other parameter. To explain this lets say we use in our system two variables. We can increase the (1) processors or the (2) size of our problem. In Strong scaling we set one with a fixed value. The we see the evolution of the other one. This kind of diagrams let us now how the system behaves with a big peak of load. Lets set this analogy: I have a computing device and a program running that is taking more load each time. The strong scaling shows me a ``vertical scalability'', i.e. whether my system is going to function without getting blocked. In this kind of diagrams we are looking forward to see a ascending line, which tell us that the system can manage the load.
	\item \textbf{\emph{Weak scaling}} Analogous to the previous one, we may vary both of the parameters. In such a way that the x-axis stays constant to that increasement: the ratio of the two increased variables stays constant. We can see this as an ``horizontal scaling'' where we can see how our system behaves against many parallel tasks. We can see this with an easy example. This would be the scalability of a ``modern cloud'' when thousand of millions of clients access at the same time. The previous scaling would be the behaviour of an individual computer system agains a very high load. When we look at this type of diagrams we expect to see a horizontal line that shows that our system is capable to manage the load and stay even.
	\item In the \textbf{\emph{time breakdown}} we will chunk into pieces the our application and visualize through bar-plots the times of each region. This will give insights of the bottlenecks of the application.  
\end{enumerate}

The next part of the chapter will be an individual exam of each of the figures. The analysis will be done afterwards, so at this point just some clarifications will be given. The figures listed below are (A) 3 Strong scaling analysis (B) two weak scaling figures (C) two time breakdown diagrams.

\begin{figure*}[htp]
  \centering
  \subfigure[The three scenarios in strong scaling.]{\includegraphics[scale=0.45,viewport=0pt 0pt 480pt 450pt,clip]{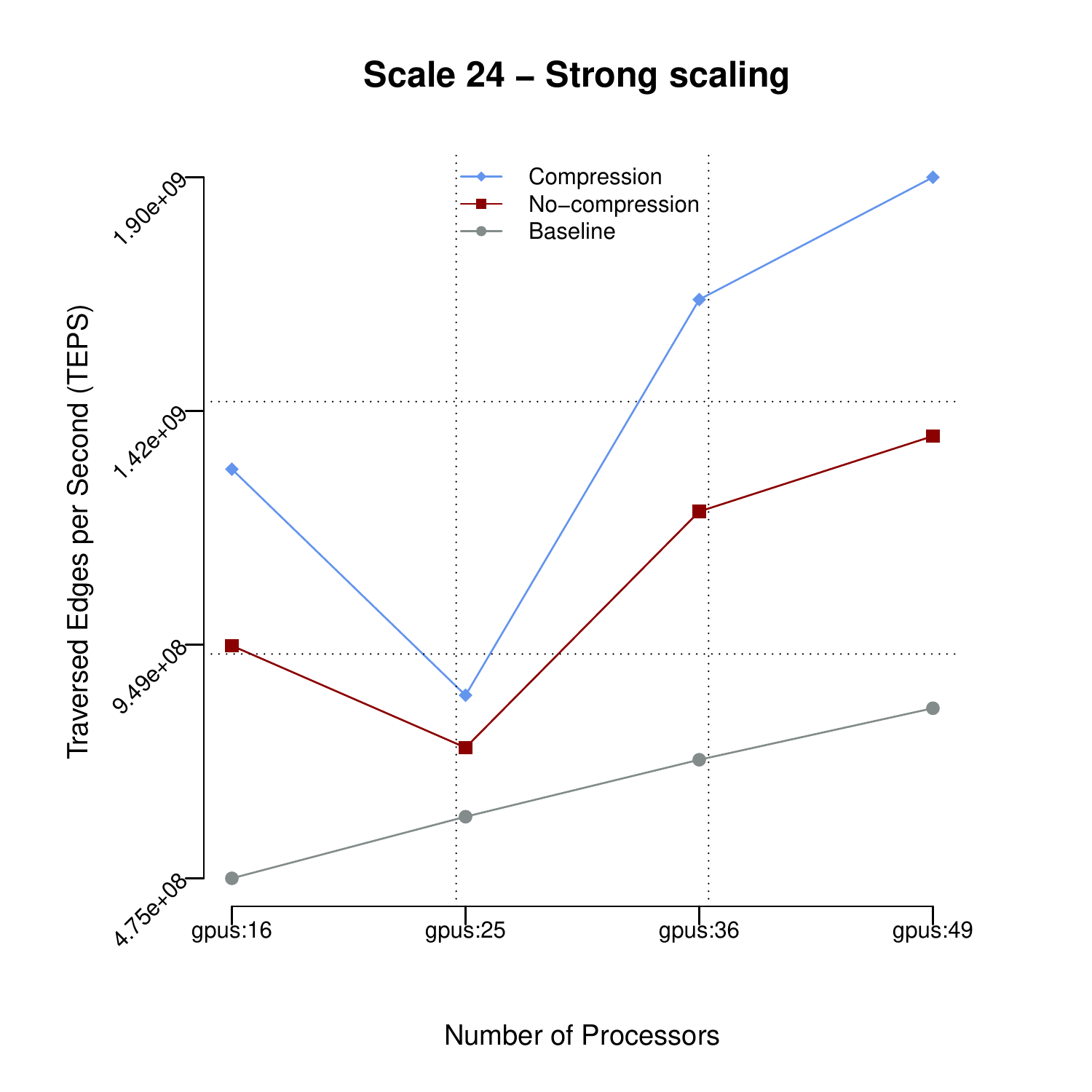}}\quad
  \subfigure[diagram showing Time in strong scaling.]{\includegraphics[scale=0.45,viewport=0pt 0pt 480pt 450pt,clip]{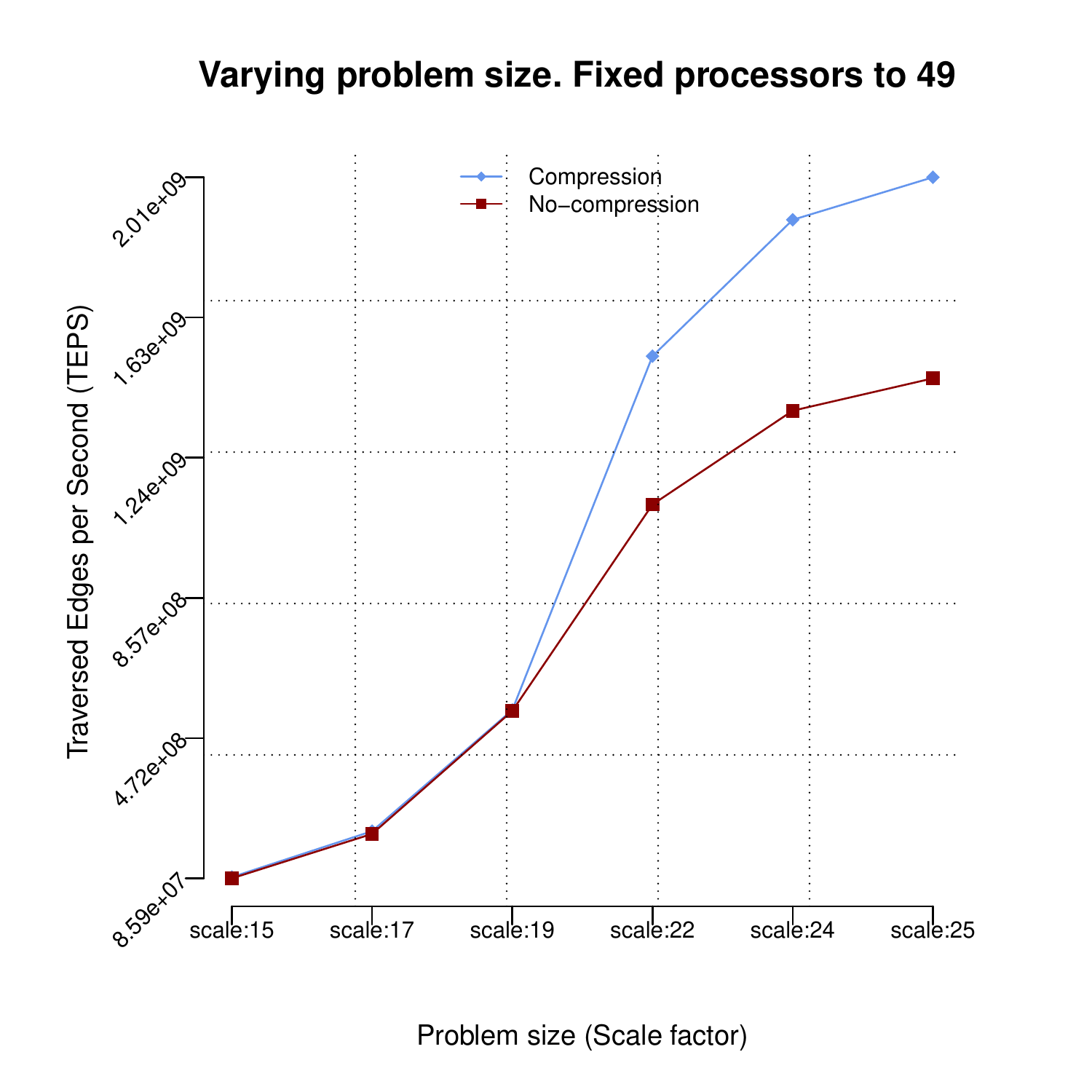}}%
  \subfigure[Number of processors stays fixed.]{\includegraphics[scale=0.45,viewport=0pt 0pt 480pt 450pt,clip]{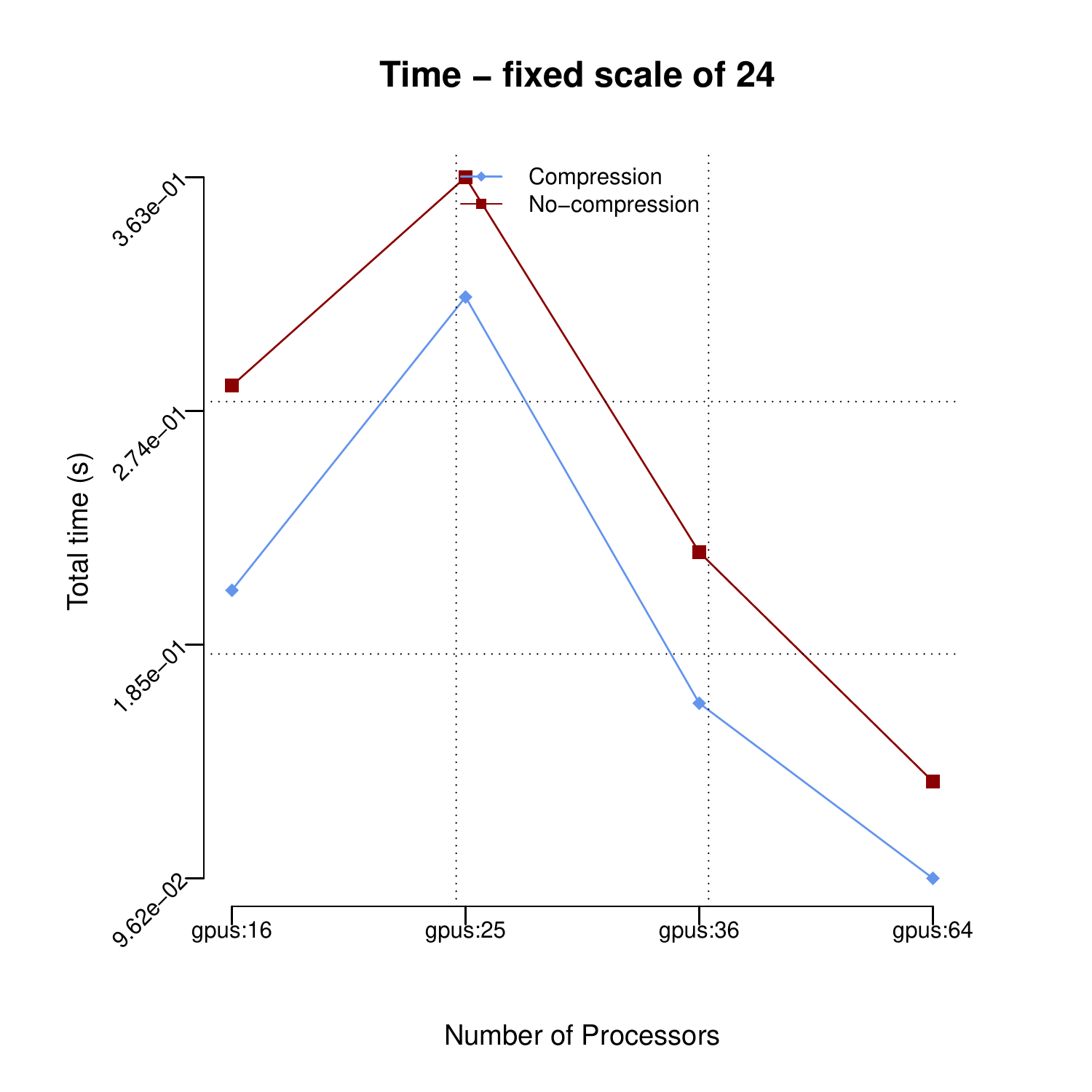}}%
  \caption{}
  \label{fig:strong1}
\end{figure*}

\begin{figure*}[htp]
  \centering
  \subfigure[Weak scaling. Times]{\includegraphics[scale=0.40,viewport=0pt 0pt 480pt 450pt,clip]{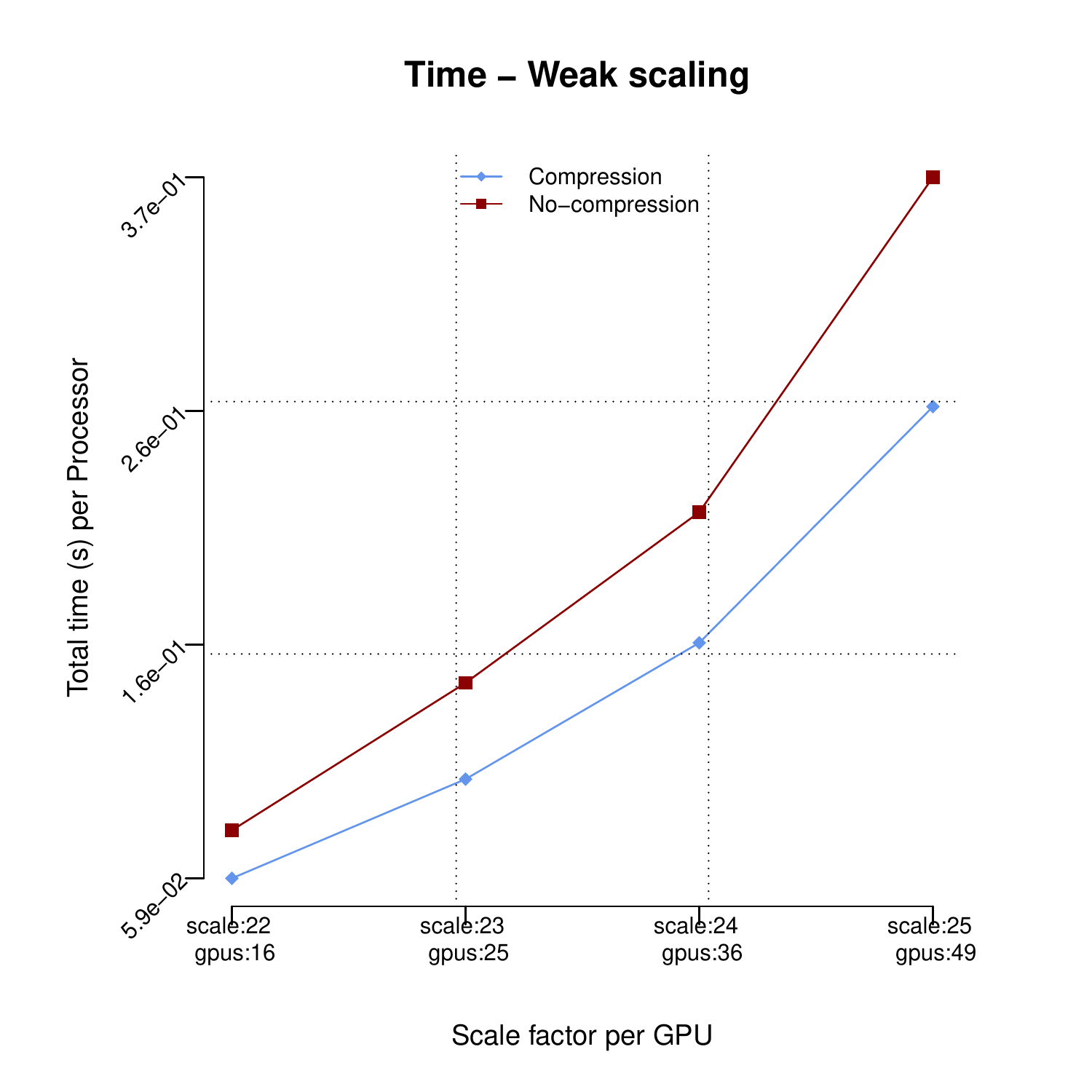}}\quad
  \subfigure[Weak scaling. TEPS]{\includegraphics[scale=0.40,viewport=0pt 0pt 480pt 450pt,clip]{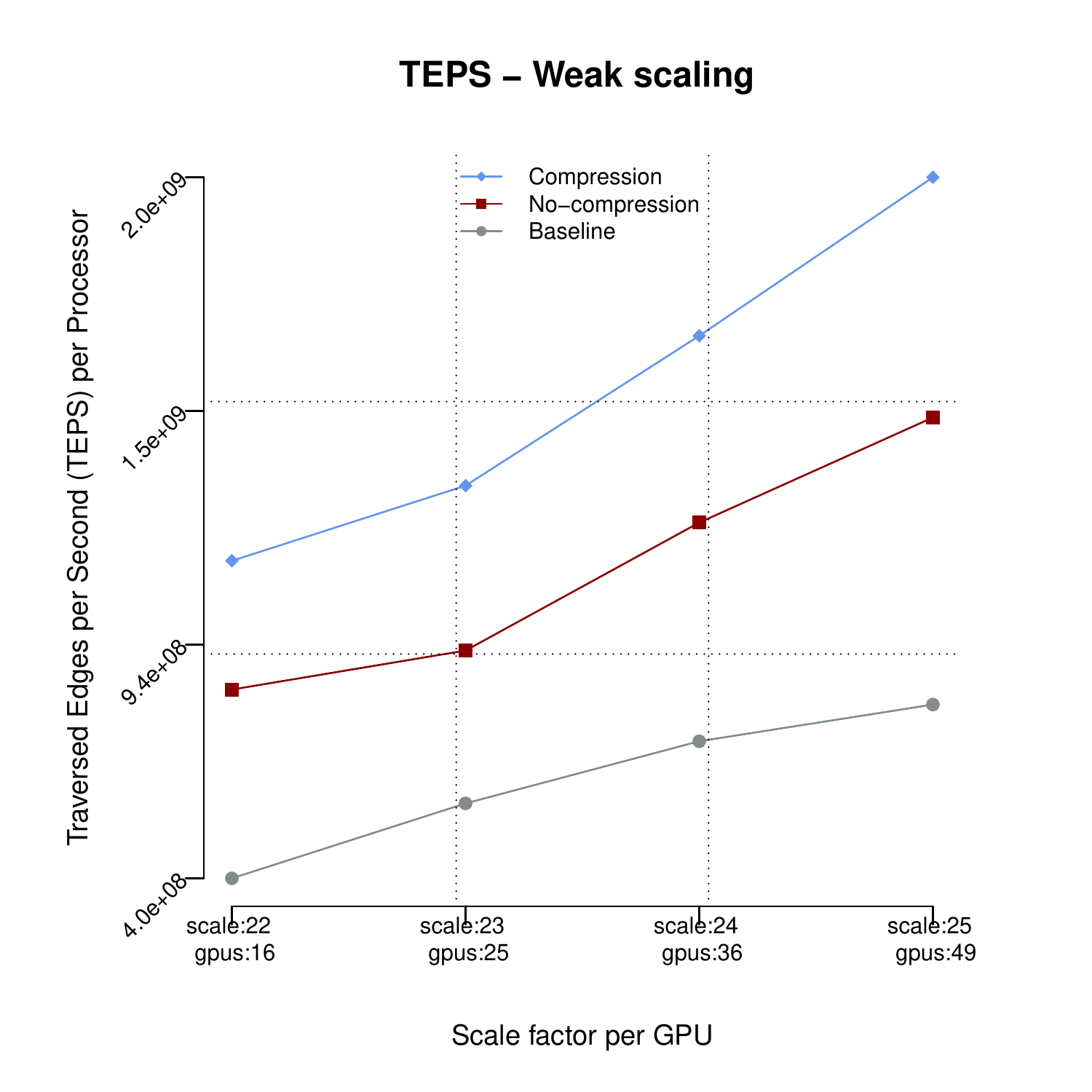}}%
  \caption{}
  \label{fig:weak1}
\end{figure*}

\begin{figure*}[htp]
  \centering
  \subfigure[Time breakdown. Compression disabled]{\includegraphics[scale=0.40,viewport=0pt 0pt 480pt 450pt,clip]{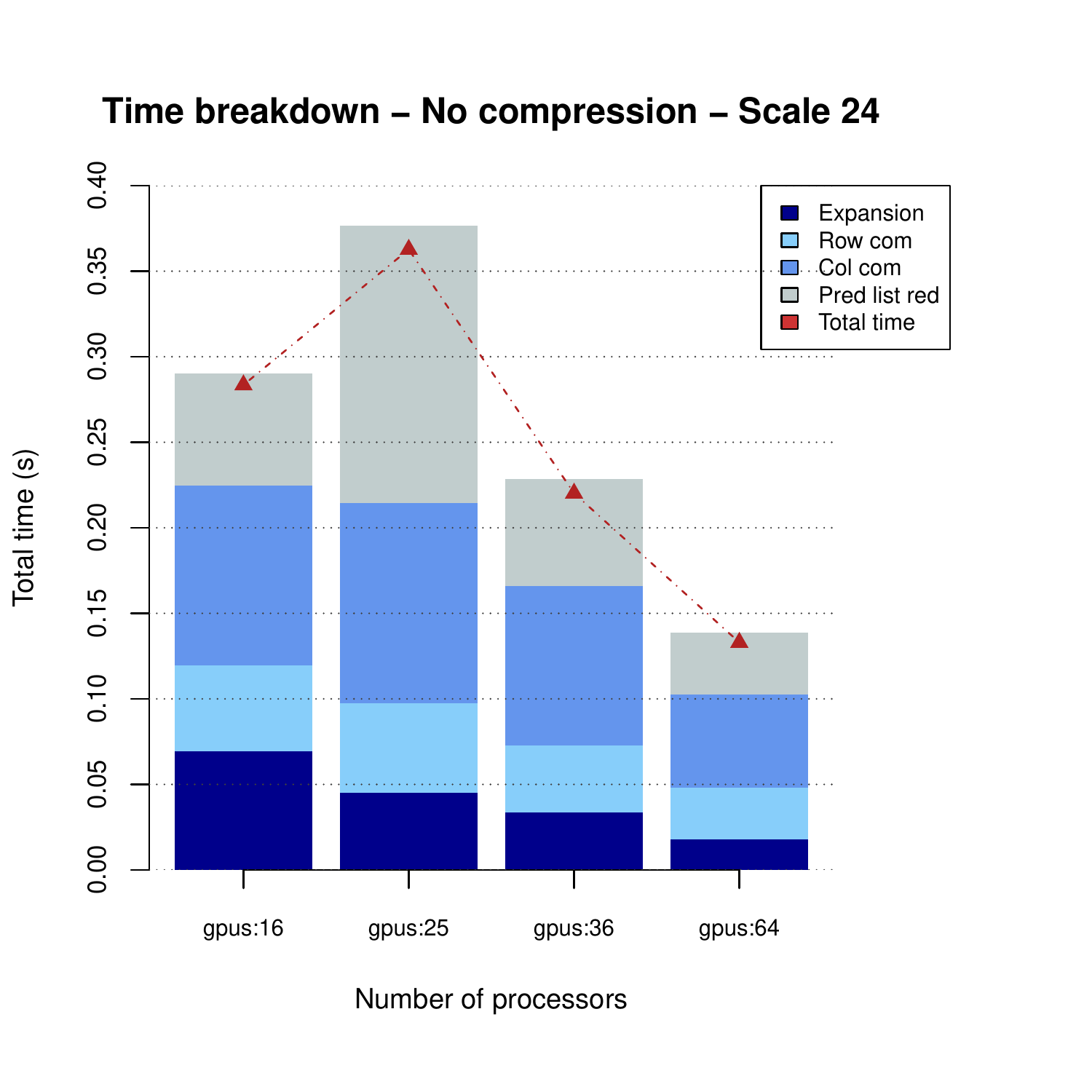}}\quad
  \subfigure[Time breakdown. Compression enabled]{\includegraphics[scale=0.40,viewport=0pt 0pt 480pt 450pt,clip]{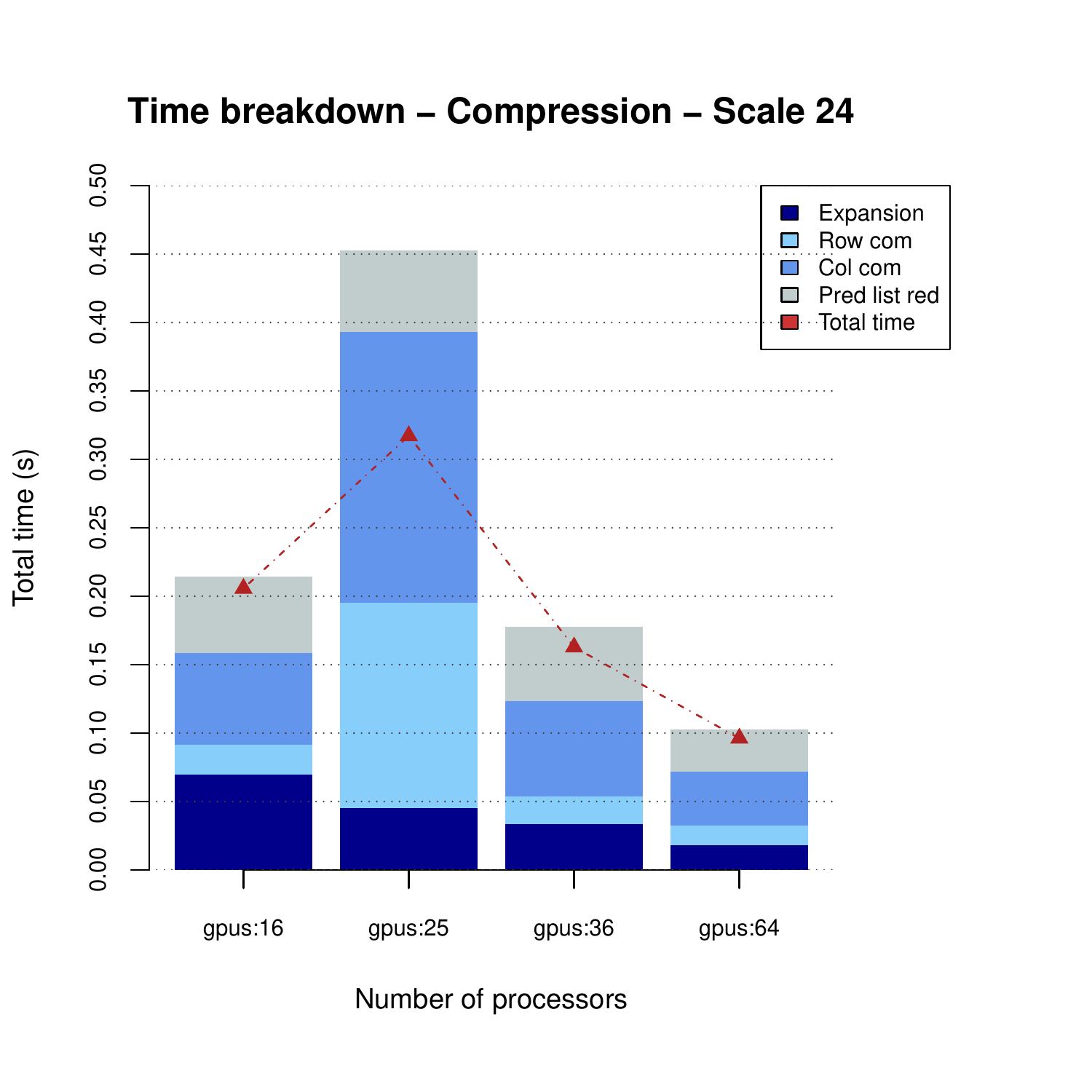}}%
  \caption{}
  \label{fig:breakdown1}
\end{figure*}

\begin{enumerate}[label={\Alph*}.]
	\item \textbf{Strong scaling diagrams}
	\begin{enumerate} 
		\item \textbf{Figure \ref{fig:strong1} (a)} 	shows the behaviour of the application for each one of the three scenarios on setting a high scale factor fixed and observing the evolution at each number of processors. In this diagram we prefer a high stepped ascending line.
		\item \textbf{Figure \ref{fig:strong1} (b)} is analogous to the previous case. The Baseline is not shown as we currently don't have data of times at these time (we do have of TEPS). The measured parameter here is the evolution of time at different processor-number scenarios. In this case we would prefer a stepped line again. However, because we are measuring time we want this to be descending.
		\item \textbf{Figure \ref{fig:strong1} (c)} shows an different picture of what is happening. We set the number of processor fixed (to one of the highest our application can reach). Then we observe the results for different problem sizes. Again, we look forward a stepped ascending line.  
	\end{enumerate}
	\item \textbf{Weak scaling diagrams}
	\begin{enumerate}
		\item \textbf{Figure \ref{fig:weak1} (a) and \ref{fig:weak1} (b)} show both, the variation of the number of processors and the variation of the problem size in the same axis (X). The scaling is used in logarithmic scale. We do not use a constant number of vertices or edges per increased number of processors. This is due to limitations in our experiments and the fact that since we are using a 2D partitioning of the graph, the required number of processors grows exponentially. Even-though we do not keep the ratio constant we approximate this thanks to the logarithmic increasement of the scale. Regarding the diagrams we measure both Time and TEPS. Again, we would prefer a plain horizontal line as result. As it has been previously noted we are not keeping our ratio constant, so son ascending skewness will be normal. 
	\end{enumerate}
	\item \textbf{Time breakdown diagrams}
	\begin{enumerate}
		\item \textbf{Figure \ref{fig:breakdown1} (a) and \ref{fig:breakdown1} (b)} will let us see one of the main goals of this work: the effect compression directly over each zone of the application 
	\end{enumerate}
\end{enumerate}

On the next section we will analyse these results and prepare the Conclusion in the next chapter.

\subsection{Scalability analysis}
\label{resultscompressionresults}

To get conclusions about scalability we are first choosing the \texttt{Compression} scenario and the \texttt{No compression} one. We will be leaving the \texttt{Baseline} for the next subsection. Regarding the diagrams, we will be looking at the Strong and Weak scalability diagrams (the time breakdown will be used deduct conclusions about compression).

As we explained at the beginning of this chapter in the strong types of diagrams we will prefer stepped ascending lines (for TEPS) and descending (for Times. The more degree in the line, the best. Looking at Figures \ref{fig:strong1} (a) and \ref{fig:strong1} (b) we see the stepped lines. The scenarios shows a better capability to perform high in the case of compression when compared to no compression. This, regardless of being the main BFS algorithm and the state-of-the-art improvements included on this, the one that will let grow more stepped in the diagram. 

We are going to start noticing an important behaviour of our application, from now own so it is going to be now when it is going to be described. We will notice that some odd scales (and also number of processors. The explanation follows) produce a very stepped low peak (in the case of TEPS) or a ascending peak in the case of Times. The reason for that lays on the management of odd units (processor ranks or matrixes with odd number of elements). We are partitioning our symmetric \texttt{N}x\texttt{N} matrix on a 2D fashion. For doing this some ranks of the initial \texttt{MPI\_COMM\_WORLD} will be designed to columns and some others to rows. Then, and from now on the communication happens using these rank partitions. The problem arises when we have to divide by 2 the rank and the number is odd. Some nodes the number of ranks on each node will be different. In our application we make use of a custom \texttt{MPI\_Allreduce} algorithm \cite{mpi-allreduce-optimized} which makes use of Point to Point MPI calls for communication. When that implementation is taken to our application and we have to deal with the odd ranks, we have to create the concept of ``residuum'' to deal with this. Within our application we make use of residuums in two points: (i) in the column communication (ii) in the predecessor list reduction step which occurs at the end of each BFS iteration implementations. The residuum approach has two mayor downsides. In a similar way matrixes where \texttt{N}x \texttt{N} is an odd number adds complexity when dividing the Matrix in Row and columns (the matrix are sparse-binary and operated through bitwise operations) 

\begin{enumerate}
	\item \textbf{Complexity}. The residuum is performed through the addition of many extra loops, branches, conditions, MPI difference cases, special cases when dealing with the bitmaps, more code in the Matrix structure class to deal with this problem, and so on and so forth. All this sums up and makes our implementation very complex and prone to crashes. In fact, we know that some scales (which turn out to be odd) will not run without crashes. (scales 17 and 25).
	\item \textbf{Low performance} The two places where these residuums are located are the column and predecessor reduction steps. In the case of the latter, due to the mechanism to deal with the residuum we can not add compression. As a result the transmitted data in this phase, which takes time only once during the BFS iteration, sums more transmitted data (11 times in for \(2^{16}\) Edges) than the whole transmitted (compressed) during the rest of the iteration.
\end{enumerate} 

We see a good scalability Strong and Weak fashions for the tested data, using compression.

\subsection{Overhead of the added compression}
\label{resultscompressionoverhead}

To measure the added overhead by integrating compression \footnote{our BFS implementation is a proof-of-the-concept implementation to test compression using 3rd party compression libraries and codecs.}, we performed tests on a third system. 

Briefly this is a multiGPU platform with a NUMA architecture: two processors. Each of the processors has attached 8x NVIDIA K80 GPUs. For this experiment we expect that the GPU-BFS kernel uses direct card-to-card communication using the network device (RDMA), avoiding that way the cpu-memory path.
As our compression is allocated in the CPU execution layer (we even use the CPU vectorization to introduce data-parallelism), we are going to be able to separate the execution of the BFS (GPU+GRAM+PCIe3+Network device+RDMA) with the rest of the computation, buffer management, BFS queues operated with the CPU, and lastly the compression calls. This latter operations will make use of the 2 CPUS, 2xRAM (one allocated to each processor), and a QPI high speed bus as interconnection. 

Because the speed of this setup is very high and the latencies of its ``network'' (will be mostly bus communication: PCIe 3 and QPI) leave not much space to gain speed by compressing data, we are dealing with the perfect scenario to test how much latency are we adding. First of all to understand the introduced overhead, they are listed and explained in further detail below. 

\begin{enumerate}
	\item the obvious computational cost of the compression (which is pretty low in the chosen compression scheme, see section \ref{compression})
	\item the data conversion before and after compression to adapt it to the requirements of the 3rd party compression libraries. This data conversions have a high memory footprint cost which may be greater than the computational cost of the compression.
	\item NUMA architectures incur in an extra cost as their processors may access banks of memory allocated in other processor. The cost of these accesses is expensive \cite{numaAwareFujisawa2015}. This add up to the expensive memory footprints from the previous buffer copies. 
	\item Due to the fact of this being a compression codecs test-bench, the compression libraries have been integrated modularily. Unfortunately, many times ``easy to use'' equals low performance. The point being that the ``modularity'' has been implemented in C++ using a kind of hierarchy called  (\texttt{virtual} keyword) which allows to choose the inherited class of other class at runtime. This prevents the compiler from knowing what class will be used, and to inline it. Each compression call using the Lemire et al. library incurs in two virtual call (the \texttt{virtual} keyword is also used in Lemire et al.). 
 \end{enumerate}
 
As expected the overhead has been detected in the previous implementation. Also, we have noticed performance loss in other situation: In communication between MPI ranks using shared memory. The performance which compressions in these cases is lower to the one between ranks allocated in different machines (rank communicating through RDMA do not enter here)

The problem could be solved easily without many changes with the use of the implemented compression \texttt{threshold} (Section \ref{compression}). At the beginning of the application it could be implemented a quick MPI\_COMM\_WORLD ranks test to detect low compression performance between ranks. We give more details aboit this idea in the Future  work section (Section \ref{futurework})

\subsection{Instruction overhead analysis}
\label{resultscompressionresults}

In this section, we will compare our \texttt{Baseline} implementation with the transformed application (leaving off the compression capabilities).

We have focused the optimizations on Vectorization. We intend to make use of the Vector Processor Unit (VPU) included in the CPU of modern systems. As an example, a x86 system with SSE\texttrademark \space a 128-bit width VPU, would be able to operate with 4x 32-bit integers at a time. The performance boost is 4\texttt{X}. 

Other aspects we have focused are the efficient memory access, and thread parallelism.

To see the successfulness of our optimizations we will be comparing also the scaling diagrams. This time we will be focusing on the Baseline and the No-compression versions. 

From what it can be seen in the results the No-compress versions outperforms the Baseline in all scenarios (low and high scales , few and many processors). Even thought these results show success in these changes, it is difficult to assert what changes produced the better (or even some) benefit.

\subsection{Compression analysis}
\label{sectionfinalresultscompressionoptimizations}

Compression is the main aim of this work and for better gathering conclusions from the results we will be using other sources of data, apart from the diagrams generated in this experiments. The diagrams responsible from giving us information about the successfulness (or not) of the compressed data movement are the time breakdowns (Figures \ref{fig:breakdown1} (a), \ref{fig:breakdown1} (b)). The zones where  the compression could be applied are three: (i) row communication, (ii) column communication and (iii) predecessor list reduction.

Zones (i), and (ii) have been made capable of compressing their data. For the case of (iii) several reasons have made it difficult to have it compressed in this work. For that matter, we present an algorithm for performing that step with compressed data transfers. 

The efficiency of the compression has also been measured using a more accurate procedure in the cluster Creek, in  University of  Heidelberg. The downside, perhaps of using this cluster is that its only 16 GPUs do not allow profiling with big scale transfers. 

Next follow the tables generated from the instrumentation results using the profiler. \medskip

\begin{table}[hbt]
\centering
\adjustbox{width=0.95\columnwidth}{%
\begin{tabular}{llll}
\hlinewd{1.2pt}
        & Initial Data (Bytes) & Compressed  data (Bytes) & Reduction (\%) \\ \hlinewd{1.2pt}
Vertex Broadcast                                    & 8192                 & 8192                     & 0,0\%          \\
Predecessor reduction  & 7.457.627.888        & 7.457.627.888            & 0,0\%          \\
Column communication         & 7.160.177.440        & 610.980.576              & 91,46\%        \\
Row communication          & 4.904.056.832        & 403.202.512              & 91,77\%          \\ \hlinewd{1.2pt}
\end{tabular}
}
\caption{Reduction with compression in terms of data volume. Measurement with Scalasca and ScoreP. Experiment on Creck platform, scale 22, 16 GPUs}
\label{compressionvolumes}
\end{table}

\begin{table}[hbt]
\centering
\adjustbox{width=0.95\columnwidth}{%
\begin{tabular}{llll}
\hlinewd{1.2pt}
      & Initial time (s) & Compression times  (s) & Reduction (\%) \\ \hlinewd{1.2pt}
Vertex Broadcast                & 0.162604         & 0.159458               & 0,0\%          \\
Predecessor reduction           & 164.169409       & 154.119694             & 0,0\%          \\
Compressed column communication & 156.615386       & 31.035062              & 80,18\%        \\
Compressed row communication    & 79.772907        & 13.497788              & 83,07\%        \\ \hlinewd{1.2pt}
\end{tabular}
}
\caption{Reduction with compression in terms of data volume. Measurement with Scalasca and ScoreP. Experiment on Creck platform, scale 22, 16 GPUs}
\label{compressiontimes}
\end{table}

\chapter{Conclusions}
\label{sectionconclusions}

The conclusions of this work are based on the subsection \emph{Analysis} in the previous section \emph{Results} (Sections \ref{sectionanalysiscommunication} and \ref{sectionanalysisinstructions}).

In this work we have reviewed (i) how does a compression based on ``Binary Packing'' (Section \ref{compression}) improves a distributed BFS algorithm (ii) The effects of applying the traditional instruction optimizations to the same distributed algorithm. 

Regarding whether or not have we successfully met our goal set in chapter \ref{problemanalysis} of improving the overall performance of the \texttt{Baseline} implementation it could be stated that yes. For the analyzed scenario, and using a relatively high scale and a relatively high number of processors we have significantly increased 
\begin{enumerate}
	\item \textbf{the ``horizontal'' scalability}: our distributed BFS distributes better the load among the nodes. This is an indirect effect of improving the communication latency.
	\item \textbf{the ``vertical'' scalability}: our algorithm can deal with bigger amounts of load on an unique processor. 
	\begin{enumerate}
		\item \textbf{the direct reason} for this are the  instruction overhead optimizations (vectorization, memory access and thread parallelism). For the particular case seen on Figure \ref{fig:strong1} (a), the direct improvement is constant and stays over the 100\% (relation between \texttt{non compressed} and \texttt{baseline}).
		\item \textbf{the indirect reason} for this to happen is the previous optimization in the compression: since the bottleneck in the algorithm is the communication, each of the processors stay still a certain amount of time awaiting to receive the Frontier Queues from its neighbours. The Figure \ref{fig:strong1} (a) in previous chapter shows a 200\% improvement of the \texttt{compression} implementation in relation to the \texttt{baseline}.  
		\end{enumerate}
\end{enumerate}

\chapter{Future work}
\label{futurework}

We believe that some good steps to follow in order to develop a good Graph 500 implementation would be:
\begin{enumerate}[label={\Alph*}]
	\item \textbf{In case we want to run our BFS on GPGPU}
		\begin{enumerate}
		\item Use the Graph 500 implementation \textbf{SC12 (Ueno et al.)} \cite{ueno-et-al} for the BFS kernel. This implementations was already top 1 in the list in the year 2013. (Section \ref{relatedwork}).
		\item The previous implementation already compress its compression. their used scheme named Variable-Length Quantity \emph{VLQ} belongs to the Varint family (Section \ref{compression}). Two characteristics of varint codecs are (i) that they do not perform well with big cardinalities, (ii) are slower and have smaller compression ration than a FOR + delta compression scheme. This latter about the FOR compression is also only valid if the data meets some criteria (as the varints codecs). The criteria for FOR codecs is ordered sequences of integers (unordered is also possible but requires an initial sorting) with low distances between them. Since the Frontier Queue in the sparse graphs that we are studying meet both criteria, this codec is ideal.
		\item  Our last step would be to add the NUMA-control\footnote{ULIBC - \url{https://bitbucket.org/yuichiro_yasui/ulibc}} library that has given the position \#1 to Fujisawa et al. to deal with the memory movements in our application (including the ones in the compression integration)
	\end{enumerate}

	\item \textbf{In case we want to run our BFS on a NUMA cluster}
	\begin{enumerate}
	\item As, to our knowledge, there is no open state-of-the-art Graph500 implementation, we would need to develop it based on the paper of the Top 1 implementation \cite{numaAwareFujisawa2015}.
	\item Since the current \texttt{Top 1} does not perform compressed movement of data, and since we also know that the algorithm includes SpMV, CSR and 2D graph partitioning, it would be a good fit for the compression scheme presented here.
	\item The last step would also be to add the NUMA control library that they feature to the memory structures in our \emph{Compression integration}
	\end{enumerate}

\end{enumerate}

As we partially described in section \ref{sectionfinalresults} when dealing with the problem of the compression-added overhead at some cases, a solution would be to learn about the ranks in the topology and know whether or not apply compression. A more optimum approach for the second case described in that section (nodes with shared memory, there the compression will not have any benefit, but only overhead) would be instead of (1) add an initial discovery cycle (2) broadcast results (3) manage structures. we could do the inverse way: since the 2D partitioning allows manually to create the partitioning for which nodes would go in the row column and the same for the row, we could keep the initial steps: (1) initial discovery of the topology (2) broadcast of the data, and now (3) would be: setup the application rank based on the map: more amount of communication \( \longrightarrow \) different MPI rank group (rows or columns) \medskip

Other proposed work related with compression, but in a different scope,   would encompass the use a Lempel Ziv fast compression algorithm to compress the visited neighborhood bitmap. This has been proved to be successful in other previous work (Huiwei-et-al \cite{compressingbitmap-sieve}) and results on a great communications transfer reduction. However, as our bitmap transfer is small (as proven in instrumentation results, Section \ref{sectioninstrumentation}) the implementation-cost vs performance-benefit balance must be previously evaluated. \medskip

The usage of an \emph{Hybrid} (MPI + OpenMP) model for parallelizing the collective operations using spare cores, shows an experimental speedup of 5\(\texttt{X}\) in other previous works in real world applications \cite{fourteen}. The revision of the dynamic task scheduling of our OpenMP thread model may also add an extra performance. \medskip 

One feature wich differes with many others is our implementation is a custom \emph{MPI\_Allreduce} function that we is used to reduce the column communication. The implementation is based on the Rabensefner-et-al algothithm \cite{mpi-allreduce-optimized} which is dated in 2005. As it was described in the section \ref{sectionfinalresults}, this algorithm uses some point-to-point communication calls which make neccessary the creation of an special case called residuum for some scenarios where \texttt{|Rows x Columns|} is odd (Section \ref{sectionfinalresults}). The usage of a more new algorithm, with the possibility of being implemented over GPGPU  may be a good improvement, as done in the work of Faraji-et-al \cite{five}. \medskip

Another possible addition to our implementation is the new acceleration framework OpenACC\footnote{\url{http://www.openacc.org/}}. It has been successfully proved to work for other high computing problems \cite{openacc}, and the results may be interesting on Graph500\footnote{\url{http://www.graph500.org/}} . \medskip

%% file: references.tex

\renewcommand\bibname{References}
\nocite{}
\def\newblock{\hskip .11em plus .33em minus .07em} 


{
  \sline
  \bibliographystyle{abbrv}
  
  \bibliography{references}{}
}

%% file: acknowledgments.tex


\thispagestyle{empty}
  \addcontentsline{toc}{chapter}{Acknowledgments}
  \thispagestyle{plain}
  {\centerline{\bfseries{\LARGE Acknowledgments}}}
  \vspace*{.4in}
  I would like thank the CEG group in the Ruprecht-Karls Universität Heidelberg for having given me this opportunity to challenge myself with this project.
  I would also like to thank all the members of the department for their helpfulness at every moment. \\
  
  Specially, I would like to thank, my advisor in this thesis; to the student previously in charge of this project and from whom I have continued the work; to the colleague and PhD student who has helped me with the german translations in this work; and to the second Professor responsible of the project. The latter, in charge of performing the tests (many times during weekends and Bank holidays).

%% file: deposition.tex
\thispagestyle{empty}
\addcontentsline{toc}{chapter}{Deposition}

\setlength{\parindent}{0em}

Erkl\"{a}rung:\par
\vspace{3\baselineskip}
Ich versichere, dass ich diese Arbeit selbstst\"{a}ndig verfasst habe und keine
anderen als die angegebenen Quellen und Hilfsmittel benutzt habe.\par
\vspace{5\baselineskip}
Heidelberg, den (Datum)\hspace{3cm}\dotfill